\documentclass[aps,prb,twocolumn,superscriptaddress,showpacs,notitlepage,a4paper,9pt]{revtex4-1}
\pdfoutput=1
\usepackage{color}
\usepackage{amssymb}
\usepackage{amsmath}
\usepackage{graphicx}

\newcommand{\bea}{\begin{eqnarray}}
\newcommand{\eea}{\end{eqnarray}}
\newcommand{\bei}{\begin{itemize}}
\newcommand{\eei}{\end{itemize}}
\newcommand{\be}{\begin{equation}}
\newcommand{\ee}{\end{equation}}
\newcommand{\bse}{\begin{subequations}}
\newcommand{\ese}{\end{subequations}}
\newcommand{\bfg}{\begin{figure}}
\newcommand{\efg}{\end{figure}}

\newcommand{\eins}{\mbox{$1 \hspace{-1.0mm} {\bf l}$}}

\DeclareRobustCommand{\sMRKsq}{{4_1^\mathrm{s}}}
\DeclareRobustCommand{\mMRKsq}{${4_1^\mathrm{s}}$}
\DeclareRobustCommand{\bMRKsq}{${4_1^\mathrm{s}}$}

\DeclareRobustCommand{\sMRKci}{{2_1^\mathrm{s}}}

\DeclareRobustCommand{\bMRKci}{${2_1^\mathrm{s}}$}

\DeclareRobustCommand{\sMRKstFi}{{8_4^\mathrm{n}}}
\DeclareRobustCommand{\mMRKstFi}{${8_4^\mathrm{n}}$}
\DeclareRobustCommand{\bMRKstFi}{${8_4^\mathrm{n}}$}

\DeclareRobustCommand{\sMRKrb}{{4_1^\mathrm{n}}}
\DeclareRobustCommand{\mMRKrb}{${4_1^\mathrm{n}}$}

\DeclareRobustCommand{\mMRKstFo}{${8_4^\mathrm{s}}$}
\DeclareRobustCommand{\sMRKstFo}{{8_4^\mathrm{s}}}

\DeclareRobustCommand{\mMRKstSi}{${16_6}$}
\DeclareRobustCommand{\sMRKstSi}{{16_6}}

\DeclareRobustCommand{\mMRKtrr}{${2_1^\mathrm{n}}$}
\DeclareRobustCommand{\sMRKtrr}{{2_1^\mathrm{n}}}

\DeclareRobustCommand{\bMRKtru}{${4_2^\mathrm{n}}$}
\DeclareRobustCommand{\mMRKtru}{${4_2^\mathrm{n}}$}

\DeclareRobustCommand{\bMRKtrd}{${4_2^\mathrm{s}}$}
\DeclareRobustCommand{\mMRKtrd}{${4_2^\mathrm{s}}$}

\newif\ifpdf\ifx\pdfoutput\undefined\pdffalse\else\pdfoutput=1\pdftrue\fi

\newcommand{\smeq}{\! = \!}

\newcommand{\smap}{\! \approx \!}
\newcommand{\smneq}{\! \neq \!}
\newcommand{\smpl}{\! + \!}

\newcommand{\ci}{\mathrm{i}}

\newcommand{\zver}{{\hat{\bf z}}}

\newcommand{\braket}[1]{\left<#1\right>}

\newcommand{\ketLR}[1]{\left|#1\right>}

\newcommand{\braLR}[1]{\left<#1\right|}

\begin{document}
\title{Dephasing and Hyperfine Interaction in Carbon Nanotubes\\ Double Quantum Dots: The Clean Limit}
\author{Andres A. Reynoso}
\affiliation{Niels Bohr International Academy, Niels Bohr
Institute, Blegdamsvej 17, 2100 Copenhagen \O, Denmark}
\affiliation{Niels Bohr Institute \& Nano-Science Center, University of Copenhagen, Universitetsparken 5, 2100 Copenhagen, Denmark}
\author{Karsten Flensberg}
\affiliation{Niels Bohr Institute \& Nano-Science Center, University of Copenhagen, Universitetsparken 5, 2100 Copenhagen, Denmark}
\date{\today}
\begin{abstract}
We consider theoretically ${}^{13}$C-hyperfine interaction
induced dephasing in carbon nanotubes double quantum dots with curvature induced
spin-orbit coupling. For two electrons initially occupying a
single dot, we calculate the average return probability after
separation into the two dots, which have random nuclear-spin
configurations. We focus on the long time saturation value of
the return probability, $P_\infty$. Because of the valley degree
of freedom, the analysis is more complex than in, for example,
GaAs quantum dots, which have two distinct $P_\infty$ values
depending on the magnetic field. Here the prepared state and the measured state is non-unique because two electrons in the same dot are allowed in six different states. Moreover, for one electron in each dot sixteen states exist and therefore are available for being mixed by the hyperfine field. The return probability experiment is found to be strongly dependent on the prepared state, on the external magnetic field---both Zeeman and orbital effects---and on the spin-orbit splitting. The lowest saturation value, being $P_\infty=1/3$, occurs at zero magnetic field for
nanotubes with spin-orbit coupling and the initial state being
the groundstate, this situation is equivalent to double dots
without the valley degree of freedom. In total, we report nine dynamically different situations that give $P_\infty=1/3$, $3/8$, $2/5$, $1/2$ and for valley anti-symmetric prepared states in an axial magnetic field, $P_\infty=1$. When the groundstate is prepared the ratio between the spin-orbit splitting and the Zeeman energy due to a perpendicular magnetic field can tune the effective hyperfine field continuously from being three dimensional to two dimensional giving saturation values from $P_\infty=1/3$ to $3/8$.
\end{abstract}
\maketitle
\section{Introduction}
Quantum dots are attractive candidates for implementing qubits;
in particular carbon-based materials such as nanotubes (CNTs) and
graphene provide the advantage of a weaker hyperfine interaction. In general the samples have high concentration of ${}^{12}$C isotopes (spin zero) and low concentration of ${}^{13}$C isotopes (spin-$1/2$), and thus the coupling between the confined electron spin and the nuclei spins is small. However,
the valley degree of freedom increases the number of available
few-particle states in the quantum dots, leading to a more
complex system, as compared to, e.g., GaAs quantum dots.
Breaking of this 4-fold spin and valley degeneracy has been
predicted for nanotubes due to a curvature-induced spin-orbit
coupling,\cite{Ando2000,HuertasGB2006,JeongJL2009,Izumida2009} which has been confirmed experimentally.\cite{Kuemmeth2008,ChurchillFlensberg2009,Jespersen2010} The spin-orbit split spectrum leads to doubly degenerate states
that can be used as qubit states. For example, by taking
advantage of the strong diamagnetic effects in an axial
magnetic field, a spin qubit at larger magnetic fields can be
defined.\cite{Bulaev2008} Alternatively, the low field
spin-orbit entangled Kramers pair can be used as the qubit,
which has been proposed as an electrically manipulatable qubit
in bend nanotubes.\cite{FlensbergMarcus2010}
\begin{figure}[!ht]
   \centering
   \includegraphics[width=.47\textwidth]{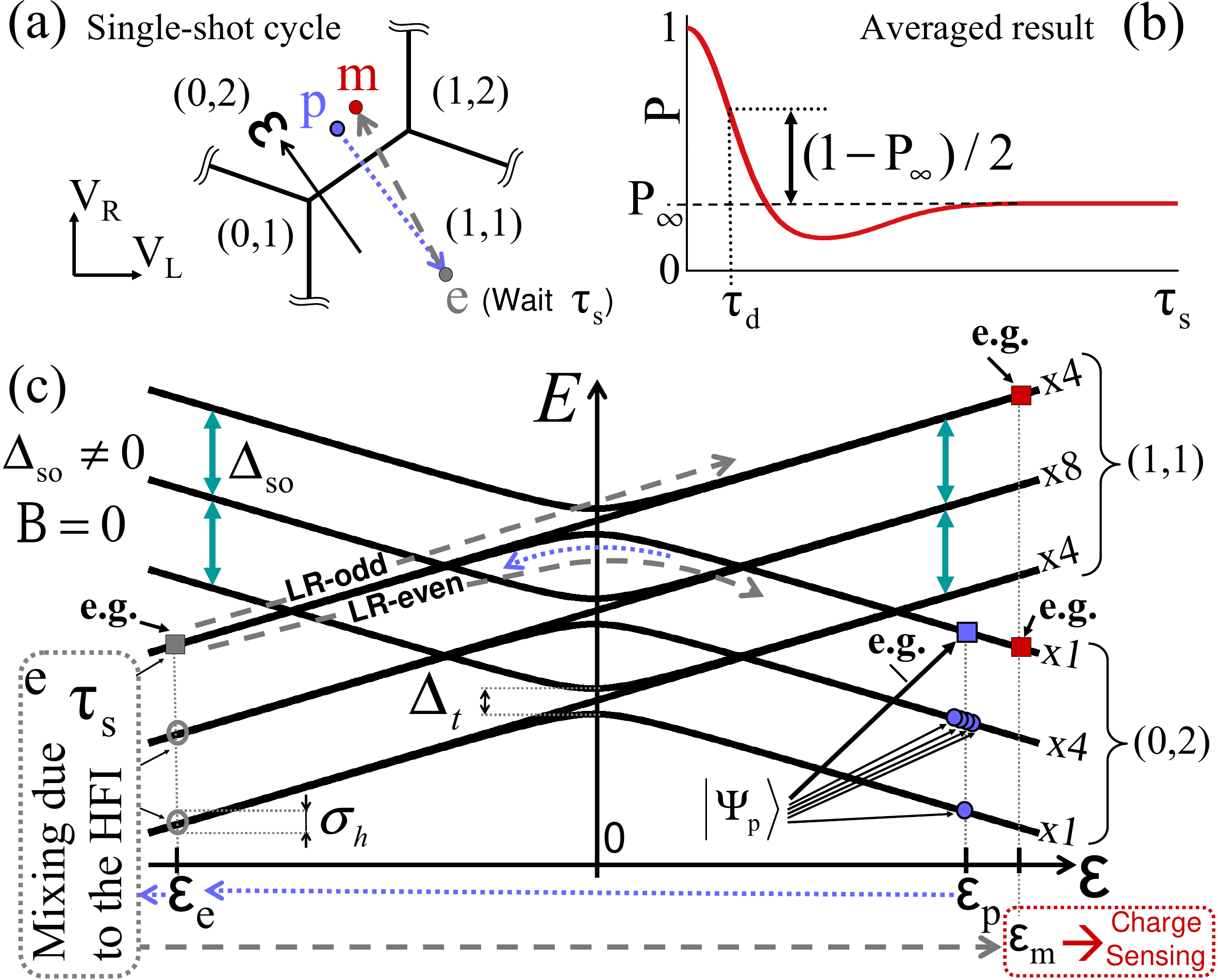}
   \vspace{-0.2cm}
   \caption{ Return probability experiment. (a)
Cycle applied in gate voltage space $(V_L,V_R)$ in each
single-shot measurement. (b) Sketch of the return probability, $P(\tau_s)$, as a function of the time waited, $\tau_s$,  at ``e"; the saturation value, $P_\infty$, and the decaying time, $\tau_{d}$, are included. (c) Spectrum of the (1,1) and (0,2) two-particle states (taken for $\Delta_\mathrm{so}\smneq 0$, tunneling gap, $\Delta_{t}$, hyperfine characteristic energy, $\sigma_h$, and $\mathbf{B}\smeq 0$) as a function of the detuning, $\varepsilon$; arrows and detuning values, $\varepsilon_\mathrm{p}$, $\varepsilon_\mathrm{e}$ and $\varepsilon_\mathrm{m}$ refer to the single-shot cycle's stages presented in (a). We label with squares (``e.g") the states that the system may occupy at the stages, ``p",``e" and ``m" if the (0,2) highest energy excited state is prepared; the possible outcomes at $\varepsilon_\mathrm{m}$ would be different if the prepared state, $\ketLR{\Psi_\mathrm{p}}$, had been one of the other five available (0,2) states.}   \label{FG:fig0}
\vspace{-0.2cm}
\end{figure}

As pointed out by Loss and DiVincenzo,\cite{LossD98} one
of the requirements for quantum dot-based quantum computation
is the ability to control the exchange interaction, which has
been successfully demonstrated in GaAs double quantum dot (DQD)
setups.\cite{Petta2005,HansonReview} Similar devices have
more recently been fabricated in CNT
systems,\cite{Watson2009,ChurchillFlensberg2009} which have
also attracted considerable theoretically attention.\cite{WunschDQDCNT2009,WunschDQDCNT2010,WeissFlensberg2010,PalyiB10}

Another key requirement is long decoherence times. Experimental
determination of the spin decoherence times in DQD is done by
converting the spin information into charge information, which
can be measured using on-chip charge detectors. The charge
state of the DQD is characterized by the number of electrons in
the left and in the right dot, $(N_L,N_R)$, which are controlled
by gate voltages applied to the left and to the right dot, $V_L$ and $V_R$, producing a map of the equilibrium charge
states of the double quantum dot, as illustrated in
Fig.\ref{FG:fig0}(a). The figure also shows the cycle that
allows measurement of the dephasing time. This is done by
initially preparing the system in a state with two electrons in
one dot, then separating them for a certain time with one
electron in each dot---while the initial state can be affected
by the environment---and finally by measuring the probability
for the electrons to return to the original dot.

In this paper, we study the return probability experiment,
which measures the characteristic decay time, $T^*_2$, for the
return probability averaged over many cycles, as well as the
saturation value, $P_\infty$, see Fig.\ref{FG:fig0}(b). In the case when the dominant
time evolution is due to hyperfine interaction with a nuclear
spin background that changes between cycles, the $T^*_2$ time
is a measure of the inhomogeneous broadening of the spin state.
This is well-studied in two-dimensional electron gases defined
quantum dots.\cite{SchultenWolynes1978,MerkulovHyperfine,CoishDQD,Petta2005,TaylorDephasingTheory2007,Cakr2008,Harju2009}

Here, due to the valley degree of freedom, the prepared state
with two electrons in one dot is not restricted to one possibility, as the spin singlet in a GaAs dot, but six states.
Furthermore, once the electrons are separated, there are sixteen available states instead of four as in a GaAs double dot (the
spin singlet and the spin triplets). As mentioned above, the
spin-orbit coupling and the diamagnetic effect of an external
magnetic field split up these manifolds of states. In Fig.\ref{FG:fig0}(c) we present, as a function of the detuning, $\varepsilon$, the spectrum of the six  (0,2) states, the sixteen (1,1) states and the mixing between them due to single-particle tunneling. The relation between the detuning and the gate voltages, the nature of the mixing of the states, and the description of the stages in the single-shot measurement cycle, are presented in detail in Sec.\ref{SC:QDmodel} and Sec.\ref{SC:experiment}.

So far one experimental study of $T^*_2$ measurements has been
reported\cite{ChurchillFlensberg2009} on samples with high
concentration of $^{13}$C. The result shows a saturation value
in the return probability of $ \approx 0.17$, which is not
understood. The theory for non-valley degenerated DQDs predicts
(as lower bounds) $1/3$ and $1/2$ in the low and high magnetic
field regimes, respectively. A multivalley case was recently
investigated for a silicon double dots.\cite{CulcerSarmaQD2010}
Depending on the prepared state, either a GaAs-like behavior or
hyperfine immune states may be found. The hyperfine interaction
considered did not, however, involve valley mixing terms in
contrast to the C-based dots we treat here.

The motivation of our detailed study is to predict the expected
return probabilities for carbon based systems with hyperfine
coupling to spin 1/2 $^{13}$C nuclei. In these graphene-based
systems, in contrast to Si double dots, one must take into
account that the hyperfine interaction affects both the spin
and the valley degrees of freedom.\cite{Fischer2009,PalyiB09}
We include those hyperfine valley mixing effects but we do not
include disorder induced spin conserving valley mixing, which
is presented in a separate publication.\cite{PosterKonstanz2010} Furthermore, we include direct Coulomb interaction, but not Coulomb exchange
which is expected to be a small effect.\cite{WeissFlensberg2010}

We also work in the limit of large detuning so that the \emph{tunneling exchange} on the (1,1) states is much smaller than the hyperfine field characteristic energy and therefore we obtain the lower bounds for $P_\infty$. On the other hand, if the tunneling exchange is important, the degeneracies for zero hyperfine are reduced diminishing the effectiveness of the hyperfine-induced mixing and therefore also increasing the saturation values of the return probability. In those situations $P_\infty$ grows continuously, as a function of the tunneling exchange, from the zero-exchange value up to one.\cite{CoishDQD}

We start from a simple model for an isolated quantum dot and
construct from this the two-electron wave functions. We study
the cases of large and small spin-orbit coupling. The result is
found to be dependent on the prepared state and on the external
magnetic field. Notably, for some situations the two-electron
wave function is almost not dephased by the hyperfine field. We
show that besides the usual saturation values of the return
probability, 1/3 and 1/2---well known in DQDs without the valley
degree of freedom and for zero tunneling exchange---other values can be observed; namely, 3/8, 0.4 and 1. In
addition, for nanotubes with spin-orbit coupling, an applied
magnetic field in a direction perpendicular to the tube axis
can tune the saturation value between $1/3$ and $3/8$.

The paper is organized as follows. In section II, we describe
the double dot model and the four special cases used in the
paper. Section III describes the experiment and the methods used to calculate the return probabilities, with results presented in Section IV.
Finally, conclusions and summary are found in Section V.

\section{Quantum dot model}
\label{SC:QDmodel}
\subsection{Single dot}
\label{SC:SingleDot}

We consider semiconducting tubes where the bandgap is due to
either chirality or, for nominally metallic tubes, to curvature.\cite{DresselhausBook,AndoReviewCNT} The
semiconducting properties allow electrons to be confined in a
gate-defined potential.\cite{Bulaev2008,WunschDQDCNT2009,WunschDQDCNT2010,WeissFlensberg2010}
This potential is assumed smooth on the scale of
the interatomic distance, conserving the valley index. Both
single\cite{Bulaev2008} and double quantum
dots\cite{WunschDQDCNT2009,WunschDQDCNT2010,WeissFlensberg2010} have been studied in this approximation. An important effect of curvature in nanotubes is that it leads to spin-orbit interaction that couples the valley index with the spin in the longitudinal
direction.\cite{Ando2000,HuertasGB2006,JeongJL2009,Izumida2009}

The Hamiltonian describing the spin and valley degrees of
freedom, excluding the hyperfine interaction, reads
\begin{equation}
    H_\mathrm{sv}=H_{\mathrm{so}}+H_{\mathrm{s}} +H_{\mathrm{orb}},
\label{EQ:H1dot}
\end{equation}
where $H_{\mathrm{so}}$ is the spin-orbit coupling term,
$H_{\mathrm{s}}$ is the Zeeman interaction and
$H_{\mathrm{orb}}$ is the diamagnetic effect of the magnetic
field. For the specific cases considered below, not all terms
in Eq.\eqref{EQ:H1dot} are present.

The spin-orbit coupling term is
\begin{equation}
H_{\mathrm{so}}=-\frac{1}{2} \Delta_\mathrm{so} \tau_3 \sigma_\parallel,
\label{EQ:Hso}
\end{equation}
where $\Delta_\mathrm{so}$ is the spin-orbit
energy splitting, $\sigma_\parallel$ is the spin operator along the direction of the tube axis and $\sigma_{x}$, $\sigma_{y}$, $\sigma_{z}$ ($\sigma _{0}$) and  $\tau_{1}$, $\tau_{2}$, $\tau_{3}$ ($\tau_{0}$) are the Pauli (identity) matrices in spin and valley space, respectively. For the valley degree of freedom, we use
$\tau=\mathrm{K},\mathrm{K}'$ to identify the $+1$ and $-1$ eigenstates of
$\tau_3$. In the following, unless otherwise stated, the spin quantization axis---$\sigma=\uparrow,\downarrow$ or equivalently, $\sigma=\pm$---is taken along the direction of the \emph{total} magnetic field. The value of the spin-orbit splitting depends on the nanotube's chiral vector and on the electron
filling.\cite{Jespersen2010}
\begin{figure}[t]
   \centering
   \includegraphics[width=.40\textwidth]{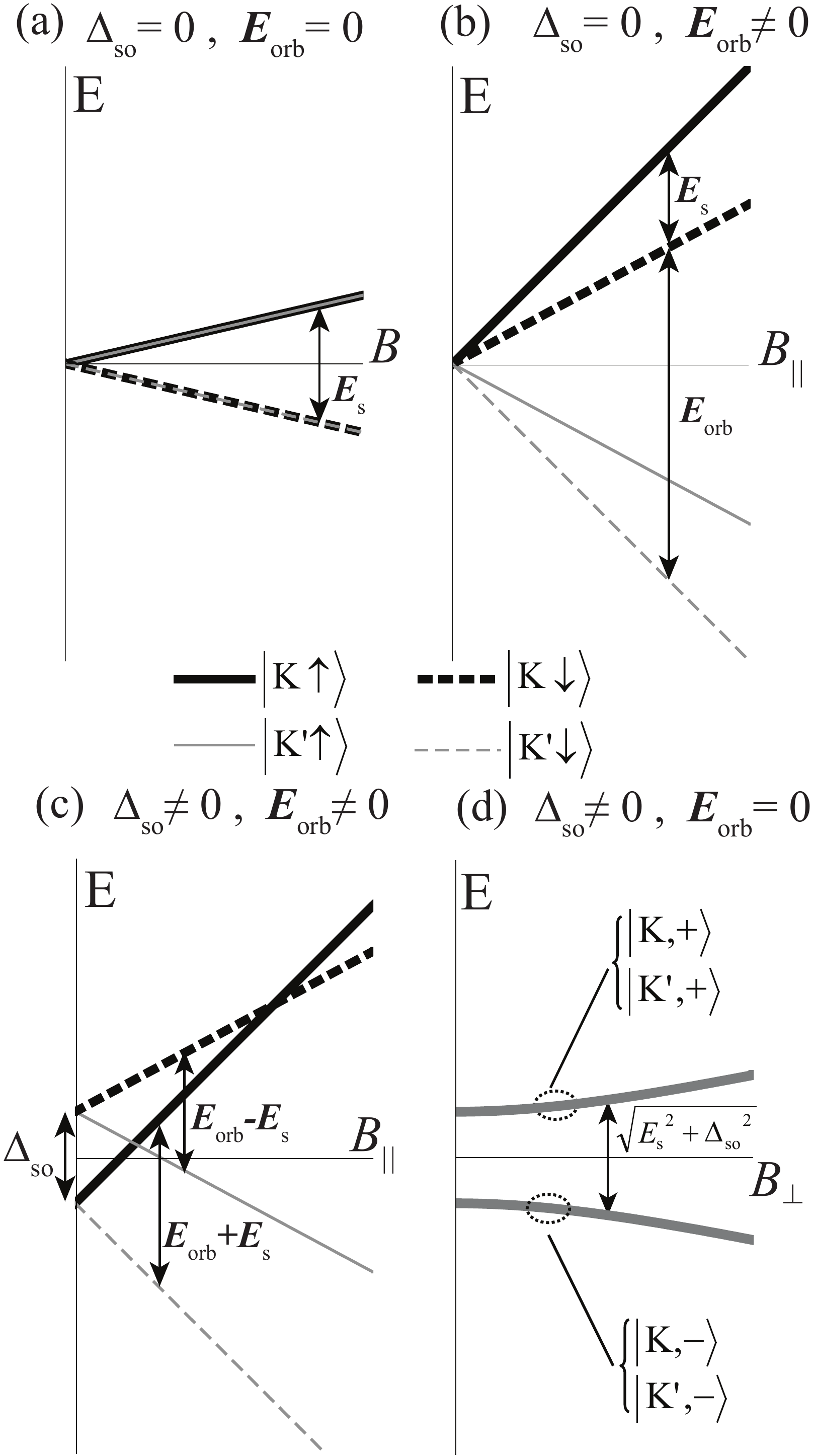}
\caption{Single-particle energies as a function of magnetic
field for the quantum dots of cases A, B, C and
D presented in the text. For A, B and C the eigenstates are $\ketLR{\tau\sigma}$ because the spin projection along the direction of the external magnetic field is a good quantum number, the corresponding eigenenergies are given in Eq.\eqref{EQ:ENER1p1d}. The parameters are: (a) $\Delta_\mathrm{so}\smeq E_\mathrm{orb}\smeq0$ and
$E_\mathrm{s}\smneq0$; (b) $\Delta_\mathrm{so}\smeq0$,
$E_\mathrm{orb}\smneq0$ and $E_\mathrm{s}\smneq0$; (c)
$\Delta_\mathrm{so}\smneq0$, $E_\mathrm{orb}\smneq0$, $E_\mathrm{s}\smneq0$ and $B_\perp\smeq 0$. (d) As in (c), $\Delta_\mathrm{so}\smneq0$, but the external magnetic field is \emph{perpendicular} to the tube's axis, so $E_\mathrm{s}\smneq0$ and $E_\mathrm{orb}\smeq 0$; the eigenfunctions, $\ketLR{\tau,\pm}$, change their spin projection according to the ratio between the Zeeman energy and the spin-orbit splitting, see details in Sec.\ref{SC:resCD}.}
\vspace{-0.4cm}
   \label{FG:fig1}
\end{figure}

The Zeeman energy due to an external magnetic field
$\mathbf{B}_\mathrm{ex}=(B_x,B_y,B_z)$ is
\begin{equation}
H_{\mathrm{s}}=\frac{1}{2}g_s\mu_B  \tau_0
\left(\mathbf{B}_\mathrm{ex}\cdot \boldsymbol{\sigma}\right),
\label{EQ:Hzeeman}
\end{equation}
where $g_s\smeq 2$ is the usual gyromagnetic factor. The
component of the magnetic field parallel to the tube axis, $B_\parallel$, gives rise to a strong diamagnetic effect:
\begin{equation}
H_{\mathrm{orb}}= g_\mathrm{orb}\mu_B B_\parallel \tau_3
\sigma_0 \label{EQ:Horb},
\end{equation}
where the orbital $g$-factor, $g_\mathrm{orb}$,
depends on the size of the nanotube\cite{Minot2004} and it is
bigger than $g_s/2$ in the typical case: nanotubes with radius
greater than 1 nanometer.\cite{Kuemmeth2008,ChurchillFlensberg2009,Jespersen2010}
To simplify the notation, we define the following two energy scales:
\begin{eqnarray}
E_\mathrm{s}&=&g_s\mu_B |\mathbf{B}_\mathrm{ex}|, \label{EQ:Bs}\\
E_{\mathrm{orb}}&=&2 g_\mathrm{orb}\mu_B |B_\parallel|.  \label{EQ:Borb}
\end{eqnarray}

\subsubsection{Four special cases}
\label{SC:subsubCA} Here we identify four special cases of single quantum dots,
representing different physical realizations, see also Fig.~2. The important classification is the splitting of the dot state compared with the hyperfine energy. The relevant energy scales are Zeeman splitting, $E_\mathrm{s}$, orbital splitting, $E_\mathrm{orb}$, and
spin-orbit splitting, $\Delta_\mathrm{so}$; we only consider each one of them when they become much bigger than the hyperfine interaction. Based on this, we define:

\textit{Case A)} No spin-orbit coupling and no orbital
magnetism, only the Zeeman energy $E_\mathrm{s}$ is considered. This is
relevant for
nanotube dots with small spin-orbit energy in a perpendicular
magnetic field (so that $E_\mathrm{orb}=0$) or, alternatively, graphene dots in an \emph{in-plane} magnetic field.\cite{RecherNBT09} The single-particle spectrum as a function of the magnetic field is shown in Fig.\ref{FG:fig1}(a). Results for valley degenerated quantum dots are presented in Sec.\ref{SC:resCA}.

\textit{Case B)} Quantum dot with no spin-orbit coupling and finite orbital
magnetism; this is the situation for nanotubes with small spin-orbit
splitting in a magnetic field with a parallel component. In this case, the
energies $E_\mathrm{s}$ and $E_{\mathrm{orb}}$ are both relevant. Figure
\ref{FG:fig1}(b) shows the spectrum as a function of the
magnetic field, which breaks the 4-fold degeneracy. Since we have assumed that $E_\mathrm{orb}>E_{s}$, which is likely the case if the total magnetic field is parallel to the tube axis,\cite{Kuemmeth2008,ChurchillFlensberg2009,Jespersen2010} the two highest energy
solutions have $\tau=\mathrm{K}$, whereas the two lowest energy solutions
have $\tau=\mathrm{K}'$. Results for the current case are presented in
Sec.\ref{SC:resCB}.

\textit{Case C)} Quantum dot in a nanotube with spin-orbit coupling and external magnetic field \emph{parallel} to the tube axis; all three energy scales $\Delta_\mathrm{so}, E_\mathrm{s}$ and $E_\mathrm{orb}$ are relevant. At zero magnetic field, in contrast to cases A and B, the spin-orbit coupling breaks the four-fold degeneracy, which results in two Kramers doublet. The energy of the two Kramers doublets are $\pm\Delta_{\mathrm{so}}/2$, as shown in \ref{FG:fig1}(c), which depicts the spectrum as a function of the magnetic field. For two finite fields some degeneracies are recovered. Results for this situation are presented in Sec.\ref{SC:resCC}.

\textit{Case D)} Here the system has a finite spin-orbit
coupling and the magnetic field is \emph{perpendicular} to the tube axis, therefore, diamagnetic effects are absent. Even for finite Zeeman energy, $E_\mathrm{s}$, the doublets are not split because the magnetic field cannot couple opposite valley, see
Fig.~\ref{FG:fig1}(d). The ratio between $E_\mathrm{s}$ and $\Delta_{\mathrm{so}}$ controls the spin projection of the solutions; the bigger the magnetic field the more similar to case A the solutions become because the spin of the eigenstates tend to align with the magnetic field. This quantum dot solutions and the return probability results are presented in Sec.\ref{SC:resCD}.

\subsection{The double dot}
The double dot single-particle Hamiltonian is
\begin{equation}
H_\mathrm{DQD}^{\rm
1p}=H_{\rm L,R}^{\rm 1p} + H_T^{\rm 1p},
\label{EQ:DQD1p}
\end{equation}
where $H_{\rm L,R}^{\rm 1p}$ includes the Hamiltonians for the
two isolated dots and $H_T^{\rm 1p}$ is the single-particle
tunneling between the two dots. In what follows we use the
superindex ${\rm 1p}$ or ${\rm 2p}$ to distinguish single-particle and two-particle operators. We introduce the Pauli (identity)
matrices $\xi_i$ ($\xi_0$) in left-right space, with $\pm 1$ eigenvalues
of $\xi_3$ for L/R, respectively. In this notation, the
tunneling part of the Hamiltonian becomes
\begin{equation}
H_T^{\rm 1p}= -t \xi_1 \tau_0 \sigma_0.\label{EQ:HT1p}
\end{equation}
This inter-dot tunneling is assumed to be valley and
spin conserving, because the gate voltage defined confining
potential is non-magnetic and it is assumed smooth on the
lattice scale. We also assume that the tunneling amplitude does not
depend on the quantum numbers $\tau$ and $\sigma$, which is
valid as long as the height of the potential barrier is much
bigger than the detuning, the spin-orbit and magnetic
splittings.\cite{WeissFlensberg2010,WunschDQDCNT2010}

The isolated left dot plus right dot single-particle Hamiltonian is
\begin{equation}
H_\mathrm{L,R}^{\rm 1p}= \tau_0\sigma_0\frac{\xi_0(\epsilon_{\rm L}
+\epsilon_{\rm R})+\xi_3(\epsilon_{\rm L}-\epsilon_{\rm R} )}{2} + \xi_0 H_\mathrm{sv},
\end{equation}
where the last term is the valley and spin Hamiltonian in
Eq.~\eqref{EQ:H1dot} which is identical in the two quantum dots. The effects of the gate voltages are
introduced as the energy shifts $\epsilon_{\rm L}$ and
$\epsilon_{\rm R}$ for the left and the right dot,
respectively.

\subsection{Two-particle basis states, no tunneling}
A single-particle basis set can be generated by the eight states $\ketLR{\xi
\tau \sigma}$, with $\xi=\mathrm{L,R},~\tau=\mathrm{K},\mathrm{K}'$ and the spin projection $\sigma=\uparrow,\downarrow$ is taken along the direction of the magnetic field. If $\Delta_\mathrm{so}\smeq 0$, or if the magnetic field is parallel to the tube's axis, $\ketLR{\xi
\tau \sigma}$ are eigenstates of the single-particle Hamiltonian $H_\mathrm{L,R}^{\rm 1p}$ with eigenenergies
\begin{equation}
E_{\xi\tau\sigma}= \left(E_{\mathrm{s}}\sigma + E_{\mathrm{orb}}\tau - \Delta_{\mathrm{so}} \tau\sigma  \right)/2. \label{EQ:ENER1p1d}
\end{equation}

Using these states we build two-particle Slater determinants
with quantum numbers $\xi\tau\sigma$ and $\xi'\tau'\sigma'$ as
follows
\begin{equation}
\ketLR{{}_{\xi\tau\sigma}^{\xi'\tau'\sigma'}}=
\frac{1}{\sqrt{2}}\left(
\ketLR{\vphantom{\xi'}\xi\tau\sigma}_{\bf
1}\ketLR{\xi'\tau'\sigma'}_{\bf
2}-\ketLR{\xi'\tau'\sigma'}_{\bf
1}\ketLR{\vphantom{\xi'}\xi\tau\sigma}_{\bf 2} \right).
\label{EQ:Sla0}
\end{equation}
Since the single-particle basis has eight elements, the
two-particle basis has 28 states ($28=8!/(6!2!)$). However, for the return probability due to
energetic reasons we do not include the states with two electrons in the
left dot, i.e., the $(2,0)$ charge configuration, which leaves 22
states.

In general, the single-particle eigenstates can differ from $\ketLR{\xi
\tau \sigma}$; we label the single-particle eigenstates of $H_\mathrm{sv}$ as $\ketLR{n}$, with energies $E_n$ using the index $n=1,\dots,4$; then, the six (0,2) eigenstates and their eigenenergies are
\begin{equation}
(0,2):\quad \ketLR{{}_{\mathrm{R}n}^{\mathrm{R}n'}},\quad
E^{(0,2)}_{n,n'}=2\epsilon_\mathrm{R}+E_n+E_{n'} + U_\mathrm{RR},
\label{EQ:02sols0En}
\end{equation}
where $n,n'=1,\dots,4$ and $n<n'$. Here we included electron-electron interaction,
represented by the right dot charging energy $U_{RR}$. We have not included Coulomb exchange since it is expected to be small.\cite{WeissFlensberg2010,Jespersen2010} With one electron in each dot the sixteen (1,1) eigenstates
and their corresponding eigenenergies are
\begin{equation}
(1,1):\quad \ketLR{{}_{\mathrm{R}n}^{\mathrm{L}n'}},
\quad E^{(1,1)}_{n,n'}=\epsilon_\mathrm{L}+\epsilon_\mathrm{R}+E_n+E_{n'}+U_\mathrm{LR},
\label{EQ:11sols0}
\end{equation}
where $U_\mathrm{LR}$ is the inter-dot Coulomb repulsion.
\subsection{Inter-dot tunneling}
\label{SC:detun}
The single-particle Hamiltonian of Eq.\eqref{EQ:HT1p} preserves the
spin and valley degrees of freedom in the inter-dot tunneling
($\mathrm{L\leftrightarrow R}$), therefore, the tunneling
Hamiltonian can be rewritten in terms of the single-particle eigenstates as:
\begin{equation}
H_T^{\rm 1p} = -t \sum_{n=1} ^4 \left( \ketLR{\mathrm{L}n}
\braLR{\mathrm{R}n} + \ketLR{\mathrm{R}n} \braLR{\mathrm{L}n}
\right),
\end{equation}
When acting with the tunneling
Hamiltonian on the (0,2) two-particle basis states it gives,
\begin{equation}
H_T^{\rm 2p} \ketLR{{}_{\mathrm{R}n}^{\mathrm{R}n'}} = -t
\left( \ketLR{{}_{\mathrm{L}n}^{\mathrm{R}n'}} +
\ketLR{{}_{\mathrm{R}n}^{\mathrm{L}n'}}\right),
\label{EQ:HTon02}
\end{equation}
i.e., a combination of (1,1) Slater determinants associated with those single-particle states. Thus, any given pair $n,n'$ gives a $3\times3$ Hamiltonian. In the basis, $\ketLR{{}_{\mathrm{R}n}^{\mathrm{R}n'}}$,
$\ketLR{{}_{\mathrm{R}n}^{\mathrm{L}n'}}$ and $\ketLR{{}_{\mathrm{L}n}^{\mathrm{R}n'}}$, this Hamiltonian
matrix becomes
\begin{equation}
H_{nn'}=E_{AV}+ E_{n}+ E_{n'}+\left(\begin{array}{ccc}-\frac{\varepsilon}{2}&-t&-t\\ -t&\frac{\varepsilon}{2}&0\\
-t&0&\frac{\varepsilon}{2} \end{array}\right),
\label{EQ:H3by3}
\end{equation}
where the detuning, $\varepsilon$, and the average energy, $E_{AV}$, have been defined as
\begin{eqnarray}
\varepsilon &\equiv&  E^{(1,1)}_{n,n'}- E^{(0,2)}_{n,n'} = \epsilon_L-\epsilon_R -U_\mathrm{RR}+U_\mathrm{LR}, \\
E_{AV}  &\equiv& \frac{1}{2} \left(E^{(1,1)}_{n,n'}+
E^{(0,2)}_{n,n'}\right) - \left(E_{n}+E_{n'}\right)
\nonumber\\&=& \frac{1}{2}\left(\epsilon_L+3\epsilon_R
+U_\mathrm{LR}+U_\mathrm{RR}\right).
\end{eqnarray}
For the mixing of the (0,2) and the (1,1) states, the global energy shift
$E_{AV}\smpl E_{n}\smpl E_{n'}$ is irrelevant, and therefore we
choose $E_{AV}=0$ from this point on.
The Hamiltonian of Eq.\eqref{EQ:H3by3} can in fact be reduced
to a 2 by 2 system because the following (1,1) combination,
\begin{equation}
\ketLR{\mathrm{LR}_\mathrm{odd}^{n,n'}}\equiv
\frac{1}{\sqrt{2}} \left(
\ketLR{{}_{\mathrm{L}n}^{\mathrm{R}n'}}-\ketLR{{}_{\mathrm{R}n}^{\mathrm{L}n'}}\right),
\end{equation}
is an eigenstate with energy independent of $t$, $E^{(1,1)}_{n,n',\mathrm{odd}} =
\frac{\varepsilon}{2} + E_{n}\smpl E_{n'}$. In the Pauli
blockade language this is a blocked state. The left-right (LR)
symmetry of this state is evident when writing it as a product in LR- and $nn'-$ spaces:
\begin{equation}
\ketLR{\mathrm{LR}_\mathrm{odd}^{n,n'}}=
\ketLR{\mathrm{S^{LR}}} \ketLR{\mathrm{T}_{n,n'}}
\label{EQ:LRodd},
\end{equation}
where we introduced the notation of a ``singlet" in LR-space. It is convenient to introduce also the triplet states in this
space, and we define
\begin{subequations}
\begin{eqnarray}
\ketLR{\mathrm{S^{LR}}}&=&\frac{\ketLR{\mathrm{L}}_{\bf 1}\ketLR{\mathrm{R}}_{\bf 2}-\ketLR{\mathrm{R}}_{\bf 1}\ketLR{\mathrm{L}}_{\bf 2}}{\sqrt{2}},\\
\ketLR{\mathrm{T^{LR}_0}}&=&\frac{\ketLR{\mathrm{L}}_{\bf 1}\ketLR{\mathrm{R}}_{\bf 2}+\ketLR{\mathrm{R}}_{\bf 1}\ketLR{\mathrm{L}}_{\bf 2}}{\sqrt{2}},\\
\ketLR{\mathrm{T^{LR}_+}}&=&\ketLR{\mathrm{L}}_{\bf 1}\ketLR{\mathrm{L}}_{\bf 2},\\
\ketLR{\mathrm{T^{LR}_-}}&=&\ketLR{\mathrm{R}}_{\bf
1}\ketLR{\mathrm{R}}_{\bf 2}. \end{eqnarray}
\end{subequations}
\begin{figure}[t]
   \centering
   \includegraphics[width=.44\textwidth]{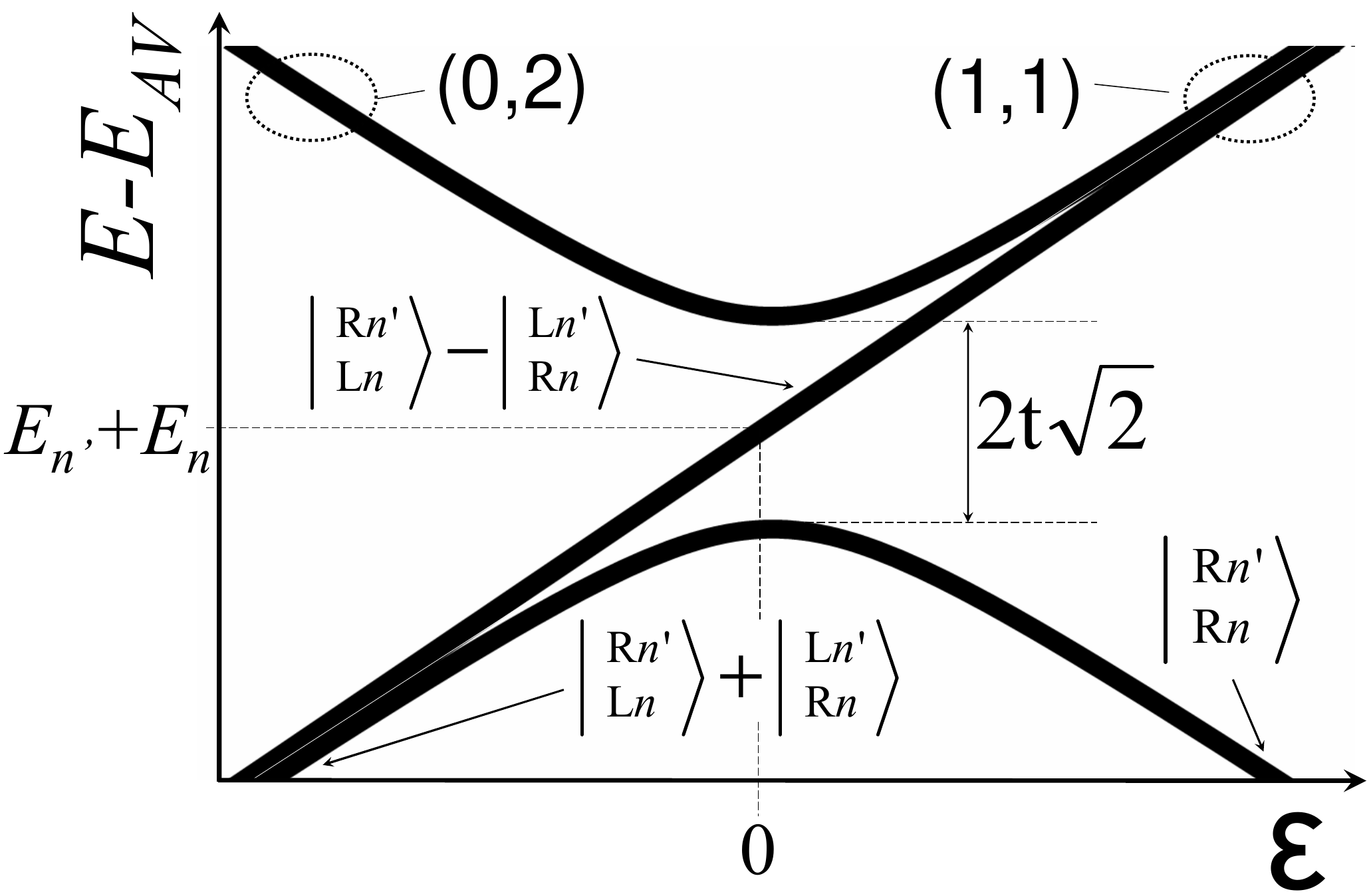}
   \vspace{-0.4cm}
\caption{Spectrum of the (0,2) and (1,1) states in the presence of inter-dot tunneling, $t$, as a function
of the detuning, $\varepsilon$. The
plot focuses on the three states based on the single-particle
states $\ketLR{n}$ and $\ketLR{n'}$ for one of the six independent cases (with $n\smneq n'$). The avoided crossing, with gap $\Delta_t=2 t\sqrt{2}$, at $\varepsilon\smeq 0$ is a signature of the mixing between (0,2) and (1,1) states. The (1,1) odd combination of Slater determinants involves a singlet in LR-space (see
Eqs.\eqref{EQ:LRodd} and \eqref{EQ:HTonSLR}) and it is not
affected by the tunneling. As shown in Eqs.\eqref{EQ:HTon02}
and \eqref{EQ:H2by2}, the (0,2) state (i.e., a
$\mathrm{T^{LR}_-}$ state) is mixed only with the even
combination of the shown (1,1) Slater determinants (i.e., a
$\mathrm{T^{LR}_0}$ state).}
   \vspace{-0.2cm}
   \label{FG:detu}
\end{figure}
Similarly, we have defined singlet-like and triplet-like functions in
valley and spin spaces (not shown). Finally, for the combined quantum number $n$, we define
\begin{subequations}
\label{EQ:SingTripnn}
\begin{eqnarray}
\ketLR{\mathrm{S}_{n,n'}} &=&\frac{\ketLR{n}_{\bf 1}\ketLR{n'}_{\bf 2}-\ketLR{n'}_{\bf 1}\ketLR{n}_{\bf 2}}{\sqrt{2}}, \\
\ketLR{\mathrm{T}_{n,n'}} &=&\frac{\ketLR{n}_{\bf
1}\ketLR{n'}_{\bf 2}+\ketLR{n'}_{\bf 1}\ketLR{n}_{\bf
2}}{\sqrt{2}}.
\end{eqnarray}
\end{subequations}
With this notation the six (0,2) $n,n'$ states can be written
as products of functions in LR- and $n$-space,
\begin{equation}
\ketLR{{}_{\mathrm{R}n}^{\mathrm{R}n'}}=
\ketLR{\mathrm{T^{LR}_-}} \ketLR{\mathrm{S}_{n,n'}}.
\label{EQ:02sols1}
\end{equation}

The tunneling Hamiltonian behaves as an identity operator in
$n$-space, whereas in LR-space it acts as follows,
\begin{subequations}
\begin{eqnarray}
H_T^{\rm 2p}  \ketLR{\mathrm{T^{LR}_-}}&=& -\sqrt{2} t \ketLR{\mathrm{T^{LR}_0}} \label{EQ:HTonTm}, \\
H_T^{\rm 2p}  \ketLR{\mathrm{T^{LR}_0}}&=& -\sqrt{2} t \left(\ketLR{\mathrm{T^{LR}_+}}+\ketLR{\mathrm{T^{LR}_-}}\right),\\
H_T^{\rm 2p}  \ketLR{\mathrm{T^{LR}_+}}&=& -\sqrt{2} t\ketLR{\mathrm{T^{LR}_0}},\\
H_T^{\rm 2p}  \ketLR{\mathrm{S^{LR}}}&=& 0. \label{EQ:HTonSLR}
\end{eqnarray}
\end{subequations}
As a consequence of Eq.\eqref{EQ:HTonSLR}, all the LR-antisymmetric solutions (the (1,1)
states with a singlet $\ketLR{\mathrm{S^{LR}}}$) are not
affected by the inter-dot tunneling, i.e., they are \emph{blocked} states. Combining
Eqs.~\eqref{EQ:HTon02},\eqref{EQ:02sols1} and~\eqref{EQ:HTonTm} it follows that the (0,2) solution
$\ketLR{{}_{\mathrm{R}n}^{\mathrm{R}n'}}$ couples only to the
\emph{even} combination of the given (1,1) Slater determinants,
\begin{equation}
\ketLR{\mathrm{LR}_\mathrm{even}^{n,n'}}= \frac{1}{\sqrt{2}} \left( \ketLR{{}_{\mathrm{L}n}^{\mathrm{R}n'}}+\ketLR{{}_{\mathrm{R}n}^{\mathrm{L}n'}}\right) =\ketLR{\mathrm{T^{LR}_0}} \ketLR{\mathrm{S}_{n,n'}}.
\label{EQ:11LRsym}
\end{equation}
In the above mentioned space, the effect of the inter-dot
tunneling can be condensed into a $2\times2$ matrix
\begin{equation}
H_{nn'}^\mathrm{even}=E_{n}+E_{n'}+\left(\begin{array}{cc}-\frac{\varepsilon}{2}
&-\sqrt{2}t\\
-\sqrt{2}t&\frac{\varepsilon}{2}
\end{array}\right).
\label{EQ:H2by2}
\end{equation}

The maximum mixture between
$\ketLR{{}_{\mathrm{R}n}^{\mathrm{R}n'}}$ and
$\ketLR{\mathrm{LR}_\mathrm{even}^{n,n'}}$ happens for
$\varepsilon\smeq 0$, where the avoided crossing occurs with a
gap given by $\Delta_t\equiv 2\sqrt{2} t$. Figure \ref{FG:detu}
shows the energies of the three two-particle states associated
with the two single-particle states $\ketLR{n}$ and
$\ketLR{n'}$. Six groups of 3 states with exactly the same
energy gap appear, one for each of the possible $n,n'$
non-equivalent pairs (see Fig.\ref{FG:fig0}(c)). The four remaining states (that complete the 22 solutions) are the (1,1) LR-antisymmetric states that
arise from Slater determinants constructed with the same
single-particle state in both dots:
\begin{equation}
\ketLR{{}_{\mathrm{L}n}^{\mathrm{R}n}}=
\ketLR{\mathrm{S^{LR}}} \left(\ketLR{n}_{\bf 1}\ketLR{n}_{\bf
2}\right).
\label{EQ:RLslaters}
\end{equation}
Using spin and valley singlet and triplets functions the states in Eq.\eqref{EQ:RLslaters} can be written  (for the quantum dots of case A, B and C: $\sigma$ is a good quantum number) as the four states with full polarization in spin and valley spaces, i.e., $\ketLR{\mathrm{S^{LR}T^{val}_\pm T^{spin}_\pm}}$ and $\ketLR{\mathrm{S^{LR}T^{val}_\pm T^{spin}_\mp}}$.

In summary, there are ten LR-antisymmetric states that do not
mix with the $(0,2)$ solutions. Six of them with $n\smneq n'$,
as it is shown in  Fig.\ref{FG:detu},  become quasi-degenerated
with their associated $n,n'$ LR-symmetric partners at the high detuning limit, i.e., when $|\varepsilon/t|\gg 1$. The six LR-symmetric states are connected by inter-dot tunneling to the corresponding $n,n'$ (0,2) states as follows,
\begin{eqnarray}
\ketLR{\mathrm{T^{LR}_0}} \ketLR{\mathrm{S}_{n,n'}}   &\leftrightarrow&  \ketLR{\mathrm{T^{LR}_-}} \ketLR{\mathrm{S}_{n,n'}},
\label{EQ:Tconnect}
\end{eqnarray}
Obviously, as it is seen in detail in the next sections, the
connectivity between (0,2) states and (1,1) states plays a key role in the return probability experiment.

\subsection{Hyperfine coupling}
\label{SC:hyp}
Taking $\boldsymbol{\sigma}$ the spin operator of the confined electron in a graphene-based quantum dot, the hyperfine interaction has the following form:\cite{PalyiB09}
\begin{equation}H_{h}= \sum_{i = 0}^3 \tau_i {\bf h}^{(i)} \cdot \boldsymbol{\sigma}.
\label{EQ:Hh1d}
\end{equation}
The values of the ${\bf h}^{(i)}$ fields can be considered fixed during every
single-shot measurement because the dynamics of the nuclei spins is much slower than the electron spin's. Within this approximation, the effect of the slow dynamics of the hyperfine field---on the average result of a experiment that is performed many times---is captured by noting that the hyperfine field components become random variables that follow zero-mean Gaussian distributions with variances\cite{PalyiB09}
\begin{eqnarray}\braket{\left({h}^{(0)}_j\right)^2}&=&
2\braket{\left({h}^{(1)}_j\right)^2}=
2\braket{\left({h}^{(2)}_j\right)^2}\equiv 2 \sigma_{h}^2,\nonumber\\
\braket{\left({h}^{(3)}_j\right)^2}&=&0,~~~~~~~~~j=x,y,z,
\label{EQ:variances}
\end{eqnarray}
where $\sigma_{h}\smeq A_{\rm iso}^2 \nu/(8 N_{QD})$, with
$N_{QD}$ being the number of atoms in the quantum dot, $\nu$
the abundance of ${}^{13}\mathrm{C}$ atoms in the dot and
$A_\mathrm{iso}$ the isotropic hyperfine coupling constant.

From Eq.\eqref{EQ:variances} it follows that Eq.\eqref{EQ:Hh1d} with $i=3$ has zero coefficients, as expected from absence of time-reversal symmetry. A $i=3$ term would correspond to a time-reversal symmetric spin-orbit interaction, which cannot originate from the hyperfine interaction. In the following sections we show that this apparent valley anisotropy of the hyperfine field have important consequences for the return probability experiment that manifest for some specific prepared states.

The Hamiltonian of Eq.\eqref{EQ:Hh1d} and the variances in Eq.\eqref{EQ:variances} are derived as follows.\cite{PalyiB09} As in Ref.~\onlinecite{Fischer2009}, we assume that the Fermi contact hyperfine interaction dominates. Furthermore, in order to minimize the number of special cases, we take the case with largest hyperfine-induced mixing, namely the case of isotropic coupling,\cite{PalyiB09} even though there is some degree of anisotropy.\cite{Fischer2009} The hyperfine coupling Hamiltonian then reads
\begin{equation}H_{hf}=A_\mathrm{iso}\sum_{l,s} \boldsymbol{\sigma} \cdot {\bf I}_{l,s} \delta({\bf r}-{\bf R}_{l,s}),
\end{equation}
where $l$ is the unit cell index, $s\in \{A,B\}$ is the
sublattice index and ${\bf I}_{l,s}$ is the nuclear spin of the carbon atom at site ${\bf R}_{l,s}$, being zero or a spin-1/2 operator for a ${}^{12}$C or a ${}^{13}$C isotope, respectively. The normalized tight-binding eigenstates,
characterized by the spin $\sigma$ and valley $\tau$ indexes
are
\begin{equation}
\left(\Psi_{\tau\sigma}\right)_{l,s} = {\rm e}^{\ci \tau \mathrm{K} \cdot {\bf R}_{l,s}} {\rm \Psi}_{s}^\tau ({\bf R}_{l,s}) \chi_\sigma,
\end{equation}
where the smoothly varying envelope functions ${\rm
\Psi}_{A}^\tau ({\bf r})$ are eigenstates of the gapped Dirac
equation in a confining potential. Taking matrix elements of these eigenstates with
respect to the hyperfine Hamiltonian then gives
\begin{eqnarray}
H_{hf}&=& A_\mathrm{iso}\boldsymbol{\sigma} \cdot \sum_{l,s} {\bf I}_{l,s} \sum_{\tau,\tau'} F_{l,s}^{\tau\tau'} \ketLR{\tau}\braLR{\tau'}, \\
F_{l,s}^{\tau\tau'}&=&  {\rm e}^{\ci (\tau-\tau') \mathrm{K} \cdot {\bf R}_{l,s}} \left[{{\rm \Psi}_{s}^{\tau'}} ({\bf R}_{l,s})\right]^* {\rm \Psi}_{s}^{\tau} ({\bf R}_{l,s}).
\end{eqnarray}
After performing averages over the nuclei fields, the
Hamiltonian of Eq.~\eqref{EQ:Hh1d} follows. The real
(imaginary) part of ${F_{l,s}^{1,-1}}$ having cosine (sine)
factors generates the terms containing $\tau_1$ ($\tau_2$). The
two terms diagonal in valley $F_{l,s}^{\tau,\tau}$, $\tau=\pm1$
are equal and do not have oscillating factors, therefore, only operators
proportional to $\tau_0$ survive and the $\tau_3$ component vanishes (leading to the important property already mentioned). The factor of $1/2$ between the
variances of the $h_j^{(1,2)}$ and the $h_j^{(0)}$ components
comes from the averages over the cosine and sine squared.

In a DQD, due to the different nuclei environments for each dot, it is
convenient to work with the left/right homogeneous and the left/right inhomogeneous
components of the hyperfine field interaction (HFI)
\begin{equation}\bar{h}^{(i)}_j\equiv\frac{{h}^{(i)}_{j,\mathrm
{L}}+{h}^{(i)}_{j,\mathrm {R}}}{2}~,~~\delta h^{(i)}_j\equiv
\frac{{h}^{(i)}_{j,\mathrm {L}}-{h}^{(i)}_{j,\mathrm {R}}}{2}.
\end{equation}
In terms of these fields, the double dot single-particle HFI
Hamiltonian can be written as
\begin{equation}H_{h,\mathrm{LR}}^{\rm 1p}=  \sum_{i = 0}^2
\sum_{j=x,y,z} \left(  { \bar{h}}^{(i)}_j \xi_0 +   {  \delta h}^{(i)}_j \xi_3 \right) \tau_i \sigma_j
\label{EQ:HFIdqd1p},
\end{equation}
which should be added to the Hamiltonian of the double dot in
Eq.\eqref{EQ:DQD1p}.
\subsubsection{Hyperfine field in single valley systems}
Here we briefly introduce the key features of the HFI found in
single valley systems, e.g., GaAs double dots. The well-known
relevant states and the action of the HFI Hamitonian in those
systems is an important reference for comparison. Because the
valley space is absent in GaAs, the single-particle Hamiltonian
of the hyperfine field interaction is
\begin{equation}H_{h,\mathrm{GaAs}}^{\rm 1p}=   \sum_{j=x,y,z} \left(  { \bar{h}}_j \xi_0 +   {  \delta h}_j \xi_3 \right) \sigma_j.
\label{EQ:HFIdqd1pGaAs}
\end{equation}
Due to the absence of the orbital degree of freedom there exist
four (1,1) states, namely, the spin singlet and triplets,
\begin{eqnarray}
\ketLR{1,\mathrm{GaAs}}=\ketLR{\mathrm{T_{0}^{LR}S^{spin}}} &,&~\ketLR{2,\mathrm{GaAs}}=\ketLR{\mathrm{S^{LR}T^{spin}_0}},~~~ \nonumber\\
\ketLR{3,\mathrm{GaAs}}=\ketLR{\mathrm{S^{LR}T^{spin}_+}} &,&~\ketLR{4,\mathrm{GaAs}}=\ketLR{\mathrm{S^{LR}T^{spin}_-}}.~~~~
\label{EQ:basisGaAs}
\end{eqnarray}
The HFI Hamiltonian of Eq.\eqref{EQ:HFIdqd1pGaAs}, written in the latter basis, becomes
\begin{subequations}
\begin{eqnarray}H_\mathrm{e}^{\mathrm{GaAs}}=\left(\begin{array}{cccc}0&\delta z&-{\delta xy}^* &{\delta xy}\\
\delta z&0&{\overline{xy}}^*&{\overline{xy}}\\
-{\delta xy}&{\overline{xy}}&\overline{z}&0\\
{\delta xy}^*&{\overline{xy}}^*&0&-\overline{z}
\end{array}\right),
\label{EQ:HFImatGaAs}~~~~~~~~~~~~\\
\delta z \mapsto 2 \delta h_z  ,~~
\overline{z} \mapsto 2 {\bar{h}}_z ~, ~~~~~~~~~~~~~~~~~\\
{\delta xy} \mapsto \sqrt{2}\left({\delta h}_x-\ci{\delta h}_y\right) ,~~
{\overline{xy}}\mapsto \sqrt{2}\left({\bar{h}}_x-\ci{\bar{ h}}_y\right).~~~~~
\label{EQ:GaAsHFIterms}
\end{eqnarray}
\end{subequations}
One sees that only the inhomogeneous HFI is able to mix the
LR-symmetric function (spin singlet) with the LR-antisymmetric
(spin-triplets) functions. In particular the $\delta h_z$
component also conserves the total $S_z$ and therefore is able
to mix the $\ketLR{\mathrm{T_{0}^{LR}S^{spin}}}$ with the
$\ketLR{\mathrm{S^{LR}T^{spin}_0}}$. In the next section we
show that some specific situations in CNT double dots can be
mapped to the Hamiltonian of Eq.\eqref{EQ:HFImatGaAs} or to a
modified version of it.
\section{Simulating the experiment}
\label{SC:experiment}
The return probability, $P(\tau_s)$, for a given
evolving time, $\tau_s$, see Fig.\ref{FG:fig0}(b), is obtained
experimentally by averaging over a set ($i=1,2,..N$) of single shot
measurements with outcomes $O_{i,\tau_s}$,
\begin{equation}
P(\tau_s)=\frac{1}{N} \sum^N_{i=1} O_{i,\tau_s}.
\end{equation}
Each single-shot measurement consists of a gate-voltage cycle with five
stages (see arrows and points ``$\mathrm{p}$'', ``$\mathrm{e}$'' and ``$\mathrm{m}$'' in Fig.~\ref{FG:fig0}(a) and the associated detuning values $\varepsilon_\mathrm{p}$, $\varepsilon_\mathrm{e}$ and $\varepsilon_\mathrm{m}$ in Fig.~\ref{FG:fig0}(c)):

(i) \emph{Preparation}, the DQD is prepared in the (0,2) region
at point ``$\mathrm{p}$''.

(ii) \emph{Separation}, by applying a voltage pulse of length
$\Delta\varepsilon$ and duration $T_S$, which is short on the
scale of the HFI interaction ($ T_S\ll
\frac{\sigma_h}{\hbar}$), but slow on the scale of the inverse
tunneling energy, the initial (0,2) state is adiabatically
moved to the point ``$\mathrm{e}$'' deep into the (1,1) region. If the detuning is changed by $\Delta \varepsilon \gg t$, the condition for adiabatic
conversion is $T_S\gg \hbar\Delta \varepsilon/t^2$.

(iii) \emph{Evolution}, the system is left to evolve at the
point ``$\mathrm{e}$'' during a time $\tau_s$; in this stage
the electron wavefunction in each dot acquires a different dynamical phase due to the hyperfine coupling with the nuclei spins; in general, the system oscillates between the initial (1,1) wavefunction and other combination of (1,1) states.

(iv) \emph{Joining}, a voltage pulse brings the system back to
the (0,2) region; with the same adiabatic condition as for the
\emph{separation} stage.

(v) \emph{Measuring},
at the ``$\mathrm{m}$'' point a nearby charge
sensing device determines the outcome $O_{i,\tau_s}$ (1 or 0) of the single shot measurement; $O_{i,\tau_s}$ is set to $1$ only when the system
has returned to an (0,2) configuration.

Depending on the preparation protocol the initial (0,2) state
can be different than the ground state. We cover all the
possibilities by assuming a prepared state in an arbitrary
superposition of the six possible zero-hyperfine (0,2)
eigenstates, $|\psi_\mathrm{p}\rangle=\sum_{l\smeq1}^6 a_l
|l,(0,2)\rangle$, where each $|l,(0,2)\rangle$ state is one of
the six Slater determinants given in Eq.\eqref{EQ:02sols0En}.
When averaging the outcomes $O_{i,\tau_s}$ of a large number
of single-shot measurements over an ensemble of random
hyperfine field, the phases of $a_l$ average out and the
resulting probability for a given initial state is simply given
by
\begin{equation}
P_{\psi_p}(\tau_s)= \sum_l|a_l|^2 P_{l}(\tau_s),
\end{equation}
where $P_{l}(\tau_s)$ is the return probability when starting
in the (0,2) eigenstate $|l,(0,2)\rangle$. Therefore, the
return probability of a general case can be evaluated by
knowing the values $|a_l|$ and the return probabilities
obtained when preparing the (0,2) zero-hyperfine eigenstates
separately. For these reasons, in each of the DQD scenarios
that we deal over the next sections, we focus on the behaviour
of the six functions, $P_{l}(\tau_s)$.

Due to its experimental importance we focus the discussion on
the saturation value of the return probability which is defined
as
\begin{equation}
P_\infty \equiv \lim_{\tau_s \rightarrow \infty} P(\tau_s).
\end{equation}
In addition, we define the decaying time, $\tau_d$, as the time
for which
\begin{equation}
  P(\tau_d) = \frac{1}{2} P_\infty + \frac{1}{2}, \label{EQ:decaytime}
\end{equation}
see  Fig.\ref{FG:fig0}(b).

In general, $P(\tau_s)$ may be different when the hyperfine field of the two dots follow Gaussian distributions
with different rms values, ${\sigma_h^\mathrm{L}}$ and
${\sigma_h^\mathrm{R}}$. In fact, this is the case when the numbers of ${}^{13}$C atoms are different for the left and right dots (see
Eq.\eqref{EQ:variances}). However, this would affect the transient of $P(\tau_s)$ but not the behavior for $\tau_s \rightarrow \infty$. In
this work we focus on the saturation values and on transient features \emph{intrinsic} to DQDs in C-based systems. For this reason we choose $\sigma_h^\mathrm{R} \smeq \sigma_h^\mathrm{L} \equiv \sigma_h$ in the following.

\subsection{Numerical evaluation of the return probabilities}

We have developed a numerical simulation of the experimental
cycle outlined above. The time evolution inside the (1,1) region (point ``$\mathrm{e}$'' in
Fig.\ref{FG:fig0}(a)) is governed by the hyperfine field alone,
i.e., it is assumed that $t^2/\Delta\varepsilon\ll \sigma_h$. Essentially, as discussed in the introduction, we are working in the large detuning limit where the effect of the tunneling exchange is negligible and therefore we are able to obtain the lower bounds of the saturation return probabilities. For solving such a time evolution we start with the LR-symmetric (1,1) state, $\ketLR{\mathrm{LR}_\mathrm{even}^{n,n'}}$, which is connected (see Eq.\eqref{EQ:Tconnect}) with the (0,2) prepared state
$\ketLR{{}_{\mathrm{R}n}^{\mathrm{R}n'}}$; this is then
decomposed into the numerically determined eigenstates of the 16 by 16 Hamiltonian  for the current hyperfine field realization (generated following the Gaussian distributions with variances given in Eq.\eqref{EQ:variances}), and
hence the time dependent (1,1) state can be computed as a
superposition of time-evolved eigenstates.

To calculate the return probability, we define an operator that
projects onto LR-symmetric states as
\begin{equation}
\hat{P}_\mathrm{LR-sym}\equiv
\ketLR{\mathrm{T^{LR}_0}}  \braLR{\mathrm{T^{LR}_0}}.
      \label{EQ:projLRsym}
    \end{equation}
The return probability after time $\tau_s$ for a given
hyperfine field realization $r_i$ is then
\begin{equation}
p_{r_i}(\tau_s)= \braLR{\Psi_{r_i}(\tau_s)}
\hat{P}_\mathrm{LR-sym}  \ketLR{\Psi_{r_i}(\tau_s)} \label{EQ:pri0}.
\end{equation}
The projection method assumes the joining stage (iv) is
performed under the same adiabatic conditions as the separating
stage (ii). After repeating this procedure for a large number of
realizations, $N_r$, the final return probability $P(\tau_s)$
is obtained by averaging
\begin{equation}P(\tau_s) = \frac{\sum_{i=1}^{N_r} p_{r_i}(\tau_s)}{N_r}.
\end{equation}

\subsection{Analytical evaluation of the return probability}

For analytic evaluation of the return probability, we use the
same set of conditions as for the numerical evaluation, namely
that the time evolution after the separation stage is only
governed by hyperfine interaction (i.e., no tunneling exchange), and that
the return probability after evolution can be computed by
projection. In this high detuning limit---excluding for the moment the HFI effect--- the LR-symmetric states can be considered degenerated with their LR-antisymmetric
partners (see Fig.\ref{FG:detu} and Eq.\eqref{EQ:11sols0}). Additional degeneracies
are determined by presence of spin-orbit coupling and applied
magnetic field. We assume that with finite spin-orbit
coupling and/or applied magnetic field the hyperfine
interaction only mixes states within the subset of quasi-degenerate
states.

Due to the LR-symmetry, since the evolution starts in a LR-even state, the degeneracies of the subspaces of evolution in (1,1) are never lower than
two. In this paper degeneracies $n_\mathrm{e}\smeq 2,3,4,8$
and $16$ appear. To determine the time evolution, we project the
Hamiltonian to the (1,1) subspace connected with the prepared (0,2) state, the reduction of the system---if any---sometimes allows analytical treatment.
Then, we focus on the form and the statistical properties (i.e., variances) of the surviving mixing terms of the HFI Hamiltonian. Those matrix elements are obtained by elementary calculations; the (1,1) basis expanded using singlet and triplet functions in LR, valley and spin spaces (instead of the Slater determinant basis) is useful for a more direct physical interpretation of the hyperfine mixing terms.

\subsection{State counting estimation of $P_\infty$}

Here we introduce a scheme for estimating the return probability based on a simple
\textit{state counting} argument. The value obtained with the procedure that follows does not always coincide with the exact value, $P_\infty$, however it is useful for visualizing special features of the exact dynamics. In order to compute the \textit{state counting} value, first, we find the number of degenerate (1,1) states, $n_\mathrm{e}$, connected with the chosen (0,2) prepared state under investigation. Second, we find the number of LR-symmetric
states, $n_{(0,2)}$, connected with that (1,1) subspace of $n_\mathrm{e}$ states, $S_\mathrm{e}$. Third, assuming a fully incoherent mixing of the initial state with all states, for $\tau_s\rightarrow \infty$ every one of the $n_\mathrm{e}$ states should have a probability $1/n_\mathrm{e}$ of being occupied, therefore, the
estimated return probability becomes
\begin{equation}
    P_\infty^{sc}\equiv\frac{n_{(0,2)}}{n_\mathrm{e}}.
\end{equation}
One can expect this to be a lower bound for $P_\infty$, because under coherent
evolution the system does not fully randomize and might
therefore maintain a larger weight on the initial state which
is connected. This is the case for example in GaAs double dots, whereas the estimation gives $P_\infty^{sc}=1/4$, the coherent evolution (averaged over many realizations) gives $P_\infty=1/3$.

\section{Results}
\label{SC:results}

\begin{table}[!b]
\caption{\label{TB:levs1} Energy levels and labels for the six LR-even/LR-odd partner states in the (1,1) configuration for cases A, B and C. We use the same labels for the connected energy levels in (0,2) to identify the prepared state.} \centering {
\begin{tabular}{l||c|c|c|c}
n,n'&Case A&Case B&Case C&Behavior\\
\hline
\hline
$\mathrm{K\uparrow},\mathrm{K\downarrow}$&$\mathrm{v_0}$&$\mathrm{h_0}$&$\mathrm{h_0}$&$E_\mathrm{orb}$\\
\hline
$\mathrm{K'\uparrow},\mathrm{K'\downarrow}$&$\mathrm{v_0}$&$\mathrm{l_0}$&$\mathrm{l_0}$&$-E_\mathrm{orb}$\\
\hline
$\mathrm{K\uparrow},\mathrm{K'\downarrow}$&$\mathrm{v_0}$&$\mathrm{c_0}$&$\mathrm{c_{0L}}$&$-\Delta_\mathrm{so}$\\
\hline
$\mathrm{K\downarrow},\mathrm{K'\uparrow}$&$\mathrm{v_0}$&$\mathrm{c_0}$&$\mathrm{c_{0H}}$&$\Delta_\mathrm{so}$\\
\hline
$\mathrm{K\uparrow},\mathrm{K'\uparrow}$&$\mathrm{v_+}$&$\mathrm{c_+}$&$\mathrm{c_+}$&$E_\mathrm{s}$\\
\hline
$\mathrm{K\downarrow},\mathrm{K'\downarrow}$&$\mathrm{v_-}$&$\mathrm{c_-}$&$\mathrm{c_-}$&$-E_\mathrm{s}$\\
\end{tabular}}
\end{table}
\subsubsection{Labeling of the energy levels}
With the exception of Sec.\ref{SC:resCD} (case D), in the following, the single-particle eigenstates in each dot are $\ketLR{\xi,n}$ with $|n\rangle=|\tau\sigma\rangle$, where $\sigma$ is the spin projection along the direction of the applied magnetic field taken, for convenience, along the $z$-direction. The two-particle functions $\ketLR{\mathrm{S}_{n,n'}}$ and
$\ketLR{\mathrm{T}_{n,n'}}$ presented in Eqs.~\eqref{EQ:SingTripnn}
can be further expanded in terms of tensor products in
spin and valley spaces. The procedure is straightforward and
states $\ketLR{\mathrm{S}_{n,n'}}$ and $\ketLR{\mathrm{T}_{n,n'}}$ are found to
be equivalent to a unique tensor product in valley and spin
except for $n,n'\equiv\mathrm{K\uparrow,K'\downarrow}$ and
$n,n'\equiv\mathrm{K\downarrow,K'\uparrow}$ in which case they are given by,
\begin{subequations}
\begin{eqnarray}
\ketLR{\mathrm{S_{K\uparrow,K'\downarrow}}}=\frac{1}{\sqrt{2}}\left(\ketLR{\mathrm{S^{val} T^{spin}_0}}+\ketLR{\mathrm{T^{val}_0 S^{spin}}}\right), \\
\ketLR{\mathrm{T_{K\uparrow,K'\downarrow}}}=\frac{1}{\sqrt{2}}\left(\ketLR{\mathrm{T^{val}_0 T^{spin}_0}}+\ketLR{\mathrm{S^{val} S^{spin}}}\right),  \\
\ketLR{\mathrm{S_{K\downarrow,K'\uparrow}}}=\frac{1}{\sqrt{2}}\left(\ketLR{\mathrm{S^{val} T^{spin}_0}}-\ketLR{\mathrm{T^{val}_0 S^{spin}}}\right), \\
\ketLR{\mathrm{T_{K\downarrow,K'\uparrow}}}=\frac{1}{\sqrt{2}}\left(\ketLR{\mathrm{T^{val}_0 T^{spin}_0}}-\ketLR{\mathrm{S^{val} S^{spin}}}\right).
\end{eqnarray}
\end{subequations}
These states are particularly important in the presence of spin-orbit coupling.\cite{Bulaev2008,PalyiB10,WunschDQDCNT2009,WeissFlensberg2010} For the sake of readability in the following we sometimes omit the \emph{ket} symbol, $\ketLR{\dots}$, when referring to a product state of triplets and singlets in left/right, valley and spin spaces.

As it is shown in detail in the next subsections, for cases A, B and C the sixteen (1,1) eigenstates for zero hyperfine are associated with 3, 9 and 10 energy levels, respectively, that sometimes cross each other or become degenerated. These situations  change drastically the outcome of the return probability experiment. In order to present the correspondence between the (1,1) states and these energy levels, we divide the states into two classes. In the first class, as in Sec.\ref{SC:detun}, we take six groups, each one corresponding to a LR-even/LR-odd pair of states, i.e., $\ketLR{\mathrm{T^{LR}_0}\mathrm{S}_{n,n'}}$ and $\ketLR{\mathrm{S^{LR}}\mathrm{T}_{n,n'}}$. These six groups---that as shown in Table \ref{TB:levs1}, may belong to the same energy level---are of great importance for the return probability experiment because they are connected to the (0,2) states, $\ketLR{\mathrm{T^{LR}_-}\mathrm{S}_{n,n'}}$, through the LR-even states. Apart from a global energy shift, these (1,1) energy levels and each associated (0,2) energy level have the same dependence with the parameters, therefore, we use the energy level labeling of Table \ref{TB:levs1} also for  identifying the (0,2) prepared states. The second class of states consists of the four fully polarized spin and valley LR-odd states, i.e., $\ketLR{\mathrm{S^{LR}}} \left(\ketLR{n}_{\bf 1}\ketLR{n}_{\bf 2}\right)$; they belong to the energy levels listed in Table \ref{TB:levs2}. These blocked states do not provide access to the (0,2) configuration neither at the preparation nor at the measurement stage, however, if their associated energy level crosses (or is degenerated with) an energy level with states of the first class, they become relevant for the effective hyperfine dynamics.

The labeling introduced in Tables \ref{TB:levs1} and \ref{TB:levs2} for the energy levels of case B and C---the less degenerated cases---is inspired by the following logic: ``l" ,``c" and ``h" denote \emph{low}, \emph{central} and
\emph{high}, respectively, referring to the valley characters
of the states associated with the energy level. Similarly, the $+$, $-$ and $0$
refer to the spin characters. Since two pairs of LR-partners states fall in level $\mathrm{c_0}$, we distinguish them (in case C because $\Delta_\mathrm{so}\smneq 0$ splits the $\mathrm{c_0}$ level) by adding the subscript ``H" or ``L" for the high and the low energy LR-partner states, respectively.
\begin{table}[!t]
\caption{\label{TB:levs2} Energy levels and labels for LR-odd full spin and valley polarized states in the (1,1) configuration for cases A, B and C.} \centering {
\begin{tabular}{l||c|c|c}
n&Case A&Case B and C&Behavior\\
\hline
\hline
$\mathrm{K\uparrow}$&$\mathrm{v_+}$&$\mathrm{h_+}$&$-\Delta_\mathrm{so}+E_\mathrm{orb}+E_\mathrm{s}$\\
\hline
$\mathrm{K'\uparrow}$&$\mathrm{v_+}$&$\mathrm{l_+}$&$\Delta_\mathrm{so}-E_\mathrm{orb}+E_\mathrm{s}$\\
\hline
$\mathrm{K\downarrow}$&$\mathrm{v_-}$&$\mathrm{h_-}$&$\Delta_\mathrm{so}+E_\mathrm{orb}-E_\mathrm{s}$\\
\hline
$\mathrm{K'\downarrow}$&$\mathrm{v_-}$&$\mathrm{l_-}$&$-\Delta_\mathrm{so}-E_\mathrm{orb}-E_\mathrm{s}$\\
\end{tabular}}
\vspace{0cm}
\end{table}

\begin{figure*}[!ht]
\includegraphics[width=.95\textwidth]{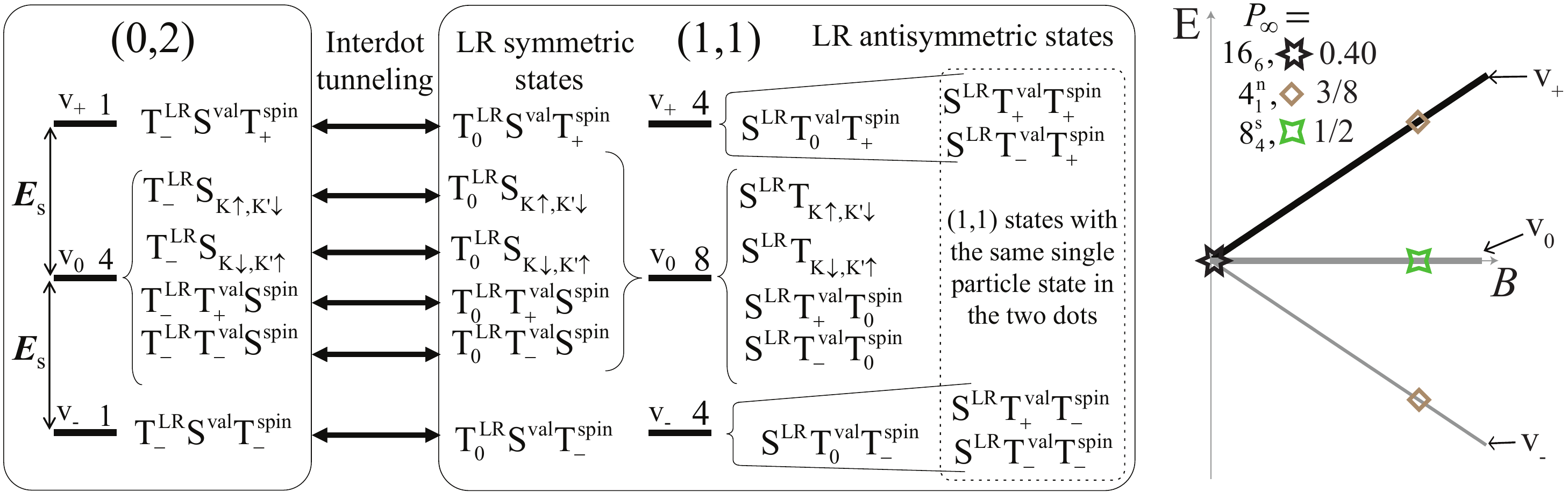}
\vspace{-0.3cm}
\caption{Case A: Saturation return probabilities, energies and states for a valley degenerated DQD; $\Delta_\mathrm{so}\smeq E_\mathrm{orb}\smeq 0$ and the Zeeman splitting is $E_\mathrm{s}\smeq g_s\mu_B B$. The associated single dot single-particle spectrum is shown in Fig.\ref{FG:fig1}(a). (Left panel) Eigenstates and spectrum of the (0,2) and (1,1) configurations; the degeneracy of each energy level at finite magnetic field is indicated. The six (0,2) states in levels $\mathrm{v_\pm}$ and $\mathrm{v_0}$ can be prepared; each one of them is connected by inter-dot tunneling, represented by double headed arrows, to a single LR-symmetric state in (1,1). (Right panel) Energy levels in the (1,1) configuration as a function of the magnetic field; as shown in the legend, $P_\infty$ is found to be 0.4, 0.50 or 3/8 depending on both the magnetic field value and the prepared (0,2) state.}
\vspace{-0.3cm}
   \label{FG:caseA}
\end{figure*}

\subsubsection{Labeling of the different physical situations}
In what follows the numbers $n_\mathrm{e}$ and $n_{(0,2)}$ are useful for distinguishing different qualitative and quantitative dynamical situations. For this reason, we label each case using the two integers as ${n_\mathrm{e}}_{n_{(0,2)}}$. As an example, with the latter convention, the case in which the (1,1) subspace is fully degenerated is to be labeled as ``$16_6$" because ${n_\mathrm{e}}=16$ and $n_{(0,2)}=6$; in such a case six independent prepared states provide access to the full subspace of evolution. As another example, a crossing between the energy levels $\mathrm{c_{0L}}$ and $\mathrm{l_0}$ (presented in Table \ref{TB:levs1}) implies a subspace of evolution with ${n_\mathrm{e}}=4$ containing two LR-symmetric states (one associated with level $\mathrm{c_{0L}}$ and the other with level $\mathrm{l_0}$) that can be prepared, therefore, this situation is to be labeled as ``$4_2$".

However, as we show below, in some cases two situations with the same ${n_\mathrm{e}}$ and $n_{(0,2)}$ numbers but different HFI dynamics appear; in such cases we use a superscript ``$\mathrm{s}$" or ``$\mathrm{n}$" to distinguish them. The ``$\mathrm{n}$" superscript is reserved for cases in which the effective HFI Hamiltonian has a zero matrix element between any of the LR-symmetric states in the evolution subspace, $\ketLR{\mathrm{T^{LR}_0}\mathrm{S}_{n,n'}}$, and its LR-antisymmetric partner state, $\ketLR{\mathrm{S^{LR}}\mathrm{T}_{n,n'}}$.

\subsection{Case A: $\Delta_\mathrm{so}\smeq 0$ and $E_\mathrm{orb}\smeq 0$}
\label{SC:resCA}

The single-particle spectrum of a single dot (in the absence of
HFI) is shown in Fig.\ref{FG:fig1}(a) as a function of the
magnetic field for this case. The magnetic field, $B \zver$, can point in any direction as long as diamagnetic effects are avoided and, therefore, the valley degeneracy is not lifted; for nanotubes this impose that $B_\parallel\smeq 0$, whereas for a graphene quantum dot the magnetic field must be in the plane.\cite{RecherNBT09} The energy levels for the (0,2) configuration are only three, $\mathrm{v}_{s_z}$ with $s_z\smeq \pm$, 0 (as listed in Table \ref{TB:levs1}), with energies $E^{(0,2)}_{\mathrm{v}_{s_z}}= s_z E_\mathrm{s}$; this follows from Eq.~\eqref{EQ:02sols0En}. In the left panel of Fig.\ref{FG:caseA}, we show the states and label them in the singlet/triplet notation for spin, valley and left/right space. Also the degeneracies at finite magnetic field are given for each group. The connections of
(0,2) and (1,1) states by tunneling are shown with double
headed arrows, such that only left/right symmetric (1,1) states
are connected to their (0,2) partners with same spin and valley
quantum numbers.

The right panel of Fig.\ref{FG:caseA} shows the energies of the (1,1) states as a
function of the magnetic field. The saturation return probability is also
given in the figure as tags to the energy versus field lines.
These symbols mean that if the (0,2) system is prepared in
$\mathrm{v}_0$ or $\mathrm{v}_\pm$ the return probability is 1/2 and 3/8,
respectively, while for $B=0$, the return probability is $0.40$.
Below we explain each of these cases in more detail.

\textbf{\emph{The Case {\mMRKstSi}}} (in Fig.\ref{FG:caseA})
is for zero magnetic field, leading to the largest possible
degeneracy, $n_\mathrm{e}\smeq 16$. All the (0,2)
eigenstates are connected to this subspace and hence the state
counting estimation for the saturation return probability is
$P_\infty^{\mathrm{sc},{\sMRKstSi}}\smeq 6/16\smeq 0.375$,
which, however, does not coincide with the actual result
$P_\infty^{{\sMRKstSi}} \smeq 0.40$. The result is found
irrespective of which of the six (0,2) states is prepared. The increment of the return probability, as compared to the
state counting estimation, is similar to the GaAs situation, where
the state counting gives 1/4, whereas coherent evolution gives
return probability 1/3. Interestingly, in both cases one goes
from $P_\infty^{\mathrm{sc}}$ to the correct result by
subtracting $1$ from the denominator, since 6/(16-1)=0.4.

\begin{figure*}[!ht]
   \centering
   \includegraphics[width=.97\textwidth]{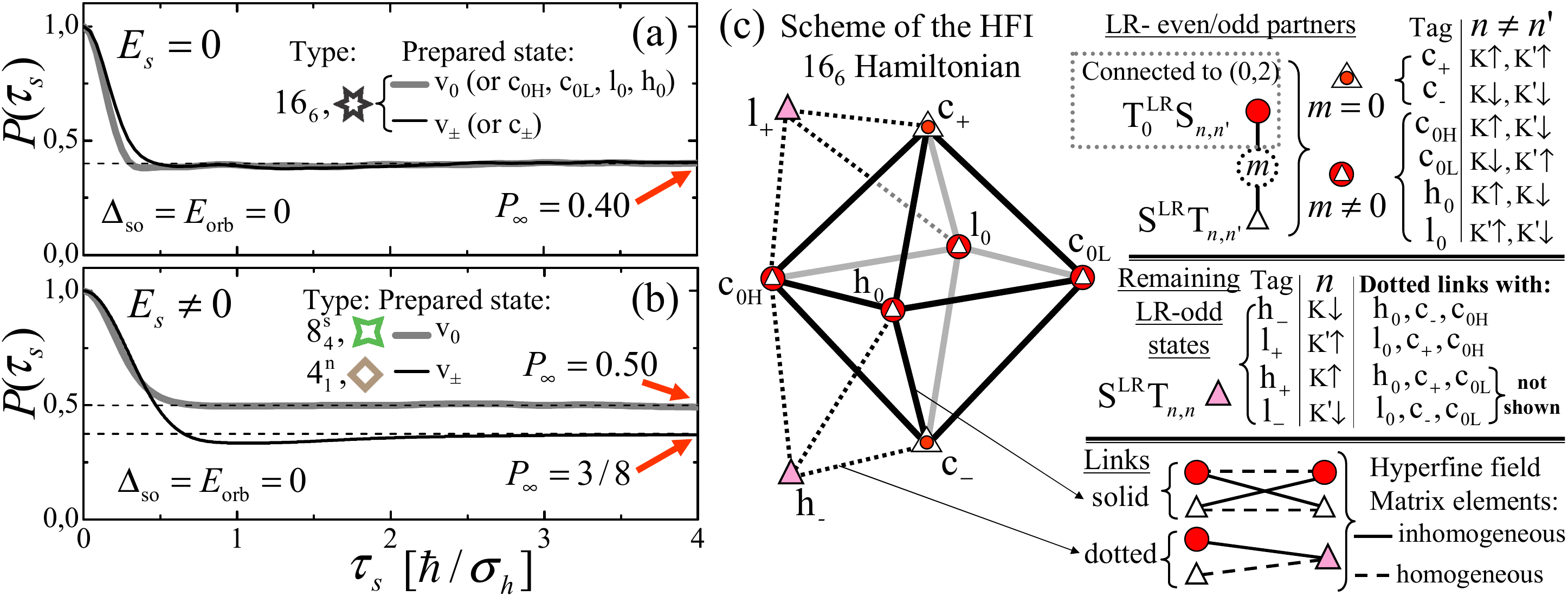}
   \vspace{-0.2cm}
   \caption{Scheme of the complete hyperfine Hamiltonian (right panel) and the return probabilities $P(\tau_s)$ for case A (left panels), i.e., $\Delta_\mathrm{so}\smeq E_\mathrm{orb}\smeq0$; see Fig.\ref{FG:caseA}. (a) Zero magnetic field, type $16_6$ dynamics, all prepared states give $P_\infty\smeq 0.40$ but $P(\tau_s)$ decays faster for total spin zero prepared states. (b) Finite magnetic field. Type $8_4^\mathrm{s}$ dynamics is found when a zero spin state is prepared, $P_\infty\smeq 0.50$. For nonzero spin prepared states one has type $4_1^\mathrm{s}$ dynamics and $P(\tau_s)$ follows Eq.\eqref{EQ:P3eights} giving $P_\infty\smeq 3/8$ (this situation also appears for the DQDs with $\Delta_\mathrm{so}\smneq 0$ of case C, see Fig.\ref{FG:caseC}).
   (c) Scheme of the full 16 by 16 HFI Hamiltonian,
$H_\mathrm{e}^{{\sMRKstSi}}$; we label the states according to the \emph{ten} energy levels they belong in case C. The six octahedron's vertices represent LR-odd/LR-even partner states of Table \ref{TB:levs1}; note that the HFI matrix element between these partners, $m$, is zero for the $\mathrm{c_+}$ and the $\mathrm{c_-}$ vertices. Outside the octahedron the four LR-odd states of Table \ref{TB:levs2} are represented (for simplicity, only two are shown). The links between nodes represent the existence of at least one nonzero matrix element between the states. Depending on the parameters---for cases A, B and C---the ten energy levels ($\mathrm{c_\pm}$, $\mathrm{c_{0H}}$, $\mathrm{c_{0L}}$, $\mathrm{h_0}$, $\mathrm{l_0}$, $\mathrm{h_\pm}$ and $\mathrm{l_\pm}$) can move together and/or cross each other, leading to the eight different effective dynamics involving less than sixteen (1,1) states.}
   \label{FG:sche8A}
\vspace{-0.3cm}
\end{figure*}

We have also studied the final state resolved saturation return
probabilities and observed that it is not the same for the
six (0,2) states. It is more probable to return to the (0,2)
state that was prepared, and this is more pronounced when the
prepared state is one of the two nonzero-spin states (the
states $\mathrm{~T^{LR}_{-}S^{val} T^{spin}_{\pm}~}$). In Fig.\ref{FG:sche8A}(a) we show that the decaying time $\tau_d$ (see Fig.\ref{FG:fig0}(b)) here depends on the prepared (0,2) state, being longer for the nonzero-spin prepared states, $0.185 \hbar/\sigma_h$, while for the remaining four prepared states it is $0.149\hbar/\sigma_h$. We discuss below, when introducing cases {\mMRKstFo} and {\mMRKrb}, the reason of such an asymmetry.

It is worth to introduce here the structure of the hyperfine field interaction Hamiltonian, $H_{h,\mathrm{LR}}^{\rm 2p}$, using the sixteen (1,1) basis states presented in Fig.\ref{FG:caseA} and Tables \ref{TB:levs1} and \ref{TB:levs2}. In the octahedral representation given in Fig.\ref{FG:sche8A}(c) each vertex stands for a LR-symmetric state and its LR-antisymmetric partner state. Therefore, the initial (1,1) state at $\tau_s=0$ has weight in one of the LR-even states located on those six vertices. The links in solid lines between vertices and the links in dotted lines with the remaining LR-antisymmetric states represent nonzero matrix elements of the HFI. Each link stands for both homogeneous and inhomogeneous elements of the hyperfine field. The absence of matrix elements between states in opposite vertices of the octahedron follows from the trivial selection rule that forbids two single-particle quantum numbers from being changed simultaneously; note that the two-particle HFI Hamiltonian is $H_{h,\mathrm{LR}}^{\rm 2p}=\eins_{\mathbf{1}}\otimes ({H_{h,\mathrm{LR}}^{\rm 1p}})\vphantom{\eins}_{\mathbf{2}}+({H_{h,\mathrm{LR}}^{\rm 1p}})\vphantom{\eins}_{\mathbf{1}}\otimes \eins_{\mathbf{2}}$. This rule also justifies that each one of the four $\ketLR{\mathrm{S^{LR}}} \left(\ketLR{n}_{\bf 1}\ketLR{n}_{\bf 2}\right)$ states (in the figure nodes $\mathrm{h_\pm}$ and $\mathrm{l_\pm}$ represented outside the octahedron, for simplicity only $\mathrm{h_-}$ and $\mathrm{l_+}$ are shown) mixes only with the three LR-even/LR-odd partners---rules given in the figure---that have one of the two electron in the $\ketLR{n}$ state.

In Fig.\ref{FG:sche8A}(c) we have used the labeling of the energy levels given in Table \ref{TB:levs1} for case C (see Fig.\ref{FG:caseC}), $\mathrm{c_\pm}$, $\mathrm{c_{0H}}$, $\mathrm{c_{0L}}$, $\mathrm{h_0}$, $\mathrm{l_0}$, $\mathrm{h_\pm}$ and $\mathrm{l_\pm}$. We show below that---depending on the parameters in cases A, B and C, as some of these \emph{ten} energy levels move together and/or cross each other---there are eight different effective HFI dynamics involving restricted subspaces of evolution. For example, in the presence of spin-orbit coupling (case C below) the \mMRKstSi~case becomes irrelevant because the hyperfine field (characterized by an energy scale, $\sigma_h$, which is smaller than $\Delta_\mathrm{so}$) is unable to mix all the sixteen states and therefore the physical situation is better captured by analyzing smaller subspaces of evolution depending on the prepared state.

\textbf{\emph{The Case {\mMRKstFo}}}. A finite Zeeman energy is applied much larger than $\sigma_h$. The initialized (0,2) state has zero spin and therefore belongs to the energy level $\mathrm{v}_0$ with degeneracy $n_\mathrm{(0,2)}\smeq 4$.
The corresponding (1,1) subspace in which the system evolves
has a degeneracy of $n_\mathrm{e}\smeq 8$, with
four LR-symmetric states. In this case, the state counting
estimation for the saturation return probability, being
$P_\infty^{sc,{\sMRKstFo}}\smeq 4/8\smeq 0.5$, coincides with
the calculated exact result. Moreover, as shown in Fig.\ref{FG:sche8A}(b), the values of $P_\infty$
and the shape of $P(\tau_s)$ do not depend on which of the four
(0,2) states is being prepared. Finally, we find that the decaying time, here $\tau_d \smap 0.238\hbar/\sigma_h $, is larger than both possible decay times for
the zero-field case, which is expected because for the \mMRKstSi~case the HFI has the ability to mix the prepared state with eight extra states.

The independence on the prepared state can be understood from the matrix elements of
the Hamiltonian in the reduced Hilbert space by using the
singlet/triplet functions. The HFI term
proportional to $\delta h^{(0)}_z$ mixes one by one the four
zero-spin pairs of LR-symmetric and LR-antisymmetric states as follows,
\begin{subequations}
\begin{eqnarray}
\braLR{\mathrm{S^{LR} T_{K\uparrow, K'\downarrow}  }} H_{h,\mathrm{LR}}^{\rm 2p}
 \ketLR{\mathrm{T^{LR}_0 S_{K\uparrow, K'\downarrow}  }} &=& 2{\delta h}^{(0)}_z
 ,~~~~\\\braLR{\mathrm{S^{LR} T_{K\downarrow, K'\uparrow}  }} H_{h,\mathrm{LR}}^{\rm 2p}
 \ketLR{\mathrm{T^{LR}_0 S_{K\downarrow, K'\uparrow}  }} &=& - 2{\delta h}^{(0)}_z
 ,~~~~\\\braLR{\mathrm{S^{LR} T^{val}_+ T^{spin}_0}} H_{h,\mathrm{LR}}^{\rm 2p}
 \ketLR{\mathrm{T^{LR}_0 T^{val}_+ S^{spin}  }} &=& 2{\delta h}^{(0)}_z
 ,\\\braLR{\mathrm{S^{LR} T^{val}_- T^{spin}_0  }} H_{h,\mathrm{LR}}^{\rm 2p}
 \ketLR{\mathrm{T^{LR}_0 T^{val}_- S^{spin}  }} &=& 2{\delta h}^{(0)}_z.
\end{eqnarray}
\label{EQ:sz0HFImixing}
\end{subequations}
The fact that the subspace
$S_\mathrm{e}^{{\sMRKstFo}}$ is composed solely by LR-symmetric
states and their LR-antisymmetric partners is a key difference
to the other cases where the state counting argument fails,
because in those cases more states are available.

An illustration of all the matrix elements $H_\mathrm{e}^{{\sMRKstFo}}$
can be extracted from the full Hamiltonian representation in Fig.\ref{FG:sche8A}(c), by considering solely the four octahedron vertices in the plane (because the $\mathrm{v_0}$ level contains all the states in nodes $\mathrm{l_0}$, $\mathrm{h_0}$, $\mathrm{c_{0H}}$ and $\mathrm{c_{0L}}$) and excluding the $\mathrm{c_+}$ and $\mathrm{c_-}$ vertices.  In accordance with Eq.\eqref{EQ:sz0HFImixing}, the HFI matrix element between the included LR-partners, $m$ in the figure, is nonzero. The remaining matrix elements (links in the figure) connect each LR-symmetric state with \emph{other two} LR-symmetric states and
their LR-antisymmetric partners. All those matrix elements follow Gaussian distributions with the same variance and thus, the average result is independent of which $\mathrm{T^{LR}_0}$ state is initialized. This symmetry is also
responsible for the distribution of the total saturation
probability (1/2) evenly among the four $\mathrm{T^{LR}_0}$
states.

\textbf{\emph{The case {\mMRKrb}}} (Fig.\ref{FG:caseA}) is also for $E_\mathrm{s}\gg\sigma_h$ but here the initial state belongs either to the energy level $\mathrm{v}_+$ or the $\mathrm{v}_-$. In
Sec.\ref{SC:resCC} the same situation is found for a particular
value of the parallel magnetic field in a nanotube with
$\Delta_\mathrm{so}\smneq 0$. Here we present an analytical
derivation of the return probability in the reduced Hilbert
space, which we have also confirmed by a full numerical
evaluation. We find that the saturation return probability is
$P_\infty^{{\sMRKrb}}\smeq 3/8=0.375$.

We restrict the analysis to the states of the energy level
$\mathrm{v_+}$, and equivalent results for level $\mathrm{v_-}$
follow by symmetry. The system is prepared in the
$\mathrm{T_{-}^{LR}S^{val}T_+^{spin}}$ (0,2) state, and after
separation the initial (1,1) state is therefore
$\mathrm{T_{0}^{LR}S^{val}T_+^{spin}}$. The degeneracy of
$\mathrm{v_+}$ in (1,1) is $4$ and the
subspace is composed by:
\begin{eqnarray}
\ketLR{1,{\sMRKrb}}=\mathrm{T_{0}^{LR}S^{val}T_+^{spin}} &,&~\ketLR{2,{\sMRKrb}}=\mathrm{S^{LR}T^{val}_0T_+^{spin}} ,\nonumber\\
\ketLR{3,{\sMRKrb}}=\mathrm{S^{LR}T^{val}_+T_+^{spin}}
&,&~\ketLR{4,{\sMRKrb}}=\mathrm{S^{LR}T^{val}_-T_+^{spin}}.
\label{EQ:rbBasis}
\end{eqnarray}
There is only one LR-symmetric
state (i.e., $n_\mathrm{(0,2)}\smeq 1$) and it is a
spin-polarized valley singlet. By restricting the full HFI Hamiltonian to the latter subspace we get the effective $4\times4$ Hamiltonian:
\begin{equation}
H_\mathrm{e}^{{\sMRKrb}}=2\bar{h}_z^{(0)} \eins_4 + H_\mathrm{e}^{\mathrm{GaAs}},
\label{EQ:He0.375}
\end{equation}
where $\eins_4$ is the 4-dimensions identity matrix and $H_\mathrm{e}^{\mathrm{GaAs}}$ is the $4\times 4$ Hamiltonian presented of Eq.\eqref{EQ:HFImatGaAs} after the following replacements,
\begin{eqnarray}
\delta z \mapsto 0 &~,~&
\overline{z} \mapsto 0 ~, \\
{\delta xy} \mapsto \sqrt{2}\left({\delta h}_z^{(1)}-\ci{\delta h}_z^{(2)}\right) &~,~&
{\overline{xy}} \mapsto \sqrt{2}\left({\bar{h}}_z^{(1)}-\ci{\bar{ h}}_z^{(2)}\right).\nonumber
\end{eqnarray}
Note that the matrix elements between the nonzero spin
LR-symmetric/LR-antisymmetric pairs is
\begin{equation}\braLR{\mathrm{S^{LR} T^{val}_0 T^{spin}_\pm   }} H_{h,\mathrm{LR}}^{\rm 2p}   \ketLR{\mathrm{T^{LR}_0 S^{val} T^{spin}_\pm }} = 0.
\label{EQ:szHFImixing}
\end{equation}
This expression, together with the four zero spin cases
presented in Eq.\eqref{EQ:sz0HFImixing}, completes the effective
mixing between the six $\mathrm{T^{LR}_0}$/$\mathrm{S^{LR}}$
partners; Fig.\ref{FG:sche8A}(c) shows the six pair of states together emphasizing that the mixing, $m$, is zero for the two pairs presented here. Remarkably, Eq.\eqref{EQ:szHFImixing} is due to the absence of terms proportional to the $\tau_3$ operator in the HFI Halmiltonian of Eq.\eqref{EQ:HFIdqd1p}; as discussed above, such terms would preserve time reversal symmetry and thus cannot originate from the hyperfine field.

It is instructive to compare $H_\mathrm{e}^{{\sMRKrb}}$ with
the effective Hamiltonian for zero field in GaAs double dots
(see Sec.\ref{SC:hyp}). The null matrix elements are those that in a
2-dimensional electron gas DQD would translate to the $\delta h_z$ and
${\bar{h}}_z$ components. In our system, the effective hyperfine field operates on valley-space (all members of the subspace in Eq.\eqref{EQ:rbBasis} share the same two-electron spin function) and there is no mixing of the valley singlet, $\mathrm{S^{val}}$, with the valley triplet, $\mathrm{T_0^{val}}$, nor any splitting of the $\mathrm{T^{val}_\pm}$ states. Hence, such a \emph{valley} double dot maps to the typical ``spin only" double dot but with an effective \emph{2-dimensional hyperfine field} instead of the usual 3-dimensional HFI.

In Appendix \ref{AP:analytics} we give the details of a standard analytical procedure for solving the time dependence dictated
by $H_\mathrm{e}^{{\sMRKrb}}$ and averaging the result over the
Gaussian fields. We obtain the return probability, shown in Fig.\ref{FG:sche8A}(b),
\begin{subequations}
\begin{eqnarray}
P^{{\sMRKrb}}(\tau_s)&=& C_\mathrm{2D}\left(\tau_s \sigma_h^\mathrm{L}/\hbar\right)C_\mathrm{2D}\left(\tau_s \sigma_h^\mathrm{R}/\hbar\right) \nonumber \\&&+ 2 S_\mathrm{2D}\left(\tau_s \sigma_h^\mathrm{L}/\hbar\right) S_\mathrm{2D}\left(\tau_s \sigma_h^\mathrm{R}/\hbar\right), \\
S_\mathrm{2D}\left(x\right)&=& \frac{1}{2}\sqrt{\frac{\pi}{2}} x {\rm e}^{-2x^2} {\rm erfi}\left(\sqrt{2} x\right) ,\\
C_\mathrm{2D}\left(x\right)&=& 1-2 S_\mathrm{2D}\left(x\right),
\end{eqnarray}
\label{EQ:P3eights}
\end{subequations}
where the imaginary error function is $\mathrm{erfi}(x)\smeq
\mathrm{erf}(\ci x)/\ci \smeq \pi^{-{1}/{2}}\int_{-x}^{x} dy
\exp(y^2)$. As already mentioned, the saturation return
probability is
\begin{figure*}[!ht]
   \centering
   \includegraphics[width=.95\textwidth]{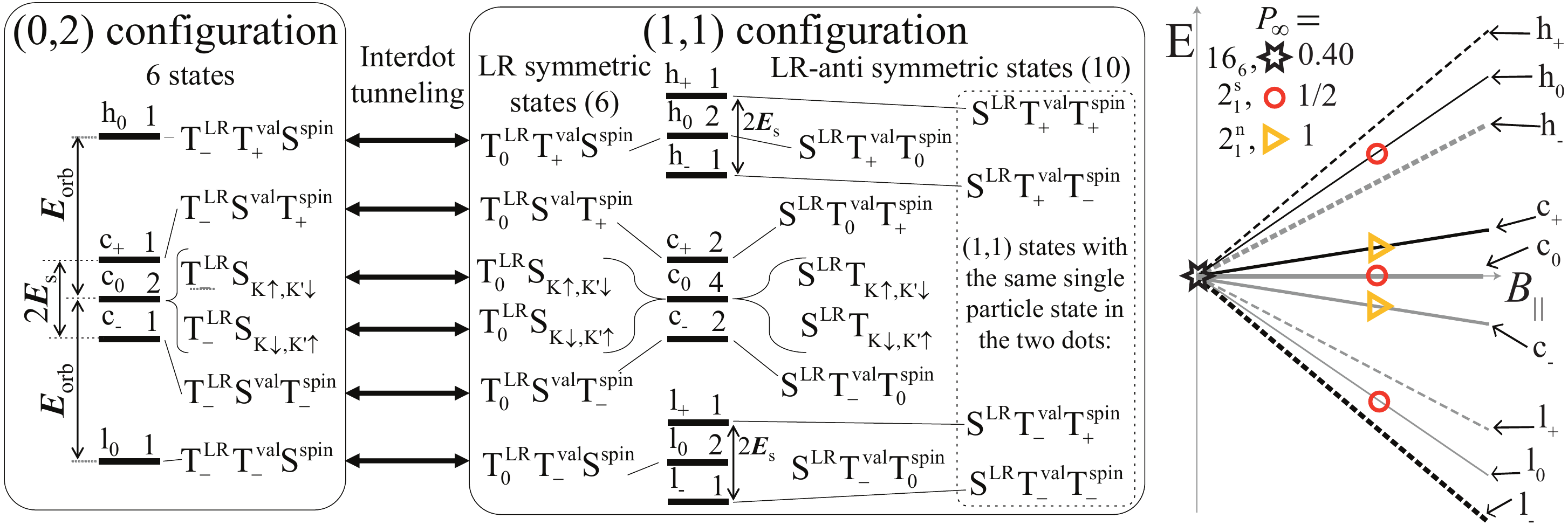}
   \vspace{-0.3cm}
   \caption{Case B: Saturation return probabilities, energies and states for a nanotube-based DQD without spin-orbit coupling in a magnetic field with a parallel component (in the figure we take $E_\mathrm{orb}>E_\mathrm{s}$), i.e., $\Delta_\mathrm{so}\smeq0$, $E_\mathrm{orb}\smneq 0$ and $E_\mathrm{s}\smneq0$; the single dot single-particle spectrum is shown in Fig.\ref{FG:fig1}(b). (Left panel) Scheme of the (0,2) and (1,1) energy levels and the associated
states' singlet/triplet characteristics---in spin, valley and
left-right space; the level degeneracies for finite magnetic field are indicated. Each one of the six (0,2) states is connected by inter-dot
tunneling to a single associated (1,1) state. For the
(1,1) charge state the six LR-symmetric and the ten LR-antisymmetric states
are distinguished. (Right panel) Energy levels in the (1,1) configuration as a function of $B_\parallel$; dashed levels contain only blocked states and therefore cannot be prepared. As shown in the legend, $P_\infty$ is found to be 0.40, 1/2 or 1 depending on both the magnetic field and the prepared (0,2) state.}
   \vspace{-0.3cm}
   \label{FG:caseB}
\end{figure*}
\begin{equation}
\lim_{x\rightarrow\infty} S_\mathrm{2D}(x)=\frac{1}{4} ~\Rightarrow~P^{{\sMRKrb}}_\infty=3/8=0.375~.
\end{equation}
For the present case, the state counting value is
$P_\infty^{\mathrm{sc},{\sMRKrb}}\smeq 1/4$, and therefore this
is another example of a case in which $P_\infty$ is greater
than the state counting value. Furthermore, here the substraction of 1 to the denominator in the state counting estimation does not provide a correct result as it does for the zero magnetic field cases in a GaAs DQD or type $16_6$ dynamics. Finally, applying
Eq.\eqref{EQ:decaytime} to the return probability
$P^{{\sMRKrb}}(\tau_s)$, we get the decaying time
$\tau_d \smap 0.3218\hbar/\sigma_h$, which is the slowest time
we find for case A; i.e., in C-based double dots with unbroken valley degeneracy.

\subsection{Case B: $\Delta_\mathrm{so}\smeq 0$ and $E_\mathrm{orb}\smneq0$}
\label{SC:resCB}
In Fig.\ref{FG:fig1}(b) we show the single-particle spectrum of a single dot as a function of the parallel component of the magnetic field, $B_\parallel$. Assuming that the total magnetic field is along the $z$-direction at an angle $\phi$ from the tube axis, we get $B_\parallel\smeq B_z \cos\phi$. The valley degeneracy is also lifted in contrast to the case without orbital magnetism studied in Sec.\ref{SC:resCA}. In the left panel of
Fig.\ref{FG:caseB} we present the spectrum for the (0,2) and (1,1) configurations including the level degeneracies at finite magnetic field and the corresponding states. In the figure, we take a ratio $E_\mathrm{orb}/E_\mathrm{s}\smeq (2 g_\mathrm{orb}/g_s)\cos\phi  $ bigger than one; which is the typical case for fully parallel magnetic field.\cite{Kuemmeth2008,ChurchillFlensberg2009,Jespersen2010} The five energy levels of the (0,2) configuration--$\mathrm{c_\pm}$, $\mathrm{c_0}$, $\mathrm{l_0}$ and
$\mathrm{h_0}$--follow from Eq.\eqref{EQ:02sols1} and are given in Tables \ref{TB:levs1} and \ref{TB:levs2}.

For zero field all the degeneracies remain and the
situation reduces to the \mMRKstSi~case already analyzed in
Sec.\ref{SC:resCA}. At finite $B_\parallel$, the return probability experiment can have two different behaviors as indicated in the
right panel of Fig.\ref{FG:caseB}. When one of the four spin-zero states is prepared, we get $P_\infty^{\sMRKci}\smeq 1/2$, whereas a nonzero spin prepared state is unaffected by the hyperfine interaction and thus $P^{\sMRKtrr}(\tau_s)\smeq 1$. We explain these two cases in the next paragraphs.

\textbf{\emph{In the case labeled {\bMRKci}}} the initial
state belongs to the $\mathrm{h_0}$, $\mathrm{l_0}$ or to the
$\mathrm{c_0}$ energy levels. The degeneracy for the (1,1) states in $\mathrm{l_{0}}$ or in
$\mathrm{h_{0}}$ is $2$, and
$n_\mathrm{(0,2)}\smeq 1$, in each case corresponding to a pair of LR symmetric/anti-symmetric states. On the other hand the $\mathrm{c_0}$ level has a 4-fold degeneracy and two
LR-even states are connected to (0,2); however, the structure of the
hyperfine interaction Hamiltonian allows us to treat the subsets of double degenerated states, $\mathrm{c_{0L}}$ and $\mathrm{c_{0H}}$ presented in Table \ref{TB:levs1}, as
two independent pairs with $n_\mathrm{e}\smeq 2$ and $n_\mathrm{(0,2)}\smeq 1$. This follows from the selection rule introduced above when explaining the absence of matrix elements for states at opposite vertices of the octahedron in Fig.\ref{FG:sche8A}(c).

\begin{figure}[!b]
     \includegraphics[width=.41\textwidth]{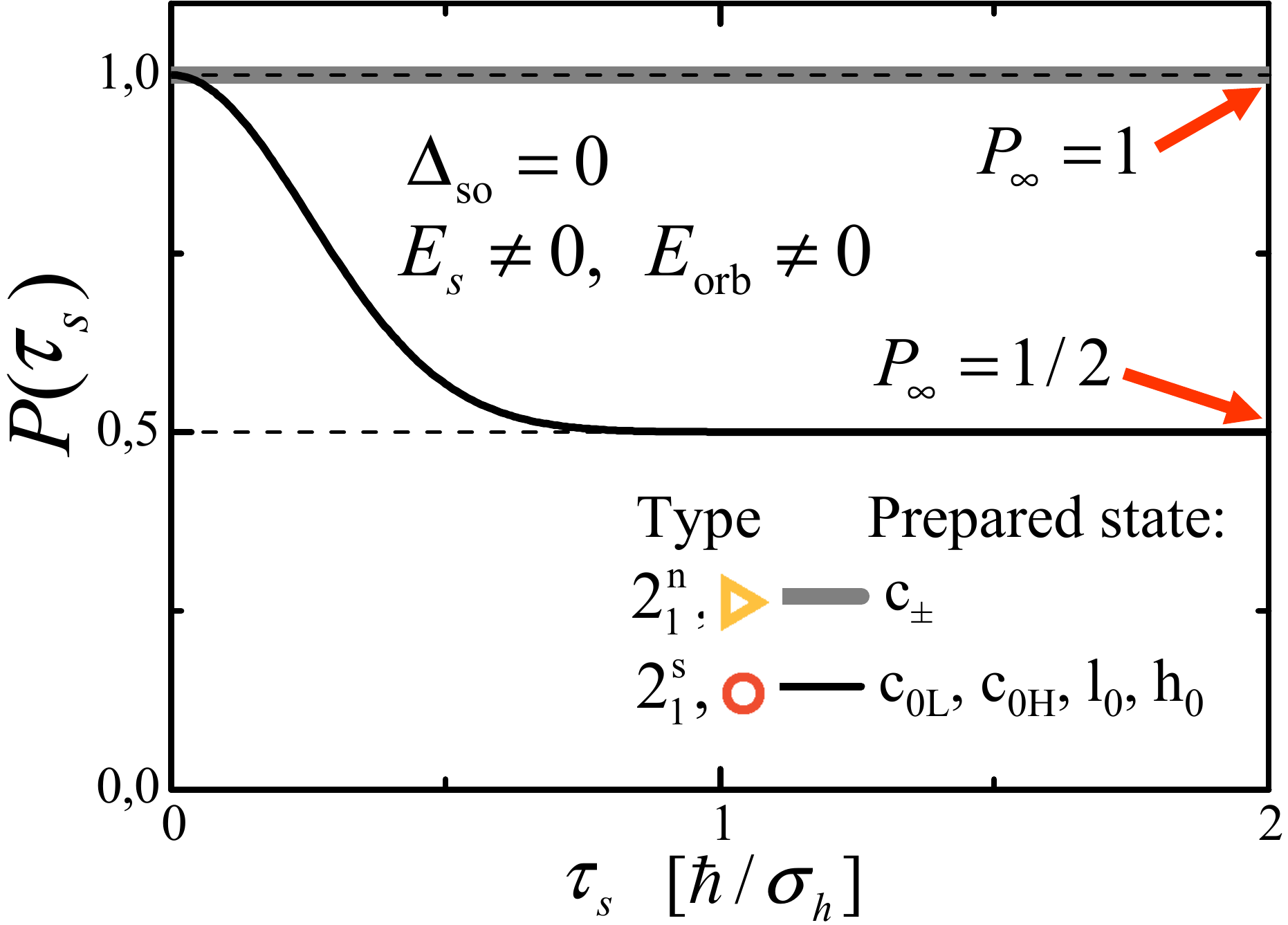}
     \vspace{-0.25cm}
   \caption{Return probabilities $P(\tau_s)$ for case B, i.e., zero spin-orbit coupling but coexistence of Zeeman and orbital effects due to the parallel component of the magnetic field (see labels in Fig.\ref{FG:caseB}). Depending on the prepared state the return probability have two behaviors: (i) type \mMRKtrr~---with $P_\infty\smeq 1$---when one of the two nonzero-spin states are prepared or (ii) type \mMRKrb~---with $P_\infty\smeq 1/2$, see Eq.\eqref{EQ:Pci}---when one of the remaining four states is prepared.}
   \label{FG:cBprob}
\end{figure}

\begin{figure*}[!t]
   \includegraphics[width=.97\textwidth]{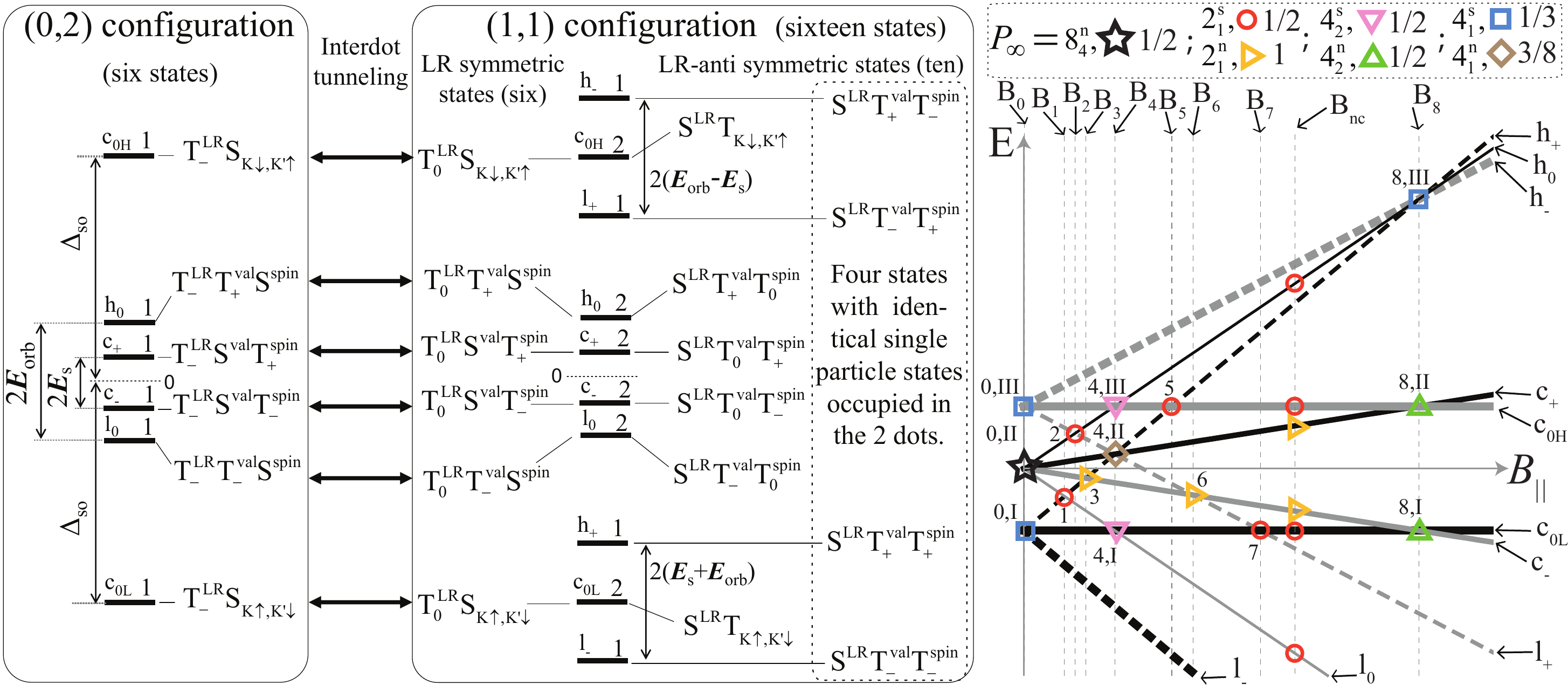}
\vspace{-0.23cm}
\caption{Case C: Saturation return probabilities, energies and states for a nanotube-based DQD with spin-orbit coupling in a parallel magnetic field, i.e., $\Delta_\mathrm{so}\smneq0$, $E_\mathrm{orb}\smneq0$ and $E_\mathrm{s}\smneq0$; the single dot single-particle spectrum is shown in Fig.\ref{FG:fig1}(c) (Left panel) Scheme of the (0,2) and (1,1) energy levels and the associated
states; the level degeneracies for finite magnetic field are indicated. Each one of the six (0,2) states is connected by inter-dot
tunneling to a single associated (1,1) state. For the
(1,1) charge state the LR-symmetric and LR-antisymmetric states
are distinguished. (Right panel) Energy levels in the (1,1) configuration as a function of the magnetic field; dashed levels contain only blocked states and therefore cannot be prepared. As shown in the legend---for the fifteen level crossings at $\mathrm{B}_i$, with $i=0,\dots,8$ and the six generic situations at any non-crossing field, $\mathrm{B_{nc}}$---the return probability presents seven different types of dynamics and $P_\infty$ is found to be $1/3$, $3/8$, $1/2$ or $1$ depending on both the magnetic field value and the prepared (0,2) state.}
\label{FG:caseC}
\vspace{-0.2cm}
\end{figure*}

Here the state counting prediction for the saturation
return probability (being $P_\infty^{sc,{\sMRKci}}\smeq
1/2$) coincides with the exact result. The $2\times 2$
effective Hamiltonian that describes the four cases follows
from Eq.\eqref{EQ:sz0HFImixing} and it is given by
\begin{equation}H_\mathrm{e}^{{\sMRKci}}=\left(\begin{array}{cc}0&2\delta h_z^{(0)}\\
2\delta h_z^{(0)}&0\end{array}\right).
\end{equation}
In Appendix \ref{AP:analytics} we derive that the return
probability is
\begin{equation}
P^{{\sMRKci}}(\tau_s)=\frac{1}{2}+\frac{{\rm e}^{-8\left(\sigma_h\tau_s/\hbar\right)^2}}{2},
\label{EQ:Pci}
\end{equation}
which is shown in Fig.\ref{FG:cBprob}. From Eq.\eqref{EQ:decaytime} one readily obtains $\tau_d\smeq\sqrt{\frac{\ln{2}}{8}}\hbar/\sigma_h\smap 0.2944 \hbar/\sigma_h$.

This type of dynamics is equivalent to the situation for a GaAs double dot in large
Zeeman field, in which case only the spin triplet
$\mathrm{S^{LR}T^{spin}_0}$ and the spin singlet
$\mathrm{T^{LR}_0 S^{spin}}$ are---neglecting tunneling exchange---degenerated: in that system the $z$-component of the hyperfine field (see Eq.\eqref{EQ:GaAsHFIterms}) is responsible for the mixing of these two states.\cite{Petta2005,CoishDQD,SchultenWolynes1978}
In the present case an equivalent physical situation is realized for four out of
the six possible prepared (0,2) states.

\textbf{\emph{In the case {\mMRKtrr}}} the prepared
state belongs to one of the two (0,2) energy levels
$\mathrm{c_+}$ and $\mathrm{c_-}$, where the states are the spin-polarized valley singlets:
\begin{equation}
\mathrm{T^{LR}_- S^{val} T^{spin}_{s_z} }~~~\mathrm{in~level~c_{s_z}}, ~\mathrm{with~s_z=\pm}.
\label{EQ:prepTRR}
\end{equation}
After separation the system is initialized at the associated LR-even state, $\mathrm{T^{LR}_0 S^{val} T^{spin}_{s_z}}$, which is degenerated with its LR-odd partner, $\mathrm{S^{LR} T^{val}_0 T^{spin}_{s_z}}$. As shown previously (see Eq.\eqref{EQ:szHFImixing}), the hyperfine interaction does not have matrix elements between these two states. The return probability is therefore $\tau_s$-independent and equal to one
\begin{equation}
P^{{\sMRKtrr}}(\tau_s)\smeq 1. \label{EQ:Ptrr}
\end{equation}
This somewhat counterintuitive result has been reported
previously in Ref.~\onlinecite{CulcerSarmaQD2010}, where the
same saturation value is found in silicon double dots---taking
into account a valley conserving HFI---for some
particular prepared states that, as here, the hyperfine field is unable to mix with other states.

\subsection{Case C: $\Delta_\mathrm{so}\smneq 0$ and $E_\mathrm{orb}\smneq 0$}
\label{SC:resCC}

Here we consider a double dot based on the single dot presented in Fig.\ref{FG:fig1}(c). Figure \ref{FG:caseC} shows the energy levels and states of the
(0,2) and (1,1) configurations, which follow from
Eqs.\eqref{EQ:02sols0En} and \eqref{EQ:11sols0}. The right
panel depicts the ten levels of the (1,1) spectrum as a function of the total magnetic field which is parallel to the tube axis. In the figure, we use solid lines for the six energy levels having LR-even states, those states are accessible from (0,2). By preparing a state in those levels the return probability experiment can be performed; at the evolution stage, it becomes accesible to the HFI each and every state belonging to others energy levels if they are degenerated with the energy level that holds the prepared state and, therefore, level crossings must be studied. As it is shown in Fig.\ref{FG:caseC}, it is intrinsic to this case the existence of level crossings for zero magnetic field and also at finite values of the magnetic field. Finite magnetic field crossings (involving at least one energy level with a LR-even state) occur at
\begin{subequations}
\be
\mathrm{B_1}=\frac{\Delta_{\rm so}}{\mu_{\rm B}(4 \mathrm{g_{orb}}+\mathrm{g_{s}})},~
\mathrm{B_2}=\frac{\Delta_{\rm so}}{\mu_{\rm B}(4 \mathrm{g_{orb}}-\mathrm{g_{s}})} ~, \ee
\be
\mathrm{B_3}=\frac{\Delta_{\rm so}}{\mu_{\rm B}( 2\mathrm{ g_{orb}}+2\mathrm{g_{s}})},~
\mathrm{B_4}=\frac{\Delta_{\rm so}}{2\mu_{\rm B} \mathrm{g_{orb}}} ~,
\ee
\be
\mathrm{B_5}=\frac{\Delta_{\rm so}}{\mu_{\rm B}( 2 \mathrm{g_{orb}}+\mathrm{g_{s}})} ,~
\mathrm{B_6}=\frac{\Delta_{\rm so}}{\mu_{\rm B}( 2 \mathrm{g_{orb}}-2\mathrm{g_{s}})} ~,
\ee
\be
\mathrm{B_7}=\frac{\Delta_{\rm so}}{\mu_{\rm B}( 2 \mathrm{g_{orb}}-\mathrm{g_{s}})} ,~
\mathrm{B_8}=\frac{\Delta_{\rm so}}{\mu_{\rm B} \mathrm{g_{s}}},
\ee
\label{EQ:Bcross}
\end{subequations}
these being positive values of the magnetic field given that we have assumed $E_\mathrm{orb}>E_\mathrm{s}$ and therefore $2g_\mathrm{orb}>g_\mathrm{s}$.
We identify twenty-one different situations, the six non-crossing cases and the fifteen level crossings. In the right panel of Fig.\ref{FG:caseC} we label each case according to its type of dynamics and we give the value of $P_\infty$ in the legend. We describe all those situations below.

\subsubsection{Zero magnetic field}

As shown in Fig.\ref{FG:fig1}(c) in the single-particle description, the spin-orbit
coupling breaks the 4-fold degeneracy resulting in
two Kramers doublets; the lowest energy Kramers doublet in the quantum dot $\xi$ (L or R) consists of the pair of time-reversal states $\ketLR{\xi
\mathrm{K}\uparrow}$ and $\ketLR{\xi\mathrm{K}'\downarrow}$,
whereas the highest energy Kramers doublet groups the states
$\ketLR{\xi\mathrm{K}'\uparrow}$ and $\ketLR{\xi
\mathrm{K}\downarrow}$. With double occupation of the right dot---states and levels shown in Fig.\ref{FG:caseC}---for the state at level $\mathrm{c_{0L}}$
($\mathrm{c_{0H}}$) the two electrons occupy the lowest
(highest) energy Kramers doublet in dot configuring
the non-degenerated ground (highest excited) state of the (0,2)
configuration. Right at the middle energy between the last two
states the levels $\mathrm{c}_+$, $\mathrm{c}_-$,
$\mathrm{l}_0$ and $\mathrm{h}_0$ are degenerated: they
correspond to the four (0,2) states with one electron in each
of the Kramers doublets.

\begin{figure}[t]
   \centering
   \includegraphics[width=.45\textwidth]{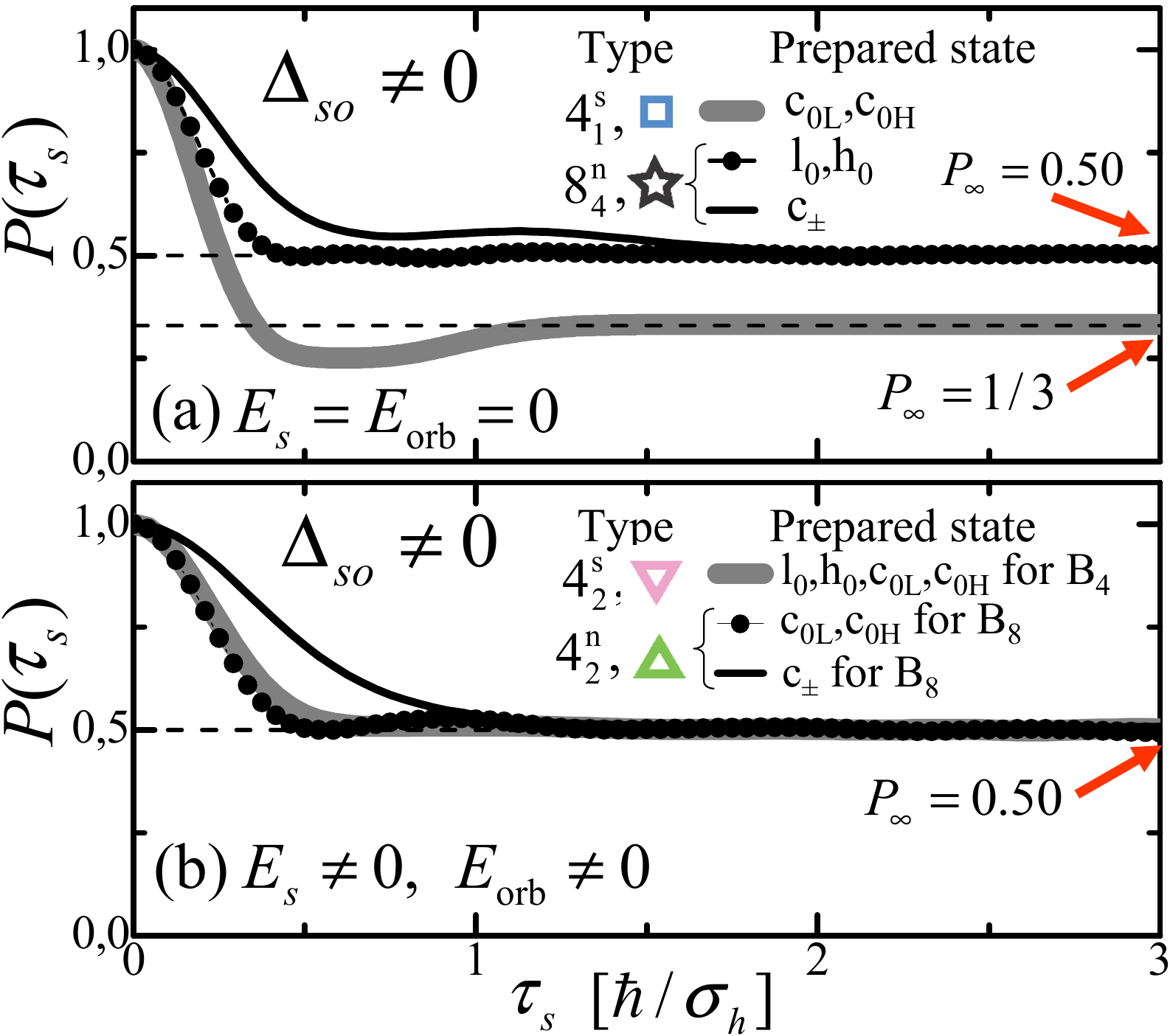}
\vspace{-0.3cm}
\caption{Return probabilities $P(\tau_s)$ that appear only for case C, i.e., $\Delta_\mathrm{so}\smneq 0$ in a parallel magnetic field. (a) At zero field one gets $P_\infty\smeq 1/3$ (type \bMRKsq) or $P_\infty\smeq 1/2$ (type \bMRKstFi). In the latter case the probability decays slower if the prepared state is a spin polarized valley singlet. (b) All cases give $P_\infty\smeq 1/2$. For $B_z\smeq\mathrm{B_4}$ (type \bMRKtrd) at the crossing (4,I) (or at the crossing (4,III)) $P(\tau_s)$ is independent of the prepared state. On the other hand (see text regarding the type \bMRKtru~case) at $B_z\smeq\mathrm{B_8}$ the shape of $P(\tau_s)$ differs depending of which state in the crossing (8,I) (or crossing (8,II)) is prepared. At the crossing (8,III) (the LR-even state at $\mathrm{h_0}$ is prepared) the behavior is the type \bMRKsq~presented in panel (a).}
   \label{FG:sche8B}
\vspace{-0.2cm}
\end{figure}

\textbf{For the type \mMRKsq}~presented in Fig.\ref{FG:caseC} the return probability behaves identically when the prepared
state is the (0,2) ground state (level $\mathrm{c_{0L}}$ identified below by $\sigma\smeq+$) or the (0,2) highest exited
state (level $\mathrm{c_{0H}}$ identified below by $\sigma\smeq-$). After separation the LR-even state becomes a member of
an (1,1) evolution subspace with $n_\mathrm{e}\smeq
4$, the four states are
\begin{eqnarray}
\ketLR{1,{\sMRKsq},\sigma}=\mathrm{T_{0}^{LR} S_{K \sigma, K'\overline{\sigma}}}~ &,&~\ketLR{2,{\sMRKsq},\sigma}=\mathrm{S^{LR} T_{K \sigma, K'\overline{\sigma}}}~, \nonumber\\
\ketLR{3,{\sMRKsq},\sigma}=\mathrm{S^{LR} T^{val}_+ T_{\sigma}^{spin}} &,&~\ketLR{4,{\sMRKsq},\sigma}=\mathrm{S^{LR}T^{val}_- T_{\overline{\sigma}}^{spin}},~~~~~
\label{EQ:sqBasis}
\end{eqnarray}
where in the spin triplets the subindex $\sigma$ is to be
interpreted as $+$ or $-$ instead of $\uparrow$ or
$\downarrow$, respectively. In this space the effective
Hamiltonian $H_\mathrm{e,\sigma}^{\sMRKsq}$ is analogous to a
double dot in GaAs described by the Hamiltonian in
Eq.\eqref{EQ:He0.375}. The equivalence with the components of
the hyperfine field given in Eq.\eqref{EQ:basisGaAs} is as
follows
\begin{subequations}
\begin{eqnarray}
{\delta h}_z \mapsto \sigma {\delta h}_z^{(0)} ~~~~~~~~~~~,&~&
\bar{h}_z \mapsto \sigma  \bar{h}_z^{(0)}~, \\
{\delta h}_x \mapsto {\delta h}_x^{(1)}-\sigma{\delta h}_y^{(2)} ~,&~& {\bar{h}}_x \mapsto {\bar{h}}_x^{(1)}-\sigma{\bar{ h}}_y^{(2)}~, \\
{\delta h}_y \mapsto {\delta h}_x^{(2)}+\sigma{\delta h}_y^{(1)} ~,&~& {\bar{h}}_y \mapsto {\bar{h}}_x^{(2)}+\sigma{\bar{ h}}_y^{(1)}.
\end{eqnarray}
\label{EQ:sqEffHFI}
\end{subequations}
From Eq.\eqref{EQ:variances} we obtain that the components of
the effective hyperfine field
\begin{equation}h_j^\mathrm{L}= {\bar{h}}_j +{\delta h}_j ~,~~h_j^\mathrm{R}= {\bar{h}}_j -{\delta h}_j~~~\mathrm{for}~j\smeq x,y,z~,
\label{EQ:dotHFIcomps}
\end{equation}
follow Gaussian distributions with identical standard
deviations $\sigma_{{\sMRKsq}}\smeq\sqrt{2}\sigma_h$. Therefore,
the situation is mapped exactly to a spin-only
double dot at zero field and the return probability is then
\begin{equation}P^{{\sMRKsq}}(\tau_s)=\frac{1}{3}\left(1+g\left(\tau_s \sigma_{{\sMRKsq}}/\hbar \right)+g^2\left(\tau_s \sigma_{{\sMRKsq}}/\hbar \right) \right),
\label{EQ:sqPanalyt}
\end{equation}
where $g(x)\equiv \mathrm{e}^{-2x^2} \left(1- 4 x^2\right)$. A
derivation of this result is presented in Appendix
\ref{AP:analytics}. This well-known shape, shown in Fig.\ref{FG:sche8B}(a), leads to
$P^{{\sMRKsq}}_\infty\smeq 1/3$ and so the state counting
estimation, which is this case is $1/4$, fails. Finally, for this shape the
decaying time defined in Eq.\eqref{EQ:decaytime} is $\tau_d \smap
0.18935\hbar/\sigma_h$.

\textbf{In the case {\mMRKstFi}} the prepared (0,2) state belong to
the energy level $\mathrm{c_+}$,
$\mathrm{c_-}$, $\mathrm{h_{0}}$ or $\mathrm{l_0}$. At
$\mathrm{B_0}\smeq 0$ the four levels have the same energy and therefore,
after the separation stage, the evolution subspace in (1,1)
includes their four LR-even partners and the associated LR-odd
states, i.e., $n_\mathrm{e}\smeq 8$. For all these prepared states we find $P_\infty^{\sMRKstFi}\smeq 0.50$, but the shape of $P^{\sMRKstFi}(\tau_s)$, shown in Fig.\ref{FG:sche8B}(a), depends on the prepared state. The decaying time is $\tau_d \smap 0.301\hbar/\sigma_h$ when the prepared state belongs to the level $\mathrm{c_+}$ or $\mathrm{c_-}$ and $\tau_d \smap 0.201\hbar/\sigma_h$ when the prepared state belongs to the level $\mathrm{h_{0}}$ or $\mathrm{l_0}$. From the full HFI Hamiltonian in Fig.\ref{FG:sche8A}(c) one can visualize the Hamiltonian of the hyperfine field for this subspace, $H_\mathrm{e}^{\sMRKstFi}$, by considering states at the octahedron vertices excluding $\mathrm{c_{0L}}$ and $\mathrm{c_{0H}}$ vertices. The difference in the decaying times arises because, as shown in Eq.\eqref{EQ:szHFImixing}, there are no direct matrix elements of the HFI Hamiltonian between the nonzero-spin LR-even/LR-odd partners; i.e., $m=0$ for $\mathrm{c_+}$ and $\mathrm{c_-}$ in Fig.\ref{FG:sche8A}(c).

\subsubsection{Finite magnetic field}

There is a set of situations---including any non-crossing value of the magnetic field, $\mathrm{B_{nc}}$---in which the return
probability is analogous to the cases without spin-orbit coupling
introduced in Sec.\ref{SC:resCB}. This happens for the
following situations (see crossings at Fig.\ref{FG:caseC}):
\begin{enumerate}
\item[(i)] When the prepared state belongs to the energy level $\mathrm{l_0}$ at $\mathrm{B_1}$ (or $\mathrm{h_0}$ at $\mathrm{B_2}$) because $\mathrm{T^{val}_+}$ and $\mathrm{T^{val}_-}$ are not mixed by the HFI. The return probability is type \bMRKci~($P_\infty\smeq 1/2$).
\item[(ii)] When the prepared state belongs to the energy level $\mathrm{c_-}$ at $\mathrm{B_3}$ (or at $\mathrm{B_6}$) because $\mathrm{T^{spin}_-}$ and $\mathrm{T^{spin}_+}$ are not mixed by the HFI. The return probability is type \mMRKtrr~($P_\infty\smeq 1$).
\item[(iii)] When the prepared state belongs either to the energy level $\mathrm{c_{0H}}$ (one electron in $\ketLR{K\downarrow}$ and the other in $\ketLR{K'\uparrow}$) at $\mathrm{B_5}$ or, to level $\mathrm{c_{0L}}$ (one electron in $\ketLR{K\uparrow}$ and the other in $\ketLR{K'\downarrow}$) at $\mathrm{B_7}$. At $\mathrm{B_5}$, the level $\mathrm{c_{0H}}$ crosses $\mathrm{h_+}$ (both electrons at the $\ketLR{K\uparrow}$ state) and, at $\mathrm{B_7}$, the level $\mathrm{c_{0L}}$ crosses $\mathrm{l_+}$ (both electros at the $\ketLR{K'\uparrow}$ state). The hyperfine field does not introduce mixing at these two crossings because in each one of them the LR-odd state have two single-particle quantum numbers different than the LR-even/LR-odd pair of states (see scheme of the full HFI Hamiltonian at Fig.\ref{FG:sche8A}(c)). The return probability is type \bMRKci~($P_\infty\smeq 1/2$).
\end{enumerate}

In addition, there are two special situations that lead to
Hamiltonians already presented (see right panel in
Fig.\ref{FG:caseC}):
\begin{enumerate}
\item[(a)] The crossing (4,II) is relevant for the
    prepared state belonging to the energy level
    $\mathrm{c_+}$ at $\mathrm{B_4}$, which crosses the
    levels $\mathrm{h_+}$ and $\mathrm{l_+}$
    simultaneously. The four states in the evolving
    subspace have spin $\mathrm{T^{spin}_+}$ function, and
    the system works as a valley double dot with type \mMRKrb~dynamics (see the Hamiltonian $H_\mathrm{e}^{\sMRKrb}$ in Eq.\eqref{EQ:He0.375}), therefore,
    the return probability follows Eq.\eqref{EQ:P3eights};
    i.e., $P_\infty\smeq3/8$.
\item[(b)] The crossing (8,III) is relevant for the prepared
    state belonging to the energy level $\mathrm{h_0}$ at
    $\mathrm{B_8}$, which crosses $\mathrm{h_+}$ and
    $\mathrm{h_-}$. The four states in the evolving
    subspace have a valley $\mathrm{T^{val}_+}$ function.
    Since only terms in the HFI proportional to the
    $\tau_0$ operator can mix them the situation can be
    mapped to a non-valley degenerated (i.e., spin-only) DQD as in GaAs. This
    is achieved by the following replacements in
    Eq.\eqref{EQ:GaAsHFIterms}
\begin{equation}
{\delta h}_j \mapsto {\delta h}_j^{(0)} ~,~~
\bar{h}_j \mapsto   \bar{h}_j^{(0)}~~\mathrm{for~}j=x,y,z. \label{EQ:sqEffHFI2}
\end{equation}

This is a type \mMRKsq~behavior and, as in crossings (0,I) and (0,III), the effective components follow a Gaussian distribution
with standard deviation $\sqrt{2}\sigma_h$. The return
probability is given in Eq.\eqref{EQ:sqPanalyt} and the
saturation value is $P_\infty\smeq1/3$.
\end{enumerate}

Finally, two double degenerated levels cross each other in the four
remaining situations; namely, cases labeled as \mMRKtrd~at (4,I) and (4,III), and
cases labeled as \mMRKtru~at (8,I) and (8,II). In each crossing the subspace of evolution has four states and two of them are connected to (0,2)
and therefore any of the two can be prepared. We find $P_\infty\smeq 0.50$ for
all the cases. However, as shown in Fig.\ref{FG:sche8B}(b), the dynamics in these two classes of crossings (see the right panel of Fig.\ref{FG:caseC}) are different: for type \mMRKtrd~the decaying time is, $\tau_d\smap 0.2585 \hbar/\sigma_h$, independently of the prepared state, while, on the other hand, for type \mMRKtru~the
decaying time is $\tau_d\smap0.2327 \hbar/\sigma_h$ if the prepared state belongs to the level $\mathrm{c_{0H}}$ or $\mathrm{c_{0L}}$, or, $\tau_d\smap0.4366 \hbar/\sigma_h$ if the prepared state belongs to the level $\mathrm{c_{-}}$ or $\mathrm{c_{+}}$. The independence on the prepared state found for type \mMRKtrd~is justified by noting that its dynamics its governed by the symmetric Hamiltonian for type \mMRKstFo---introduced in case A---in a smaller subspace that preserves its original symmetry. On the other hand, for type \mMRKtru, the strong dependency of the transient on the prepared state arises because the crossings (8,I) and (8,II) involve the LR-even spin polarized valley singlet states at level $\mathrm{c_{+}}$ or level $\mathrm{c_{-}}$. Those states dephase more slowly because, in accordance with Eq.\eqref{EQ:szHFImixing}, the HFI matrix elements with their LR-odd partners are zero.

\subsection{Case D: $\Delta_\mathrm{so}\smneq 0$ and $B_\mathrm{\perp}\smneq 0$}
\label{SC:resCD}

Here, in contrast to cases A, B and C, the spin projection along the direction of the magnetic field is not a good quantum number. The perpendicular
field, $B_\perp$, introduces a Zeeman energy $E_\mathrm{s}\smeq g\mu_B
B_\perp$ and zero diamagnetic effects. Due to the competition
between the Zeeman interaction and the spin-orbit coupling the
single-particle and single dot problem has eigenstates with
spin projection in the plane generated by the tube axis and the
direction of the magnetic field. In the following the tube axis is chosen along the $z$-direction and the magnetic field is applied along the $x$-direction. The solutions are
\begin{subequations}
\begin{eqnarray}
\ketLR{\mathrm{K}\vphantom{'},+}&=&\ketLR{\mathrm{K}} \otimes\left(\cos \frac{\eta}{2} \ketLR{\downarrow} + \sin \frac{\eta}{2} \ketLR{\uparrow} \right), \\
\ketLR{\mathrm{K}',+}&=&\ketLR{\mathrm{K}'} \otimes\left(\cos \frac{\eta}{2} \ketLR{\uparrow} + \sin \frac{\eta}{2} \ketLR{\downarrow} \right), \\
\ketLR{\mathrm{K}\vphantom{'},-}&=&\ketLR{\mathrm{K}} \otimes\left(\cos \frac{\eta}{2} \ketLR{\uparrow} - \sin \frac{\eta}{2} \ketLR{\downarrow} \right), \\
\ketLR{\mathrm{K}',-}&=&\ketLR{\mathrm{K}'} \otimes\left(\cos \frac{\eta}{2} \ketLR{\downarrow} - \sin \frac{\eta}{2} \ketLR{\uparrow} \right),
\end{eqnarray}
\label{EQ:eigvecBperp}
\end{subequations}
where $\eta\equiv \arctan E_\mathrm{s}/\Delta_\mathrm{so}$. The
eigenenergies  $E_\pm\smeq \pm \frac{1}{2}
\sqrt{\Delta_\mathrm{so}^2+E_\mathrm{s}^2}$ are shown in
Fig.\ref{FG:fig1}(d). From hereon we use the doublet index,
$\mathrm{d}=\pm$, to identify the two doublets. The
tunneling Hamiltonian is still diagonal in this basis, which
means that each (0,2) state mixes with only one (1,1) state and
the tunneling energy gap is $\Delta_t\smeq 2\sqrt{2}t$, as
before.

We now show that the perpendicular field situation reduces to a modified version of
cases already considered in the paper. When the single-particle states are well separated on the scale of the HFI
only the matrix elements of the HFI Hamiltonian between the
$\mathrm{d}=+$ solutions or in between the $\mathrm{d}=-$ solutions enter. The
$2\times2$ effective hyperfine field for the dot $\xi$ and the
doublet $\mathrm{d}$ can be writing as
\begin{equation}H_h^{\xi,\mathrm{d}}= (h_0^{\xi,\mathrm{d}} \sigma_0^{\mathrm{d}}  +h_x^{\xi,\mathrm{d}} \sigma_x^{\xi,\mathrm{d}}  + h_y^{\xi,\mathrm{d}} \sigma_y^{\xi,\mathrm{d}}  +h_z^{\xi,\mathrm{d}} \sigma_z^{\xi,\mathrm{d}}),
\label{EQ:doubletHFI}
\end{equation}
where $\xi=L,R$ and $\sigma_j^{\xi,\mathrm{d}}$ are Pauli
and identity matrices in the doublet space. The
coefficients are given by
\begin{subequations}
\begin{eqnarray}
h_0^{\xi,\mathrm{d}}&=& h_{x,\xi}^{(0)} \mathrm{d} \sin\eta,   \\
h_x^{\xi,\mathrm{d}}&=&h_{x,\xi}^{(1)}+h_{y,\xi}^{(2)} \mathrm{d} \cos\eta, \\
h_y^{\xi,\mathrm{d}}&=&h_{x,\xi}^{(2)}-h_{y,\xi}^{(1)} \mathrm{d} \cos\eta,\\
h_z^{\xi,\mathrm{d}}&=& -h_{z,\xi}^{(0)} \mathrm{d} \cos\eta.
\end{eqnarray}
\label{EQ:HFIperp}
\end{subequations}
The Hamiltonian $H_h^{\xi,\mathrm{d}}$ is equivalent to a spin
in a Zeeman field, plus an energy shift, $h_0^{\xi,\mathrm{d}}$,
which is irrelevant for the dynamics of the 2-level system. The
values of the effective field components depend on the angle
$\eta$, i.e., on the external perpendicular magnetic field.
Using the variances of the hyperfine components
$h_{j,\xi}^{(i)}$ (for $j\smeq x,y,z$ and $i\smeq0,1,2$) given
in Eq.\eqref{EQ:variances}, it follows that the variances of
the effective components in $H_h^{\xi,\mathrm{d}}$ are:
\begin{subequations}
\begin{eqnarray}
\widetilde{\sigma}_\perp^2 &\equiv&\braket{\left(h_x^{\xi,\mathrm{d}}\right)^2}=\braket{\left(h_y^{\xi,\mathrm{d}}\right)^2}=
\sigma_h^2 \left(1+\cos^2{\eta}\right),~~~~~~~ \\
\widetilde{\sigma}_\parallel^2 &\equiv&\braket{\left(h_z^{\xi,\mathrm{d}}\right)^2}= 2 \sigma_h^2 \cos^2{\eta}.
\label{EQ:variances2}
\end{eqnarray}
\end{subequations}

In the following cases the return probability behaves as situations already investigated:
\begin{enumerate}
\item[(a)]When $E_\mathrm{s}\smeq 0$ (i.e., $\eta\smeq0$) the results are presented in Sec.\ref{SC:resCC} for $B_\parallel \smeq 0$.
\item[(b)] When $E_\mathrm{s} \gg \Delta_\mathrm{so}$ (i.e., $\eta\rightarrow\pi/2$) the behavior is as the situation presented in Sec.\ref{SC:resCA} for nonzero magnetic field.
\end{enumerate}

Away from these two limits, for any of the four (0,2) prepared states having one electron in each doublet ($\mathrm{R},+$ and $\mathrm{R},-$) the return
probability goes smoothly from type \mMRKstFi~(at $E_\mathrm{s} \smeq
0$) to type \mMRKstFo~(at $E_\mathrm{s} \gg \Delta_\mathrm{so}$). The
situation is not so interesting since the saturation value
$P_\infty$ is always 0.50. The shape of $P(\tau_s)$ and the
decaying times $\tau_d$ for each prepared state depend on the
effective $8\times 8$ evolving (1,1) Hamiltonian. The result falls in
between the two above mentioned limits (a) and (b). As shown in the previous sections the Hamiltonian for $E_\mathrm{s}\smeq0$ is nonsymmetric, while it is symmetric for
$E_\mathrm{s}\gg\Delta_\mathrm{so}$. In the latter case, $P(\tau_s)$ is independent of the prepared state and a smaller decaying time is observed.
\begin{figure}[t]
   \centering
   \includegraphics[width=.47\textwidth]{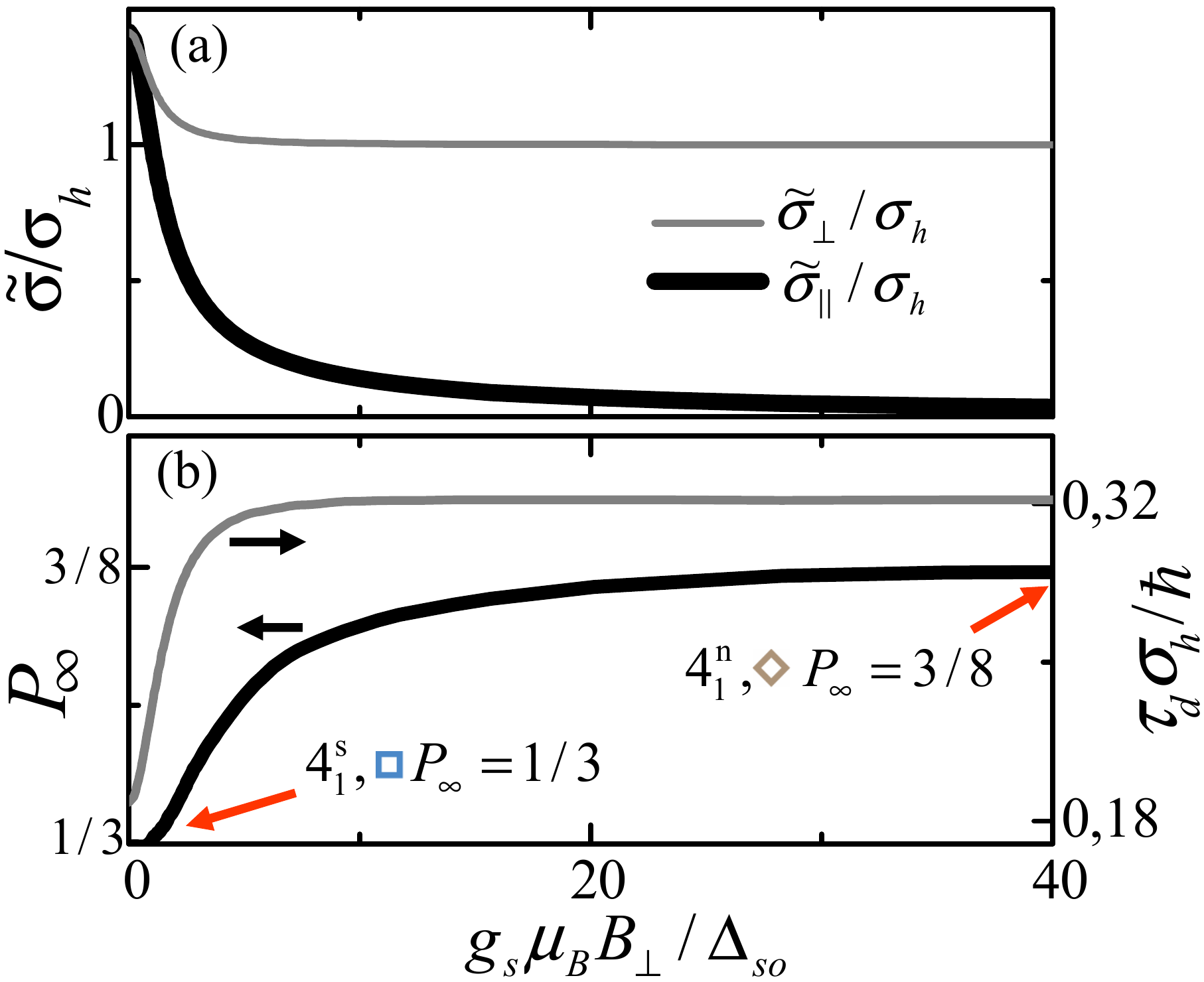}
   \vspace{-0.3cm}
   \caption{Case D. Results for a double dot with $\Delta_\mathrm{so}\smneq0$ as a
function of the Zeeman energy produced by the perpendicular
magnetic field. (a) Variances of the effective hyperfine field. (b) Saturation
value of the return probability, $P_\infty$, and decaying time,
$\tau_d$, for the cases when the prepared state is the ground
state or the highest energy excited state.}
   \label{FG:cCbx}
\vspace{-0.3cm}
\end{figure}

On the other hand, the value of the saturation return
probability changes if the (0,2) ground state or the (0,2)
highest excited state is prepared. The ground state is the
following Slater determinant:
\begin{eqnarray}
\ketLR{\Psi^{(0,2)}_\perp(\eta)}&=&\ketLR{{}_{\mathrm{R,K},-}^{\mathrm{R,K}',-}} \nonumber \\
&=&\frac{1}{\sqrt{2}}\ketLR{\mathrm{ T_{-}^{LR} S^{val} T^{spin}_0}}
\nonumber\\&&+\frac{1}{\sqrt{2}} \cos\eta \ketLR{\mathrm{ T_{-}^{LR} T^{val}_0S^{spin}}}\nonumber\\&&
 -\frac{1}{2}\sin\eta \ketLR{\mathrm{ T_{-}^{LR} S^{val} \left(T_+^{spin}+T_-^{spin}\right)}},~~~~~~
\label{EQ:gsPerp}
\end{eqnarray}
At zero-field $\eta\smeq0$ and the prepared state is $\mathrm{ T_{-}^{LR} S_{K\uparrow,K'\downarrow}}$, and we find type \mMRKsq~behavior with $P_\infty\smeq1/3$. For a dominant Zeeman energy $\eta\smeq\pi/2$ and the prepared state is the spin polarized valley singlet, $\mathrm{ T_{-}^{LR} S^{val} T_{-,x}^{spin}}$, with the spin triplet along the
$x$-direction; we then find type \mMRKrb~behavior with
$P_\infty\smeq3/8$.

For the intermediate magnetic field regime, with the (0,2)
state having two electrons in one doublet, the problem is
mapped to a double dot without the valley degree of freedom in
a hyperfine field with the variances of
Eq.\eqref{EQ:variances2}. Following Appendix
\ref{AP:analytics}, we obtain the return probability by
computing the averages of Eq.\eqref{EQ:analytP0} with the
probability distribution of Eq.\eqref{EQ:pdf3D} providing that
$\sigma_{h_{x,\xi}}\smeq
\sigma_{h_{y,\xi}}\smeq\widetilde{\sigma}_\parallel$ and
$\sigma_{h_{z,\xi}}\smeq\widetilde{\sigma}_\perp$.

We plot the standard deviations $\widetilde{\sigma}_\perp$ and
$\widetilde{\sigma}_\parallel$ in Fig.\ref{FG:cCbx}(a) as a
function of $E_\mathrm{s}/\Delta_\mathrm{so}$. There is an overall
reduction of the HFI when the magnetic field increases, which
explains the larger decaying time $\tau_d$ plotted in
Fig.\ref{FG:cCbx}(b). Moreover, the standard deviation of the
perpendicular component goes to zero, and in this limit the HFI
becomes 2-dimensional, approaching type \mMRKrb~behavior.
Thus, the initial state in Eq.\eqref{EQ:gsPerp} becomes a
valley singlet and the hyperfine interaction is unable to mix
directly with the $\mathrm{T^{val}_0}$ partner. The saturation value of the return probability is presented in
Fig.\ref{FG:cCbx}(b). $P_\infty$ can take any value between 1/3
and 3/8 as a function of the magnetic field. This interesting result allows a direct measurement of the spin-orbit coupling splitting and the hyperfine inter-valley mixing (see $x$ and $y$-components of the effective HFI in Eq.\eqref{EQ:HFIperp}) and its relation with the valley conserving hyperfine mixing ($z$-component of the effective HFI).

\section{Conclusions}
We have analyzed the expected return probabilities for a
dephasing measurement in clean carbon nanotube based double
quantum dots. We have focused on the intrinsic properties and
therefore neglected disorder induced valley mixing and also
Coulomb exchange, which are
predicted\cite{WeissFlensberg2010} and
measured\cite{Jespersen2010} to be small in multi-electron
dots. In a forthcoming publication, we study the influence of
valley mixing.

We have shown that a multiple number of scenarios exists for the return probability experiment, due to the valley degree of freedom (as in Si-based DQDs,\cite{CulcerSarmaQD2010}) which makes the system very different from a double dot in a 2-dimensional electron gas (2DEG). Here, more
specifically, these scenarios are due to: (i) the non-trivial structure of the hyperfine coupling with the $^{13}$C nuclei that affects both the electron spin and valley degrees of freedom; (ii) the experimental preparation protocol that determines which of the six (0,2) states is prepared; (iii) the
availability of sixteen (1,1) states for the system in the evolution stage; (iv) the change (for every possible prepared state) of the subset of (1,1) states accessible in the evolution stage, and (v) the manifold of six possible return (0,2) states. The last point is an important difference to the 2DEG-based double dots, where only spin singlet returns to
(0,2). Here the projection onto (0,2) is more generally determined by symmetry of the wavefunction, allowing only even left-right components to return. The level structure of the sixteen (1,1) and six (0,2) states depends on the values of the spin-orbit coupling and of the external magnetic field, through
the Zeeman interaction, diamagnetic effects, or both.

In a 2DEG-based double
dot the return probability shows two different behaviors being,
type \mMRKsq~dynamics ($P_\infty\smeq1/3$) for zero field,
or, type \bMRKci~dynamics ($P_\infty\smeq1/2$) in the high
magnetic field limit (we use the labeling introduced in
Sec.\ref{SC:results}). Here, depending on the parameters we
find seven additional types of dynamics leading to saturation
values $P_\infty\smeq 3/8$, $0.4$, $1/2$ and $1$. The results for all the nine situations are presented in Table \ref{TB:summary}.

Type \mMRKstSi~dynamics can be found for zero magnetic field in the
absence of spin-orbit coupling. In the cases with
$\Delta_\mathrm{so}\smeq 0$, the behavior of a  Zeeman
interaction only system (type \mMRKstFo~and \mMRKrb) is
very different from the situation with both Zeeman and
diamagnetic effects (type \bMRKci~and \mMRKtrr). For
nonzero spin-orbit, the breaking of the spin degeneracy
replaces, for zero magnetic field, type \mMRKstSi~behavior
with type \mMRKstFi~and the well known type \mMRKsq. At
finite magnetic fields (when considering both Zeeman and
diamagnetic effects) situations \bMRKci, \mMRKtrr,
\mMRKtrd, \mMRKtru, \mMRKrb~and once again
\mMRKsq~can be obtained depending on the value of the
magnetic field and on the prepared state.

\begin{table}[!t]
\caption{\label{TB:summary} Summary of the return probability results for the nine situations investigated.} \centering {\scriptsize
\begin{tabular}{l|c|c|c|c|c|c}
$P_\infty$&Type&Cases&$n_\mathrm{e}$&$n_\mathrm{(0,2)}$&\multicolumn{2}{c}{$\tau_d \sigma_h/\hbar$}\\ &&&&&for $\mathrm{S^{val} T^{spin}_{\pm}}$&for others\\
\hline
\hline
$1/3$&{\mMRKsq}&C&4&1&-&0.18935\\
\hline
$3/8$&{\mMRKrb}&A and C, $\mathbf{B}\smneq 0$&4&1&0.3218&-\\
\hline
$0.40$&{\mMRKstSi}&A and B, $\mathbf{B}\smeq 0$&16&6&0.185&0.149\\
\hline
$1/2$&{\bMRKci}&B and C, $\mathbf{B}\smneq 0$&2&1&-&0.2944\\
$0.5$&{\mMRKstFo}&A, $\mathbf{B}\smneq 0$&8&4&-&0.238\\
$0.5$&{\mMRKstFi}&C, $\mathbf{B}\smeq 0$&8&4&0.3014&0.2011\\
$0.5$&{\mMRKtrd}&C, $\mathbf{B}\smeq \mathrm{B_4}$&4&2&-&0.2585\\
$0.5$&{\mMRKtru}&C, $\mathbf{B}\smeq \mathrm{B_8}$&4&2&0.4366&0.2327\\
\hline
$1$&{\mMRKtrr}&B and C, $\mathbf{B}\smneq 0$&2&1&$\infty$&-\\
\end{tabular}}
\vspace{-0.2cm}
\end{table}

In only two out of these seven novel situations (types \mMRKtrr~and \mMRKrb), the return probability is associated with the system returning to the original prepared (0,2) state (see point (v) above). In all the remaining cases the system can be measured and also prepared in \emph{more than one} (0,2) state, and therefore the functional dependence return of
the probability on $\tau_s$ depends on both the prepared state and the dynamics type. We have defined a shape-independent decaying time and we find, $\tau_d\smap 0.149 \hbar/\sigma_h$,
in the fastest case (for zero spin prepared states in type \mMRKstSi~dynamics) and a infinite decaying time (since the system does not decay for type \mMRKtrr~dynamics) in the slowest case. It should be noted that we have assumed throughout that the time scale of the experiment is much smaller than $T_1$, the inelastic dephasing time; otherwise inelastic processes would relax the system to the ground state invalidating the investigation of the
dephasing in the return probability experiment as purely due to the hyperfine interaction. Therefore, the transient that defines the decaying time presented in Table \ref{TB:summary}, $\tau_d$, is to be understood as valid only for evolving times, $\tau_s$, smaller than $T_1$.

In addition to the robustness of case \mMRKtrr~we find, for types \mMRKstSi, \mMRKstFi, \mMRKrb~and \mMRKtru, asymmetries and/or long
decaying times if the prepared state is a
spin polarized valley singlet. The reason is that the hyperfine coupling of
Eq.\eqref{EQ:HFIdqd1p} does not introduces direct matrix
elements between the LR-even spin polarized valley singlets and
their LR-odd partners (see Eq.\eqref{EQ:szHFImixing}); these two states can only be mixed by an inhomogeneous (in LR space) time-reversal symmetric term (i.e., spin-orbit coupling like), which does not appear in the HFI. In Ref.~\onlinecite{PalyiB09} it has been shown that this intrinsically anisotropic hyperfine field gives rise to a dip in the spin-blockade signal as a function of the orbital field.

Here, we have shown that this property of the hyperfine field
also leads to an interesting behavior of the return probability
when the (0,2) ground (or the highest excited) state is
prepared for the case with nonzero spin-orbit coupling and the
magnetic field is perpendicular to the tube axis, $B_\perp$. As a function of
the Zeeman energy, $E_\mathrm{s}$, the groundstate changes from a
spin-unpolarized state (for $E_\mathrm{s}\smeq0$) to a spin-polarized
valley singlet (for $E_\mathrm{s}/\Delta_\mathrm{so}\gg 1$ ) and the
saturation return probability goes from 1/3 (type \mMRKsq, i.e., effective 3-dimensional HFI) to 3/8 (type \mMRKrb, i.e., effective 2-dimensional HFI in the \emph{valley} double dot). Measurement of $P_\infty$ and $\tau_d$ as a function of $B_\perp$ would test the
validity of the hyperfine Hamiltonian in Eq.\eqref{EQ:HFIdqd1p}
allowing, in principle, for the determination of the spin-orbit
coupling and the hyperfine strength $\sigma_h$.

Only a single return probability experiment\cite{ChurchillFlensberg2009}
has been reported in a carbon-based double dot. The result, only available for zero magnetic field, was an
unexpected small return probability $P_\infty\smap 1/6$, that
cannot be explained within the model presented here. We have
shown that the minimum saturation return probability for coherent mixing is
1/3, similar to the situation in a spin-only double dot. Incoherent mixing will also not explain the experimental findings, since there the minimum return
probability is 1/4, which could happen for crossings type
\mMRKrb~or \mMRKsq. We also note that by having
worked in the high detuning limit in which the tunneling exchange is negligible we have obtained lower
bounds of $P_\infty$, since it is known that
this coupling reduces the effectiveness of the
hyperfine mixing and thus increases $P_\infty$.\cite{CoishDQD} One could
speculate that valley mixing is responsible for the
discrepancy. In a forthcoming publication, we discuss the role
of such mixing, which however also cannot explain the small
ratio between $P(0)$ and $P_\infty$ seen in experiment.

Clearly more experimental work is needed to better understand
the rather rich structure of the carbon based double dots
system, including the dependence of $P(\tau_s)$ on magnetic
field. One interesting aspect would be to design alternative
preparation protocols for being able to select different
initial (0,2) states.

\acknowledgments We acknowledge useful discussions with S. Weiss, K. Grove-Rasmussen, H. Churchill, F. Kuemmeth, M. Leijnse, C. Marcus, B. Trauzettel and G. Burkard.
\appendix

\section{Analytical calculation of $P(\tau_s)$}
\label{AP:analytics}

\subsection{Mixing of a $\mathrm{T^{LR}_0}$, $\mathrm{S^{LR}}$ pair}
This case (type {\bMRKci} in Sec.\ref{SC:results}) is valid whenever a LR-symmetric state and its partner LR-antisymmetric are mixed by the hyperfine interaction and no other states are involved the evolution Hamiltonian $H_\mathrm{e}$. In such a case the dynamics in the evolution subspace $S_\mathrm{e}$ is governed by the simple Hamiltonian:
\begin{equation}H_\mathrm{e}^{{\sMRKci}}=\left(\begin{array}{cc}0&2\delta h_z^{(0)}\\
2\delta h_z^{(0)}&0\end{array}\right).
\end{equation}This Hamiltonian is valid for four out of the six $\mathrm{T^{LR}_0}$, $\mathrm{S^{LR}}$ pairs, specifically the zero-spin cases (see Eq.\eqref{EQ:sz0HFImixing} and Eq.\eqref{EQ:szHFImixing}). Here we use the notation of Eq.\eqref{EQ:11LRsym} and Eq.\eqref{EQ:LRodd} for two single-particle single dot eigenstates with quantum numbers $n$ and $n'$ ($n$ and $n'$ must correspond to solutions with opposite spin projections).

Following Eq.\eqref{EQ:Tconnect} for $\tau_s\smeq0$ the state $\ketLR{\mathrm{LR}_\mathrm{even}^{n,n'}}$ is initialized. The system evolves as,
\begin{eqnarray}
  \ketLR{\mathrm{e}(\tau_s)} &=& \cos \left(\frac{2\delta h_z^{(0)}\tau_s}{\hbar}\right) \ketLR{\mathrm{LR}_\mathrm{even}^{n,n'}} \nonumber \\&& + \sin \left(\frac{2\delta h_z^{(0)}\tau_s}{\hbar}\right) \ketLR{\mathrm{LR}_\mathrm{odd}^{n,n'}}.
\end{eqnarray}
Therefore, for the realization $r_i$ of the hyperfine field, the probability to find the system in the LR-even combination (i.e., to measure an (0,2) charge state after the adiabatical joining stage) is just:
\begin{equation}p_{r_i}^{{\sMRKci}}(\tau_s) = \cos^2 \left(\frac{2\delta h_z^{(0)}\tau_s}{\hbar}\right).
\end{equation}
We have to average the last oscillating function over the normal distribution that describes the hyperfine field. The inhomogeneous HFI component is given by $\delta h_z^{(0)}\smeq (h_{z,\mathrm{L}}^{(0)}-h_{z,\mathrm{R}}^{(0)})/2$ and the standard deviations for the components $h_{z,\mathrm{L}}^{(0)}$ and $h_{z,\mathrm{R}}^{(0)}$ are both $\sqrt{2} \sigma_h$. Then the standard deviation of the Gaussian distribution for the frequency variable $\omega\equiv 2\delta h_z^{(0)}/\hbar$ is $\sigma_\omega \equiv 2\sigma_h /\hbar$. The final result is,
\begin{eqnarray}
P^{{\sMRKci}}(\tau_s)&=&\frac{1}{\sqrt{2\pi}\sigma_\omega}\int_{-\infty}^{\infty}d\omega \cos^2\left(\omega \tau_s \right){\rm e}^{-\frac{\omega^2}{2{\sigma_\omega}^2}} \nonumber\\
&=&\frac{1+{\rm e}^{-2\left(\sigma_\omega\tau_s\right)^2}}{2}=\frac{1}{2}+\frac{{\rm e}^{-8\left(\sigma_h\tau_s/\hbar\right)^2}}{2}.
\end{eqnarray}

\subsection{Mixing of a $\mathrm{T^{LR}_0}$ state with three $\mathrm{S^{LR}}$ states - Analytical approach}
The effective HFI Hamiltonians $H_\mathrm{e}^{{\sMRKsq}}$ and $H_\mathrm{e}^{{\sMRKrb}}$ presented above (and also the intermediate situations we find for $\Delta_\mathrm{so}$ in a perpendicular magnetic field) can be mapped to the problem of dephasing in a non-valley degenerated DQD as the one given in Eq.\eqref{EQ:HFIdqd1pGaAs} and Eq.\eqref{EQ:GaAsHFIterms}. Here we present the derivation of the latter case and then we particularize for the three mentioned cases.

The electron spin in each dot ($\xi\smeq\mathrm{L,R}$) follows the evolution operators,
\begin{equation}U^\xi(\tau_s)=\cos\left(\omega_\xi \tau_s\right)\sigma_0 - \ci~ \boldsymbol{\sigma} \cdot \hat{{\bf n}}_\xi \sin\left(\omega_\xi \tau_s\right),
\label{EQ:EVopLOC}
\end{equation}
that describe precession around the direction of the hyperfine field with frequencies,
\begin{equation}\omega_\xi=\frac{1}{\hbar}\sqrt{\left(h_{x,\xi}\right)^2+\left(h_{y,\xi}\right)^2+\left(h_{z,\xi}\right)^2}. \end{equation}
The normalized vectors in Eq.\eqref{EQ:EVopLOC} point in the direction of the local hyperfine field.
\begin{equation}\hat{{\bf n}}_\xi =\frac{1}{\hbar \omega_\xi} \left(h_{x,\xi},h_{y,\xi},h_{z,\xi} \right).
\end{equation}
Using the former evolution operators it follows that an (1,1) Slater determinant in the double dot evolves as,
\begin{equation}U(\tau_s) \left|{}_{\mathrm{L},\sigma_L}^{\mathrm{R},\sigma_R}\right>= \sum_{\sigma_1 \sigma_2} U^{\mathrm{L}}_{\sigma_1 \sigma_L}(\tau_s) U^{\mathrm{R}}_{\sigma_2\sigma_R}(\tau_s) \left|{}_{\mathrm{L},\sigma_1}^{\mathrm{R},\sigma_2}\right>.
\label{EQ:TimeEvolOp0}
\end{equation}

At $\tau_s\smeq 0$ the system is initialized in $\mathrm{T_{0}^{LR}S^{spin}}$, the only available LR-symmetrical state in Eq.\eqref{EQ:basisGaAs}. In order to time evolve the last two-particle state, we use its Slater determinant version, $\left( \ketLR{{}_{\mathrm{L}\uparrow}^{\mathrm{R}\downarrow}}+\ketLR{{}_{\mathrm{R}\uparrow}^{\mathrm{L}\downarrow}}\right)/\sqrt{2}$. We apply the evolution operator of Eq.\eqref{EQ:TimeEvolOp0} and project the result back to the initial LR-even state,
\begin{widetext}
\begin{eqnarray}
\sqrt{p_{r_i}(\tau_s)}=\braLR{\mathrm{T_{0}^{LR}S^{spin}}}U(\tau_s)\ketLR{\mathrm{T_{0}^{LR}S^{spin}}}&=&\frac{1}{2}\left(U^{\mathrm{R}}_{\uparrow\uparrow}U^{\mathrm{L}}_{\downarrow \downarrow}+U^{\mathrm{R}}_{\downarrow\downarrow}U^{\mathrm{L}}_{\uparrow \uparrow}-U^{\mathrm{R}}_{\uparrow\downarrow}U^{\mathrm{L}}_{\downarrow \uparrow}+U^{\mathrm{R}}_{\downarrow\uparrow}U^{\mathrm{L}}_{\uparrow \downarrow}\right) \\
&=& \cos\left(\omega_\mathrm{R} \tau_s\right)\cos\left(\omega_\mathrm{L} \tau_s\right) \nonumber \\
&&+\frac{\sin\left(\omega_\mathrm{L} \tau_s\right) }{\hbar \omega_\mathrm{L}}\frac{\sin\left(\omega_\mathrm{R} \tau_s\right)}{\hbar \omega_\mathrm{R}} \left(h_{x,\mathrm{R}}h_{x,\mathrm{L}}+h_{y,\mathrm{R}}h_{y,\mathrm{L}}+h_{z,\mathrm{R}}h_{z,\mathrm{L}}\right). \nonumber
\end{eqnarray}
The probability of finding the system in the original state---i.e., of measuring its (0,2) partner state after the adiabatical joining stage---is then the square of the former amplitude.

Since the probability density functions of the hyperfine field components (Gaussian distributions with zero mean) are even, the odd powers terms in those components within ${p_{r_i}(\tau_s)}$ do not contribute to the average. We arrive to the well known expression:
\begin{equation}P(\tau_s)= \braket{\cos^2\left(\omega_\mathrm{L} \tau_s\right)}_{\rm L-HF}\braket{\cos^2\left(\omega_\mathrm{R} \tau_s\right)}_{\rm R-HF}+\sum_{j=x,y,z}\left[\braket{\left(\frac{h_{j,\mathrm{L}}}{\hbar \omega_\mathrm{L}}\right)^2\sin^2\left(\omega_\mathrm{L} \tau_s\right)}_{\rm L-HF}\braket{\left(\frac{h_{j,\mathrm{R}}}{\hbar \omega_\mathrm{R}}\right)^2\sin^2\left(\omega_\mathrm{R} \tau_s\right)}_{\rm R-HF}\right],
\label{EQ:analytP0}
\end{equation}
where $\braket{ Q\left(h_{x,\xi},h_{y,\xi},h_{z,\xi}\right) }_{\xi\mathrm{-HF}}$ stands for the average of the function $Q\left(h_{x,\xi},h_{y,\xi},h_{z,\xi}\right)$  over the hyperfine fields of the $\xi$ (L or R) dot. The probability density function (in the 3-dimensional space of $h_{x,\xi}$,$h_{y,\xi}$ and $h_{x,\xi}$) is
\begin{equation}
F_{\rm 3D}\left(h_{x,\xi},h_{y,\xi},h_{z,\xi}\right)=\frac{1}{\sqrt{2\pi} \sigma_{h_{x,\xi}}} \frac{1}{\sqrt{2\pi} \sigma_{h_{y,\xi}}}  \frac{1}{\sqrt{2\pi} \sigma_{h_{z,\xi}}} \exp\left[-\frac{1}{2}\left(\left(\frac{h_{x,\xi}}{\sigma_{h_{x,\xi}}}\right)^2
+\left(\frac{h_{y,\xi}}{\sigma_{h_{y,\xi}}}\right)^2+\left(\frac{h_{z,\xi}}{\sigma_{h_{z,\xi}}}\right)^2\right)\right].
\label{EQ:pdf3D}
\end{equation}

Where we have used the following standard deviations of the Gaussian distributions for the HFI components,
\begin{equation}\sigma_{h_{x,\xi}}\equiv{\rm Std}\left(h_{x,\xi}\right)~,~\sigma_{h_{y,\xi}}\equiv{\rm Std}\left(h_{y,\xi}\right)~,~\sigma_{h_{z,\xi}}\equiv{\rm Std}\left(h_{z,\xi}\right).
\end{equation}
We see below that the degree of anisotropy arising from a difference in the last quantities affects the averages in Eq.\eqref{EQ:analytP0} and therefore the return probability.

\subsubsection{Statistical isotropic 3-dimensional effective hyperfine field} When the effective Hamiltonian is $H_\mathrm{e}^{{\sMRKsq}}$ the three hyperfine components share the same standard deviation, $\sigma^\xi_{{\sMRKsq}}\smeq\sigma_{h_{x,\xi}}\smeq\sigma_{h_{y,\xi}}\smeq\sigma_{h_{z,\xi}}$ and
therefore,
\begin{eqnarray}
\braket{Q\left(h_{x,\xi},h_{y,\xi},h_{z,\xi}\right)}_{\rm \xi-HF}&=&\int_{-\infty}^{\infty} dh_{x,\xi}\int_{-\infty}^{\infty} dh_{y,\xi}\int_{-\infty}^{\infty} dh_{z,\xi} Q\left(h_{x,\xi},h_{y,\xi},h_{z,\xi}\right)  F_{\rm 3D}\left(h_{x,\xi},h_{y,\xi},h_{z,\xi}\right) \nonumber \\
&=& \int^{\pi}_{0} d\theta \int^{2\pi}_{0} d\varphi  \int^{+\infty}_{0} d r
Q\left(r,\theta,\varphi\right)  F_{\rm 3D} \left(r,\theta,\varphi\right),  \\
F_{\rm 3D} \left(r,\theta,\varphi\right) &=&  \frac{r^2 \sin(\theta)}{\left(\sqrt{2\pi} {\sigma_{{\sMRKsq}}^\xi}\right)^3}  \exp\left({-\frac{r^2}{2 \left({\sigma_{{\sMRKsq}}^\xi}\right)^2}}\right),
\end{eqnarray}
\end{widetext}
where we have made a change to spherical coordinates with $r^2\smeq \left(h_{x,\xi}\right)^2+\left(h_{y,\xi}\right)^2+\left(h_{z,\xi}\right)^2$.

The two types of averages that appear in Eq.\eqref{EQ:analytP0} are obtained by integrating,
\begin{eqnarray}
C_\mathrm{3D}(\tau_s)&\equiv& \braket{\cos^2\left(\omega_\xi \tau_s\right)}_{\rm \xi-HF} \nonumber \\
 &=& \frac{1}{2}\left(1 + g\left(\tau_s {\sigma_{{\sMRKsq}}^\xi}/\hbar  \right)\right),\\
S_\mathrm{3D}(\tau_s)&\equiv& \braket{\left(\frac{h_{j,\xi}}{\hbar\omega}\right)^2 \sin^2\left(\omega_\xi \tau_s\right)}_{\rm \xi-HF} \nonumber \\
 &=& \frac{1}{6}\left(1 - g\left(\tau_s {\sigma_{{\sMRKsq}}^\xi}/\hbar  \right)\right),\\
g(x)&\equiv& \mathrm{e}^{-2x^2} \left(1- 4 x^2\right).
\end{eqnarray}
Since $\hbar\omega_\xi \smeq r$ the arguments of the sinusoidal functions in spherical coordinates is $r \tau_s /\hbar$. Note that in $S_\mathrm{3D}$ the average is independent of the direction of the hyperfine component $(j\smeq x,y,z)$. This is valid here because the effective hyperfine field is statistically isotropic. Then, for simplicity, the integral is computed using the $z$-component, $h_{j,\xi}\smeq r\cos(\theta)$.

In the investigated situations in Sec.\ref{SC:results} the standard deviations are equal in the two dots, $\sigma_{{\sMRKsq}}\smeq{\sigma_{{\sMRKsq}}^\mathrm{R}}\smeq{\sigma_{{\sMRKsq}}^\mathrm{L}}$; the return probability becomes
\begin{equation}P^{{\sMRKsq}}(\tau_s)=\frac{1}{3}\left(1+g\left(\tau_s \sigma_{{\sMRKsq}}/\hbar \right)+g^2\left(\tau_s \sigma_{{\sMRKsq}}/\hbar \right) \right).
\end{equation}
\subsubsection{Statistical isotropic 2-dimensional effective hyperfine field} As discussed in Sec.\ref{SC:subsubCA} the effective Hamiltonian $H_\mathrm{e}^{{\sMRKrb}}$ can be mapped to the GaAs zero-field double dot but it must be assumed that the effective $z$-component of the hyperfine field is absent. As $h_{z,\xi}$ is identically zero we must not average over it, therefore, instead of the probability density function given in Eq.\eqref{EQ:pdf3D} a two dimensional probability density function must be used. Adding the fact that the standard deviations of the in-plane components are identical, $\sigma^\xi_{{\sMRKrb}}\smeq\sigma_{h_{x,\xi}}\smeq\sigma_{h_{y,\xi}}$, it becomes useful to work in polar coordinates. The averages are obtained as follows,
\begin{widetext}
\begin{eqnarray}
\braket{Q\left(h_{x,\xi},h_{y,\xi}\right)}_{\rm \xi-HF}&=&\iint dh_{x,\xi} dh_{y,\xi} Q\left(h_{x,\xi},h_{y,\xi}\right)  F_{\rm 2D}\left(h_{x,\xi},h_{y,\xi}\right) \nonumber \\
&=& \int^{2\pi}_{0} d\varphi  \int^{+\infty}_{0} d r
Q\left(r,\varphi\right)  F_{\rm 2D} \left(r,\varphi\right),  \\
F_{\rm 2D} \left(r,\varphi\right) &=&  \frac{r}{\left(\sqrt{2\pi} {\sigma_{{\sMRKrb}}^\xi}\right)^2}  \exp\left({-\frac{r^2}{2 \left({\sigma_{{\sMRKrb}}^\xi}\right)^2}}\right).
\end{eqnarray}
\end{widetext}
Then we define the 2-dimensional averages needed to compute the return probability as:
\begin{eqnarray}
C_\mathrm{2D}(\tau_s)&\equiv& \braket{\cos^2\left(\omega_\xi \tau_s\right)}_{\rm \xi-HF},  \label{EQ:defS2C2} \\
S_\mathrm{2D}(\tau_s)&\equiv& \braket{\left(\frac{h_{j,\xi}}{\hbar\omega}\right)^2 \sin^2\left(\omega_\xi \tau_s\right)}_{\rm \xi-HF}~~j=x,y.\nonumber
\label{EQ:defS2C2}
\end{eqnarray}
As in $S_\mathrm{3D}$ the average in $S_\mathrm{2D}$ is independent of the direction of the hyperfine component $(j\smeq x,y)$. This is valid here because the effective in-plane hyperfine field is statistically isotropic. Then, for simplicity, the integral is computed using the $x$-component, $h_{j,\xi}\smeq r\cos(\varphi)$.
The obtained results are presented and discussed in Sec.\ref{SC:subsubCA}.
\bibliographystyle{apsrev4-1}
\bibliography{baseCNT}

\begin{thebibliography}{31}%
\makeatletter
\providecommand \@ifxundefined [1]{%
 \@ifx{#1\undefined}
}%
\providecommand \@ifnum [1]{%
 \ifnum #1\expandafter \@firstoftwo
 \else \expandafter \@secondoftwo
 \fi
}%
\providecommand \@ifx [1]{%
 \ifx #1\expandafter \@firstoftwo
 \else \expandafter \@secondoftwo
 \fi
}%
\providecommand \natexlab [1]{#1}%
\providecommand \enquote  [1]{``#1''}%
\providecommand \bibnamefont  [1]{#1}%
\providecommand \bibfnamefont [1]{#1}%
\providecommand \citenamefont [1]{#1}%
\providecommand \href@noop [0]{\@secondoftwo}%
\providecommand \href [0]{\begingroup \@sanitize@url \@href}%
\providecommand \@href[1]{\@@startlink{#1}\@@href}%
\providecommand \@@href[1]{\endgroup#1\@@endlink}%
\providecommand \@sanitize@url [0]{\catcode `\\12\catcode `\$12\catcode
  `\&12\catcode `\#12\catcode `\^12\catcode `\_12\catcode `\%12\relax}%
\providecommand \@@startlink[1]{}%
\providecommand \@@endlink[0]{}%
\providecommand \url  [0]{\begingroup\@sanitize@url \@url }%
\providecommand \@url [1]{\endgroup\@href {#1}{\urlprefix }}%
\providecommand \urlprefix  [0]{URL }%
\providecommand \Eprint [0]{\href }%
\providecommand \doibase [0]{http://dx.doi.org/}%
\providecommand \selectlanguage [0]{\@gobble}%
\providecommand \bibinfo  [0]{\@secondoftwo}%
\providecommand \bibfield  [0]{\@secondoftwo}%
\providecommand \translation [1]{[#1]}%
\providecommand \BibitemOpen [0]{}%
\providecommand \bibitemStop [0]{}%
\providecommand \bibitemNoStop [0]{.\EOS\space}%
\providecommand \EOS [0]{\spacefactor3000\relax}%
\providecommand \BibitemShut  [1]{\csname bibitem#1\endcsname}%
\let\auto@bib@innerbib\@empty
\bibitem [{\citenamefont {Ando}(2000)}]{Ando2000}%
  \BibitemOpen
  \bibfield  {author} {\bibinfo {author} {\bibfnamefont {T.}~\bibnamefont
  {Ando}},\ }\href@noop {} {\bibfield  {journal} {\bibinfo  {journal} {J. Phys.
  Soc. Jpn.}\ }\textbf {\bibinfo {volume} {69}},\ \bibinfo {pages} {1757}
  (\bibinfo {year} {2000})}\BibitemShut {NoStop}%
\bibitem [{\citenamefont {Huertas-Hernando}\ \emph {et~al.}(2006)\citenamefont
  {Huertas-Hernando}, \citenamefont {Guinea},\ and\ \citenamefont
  {Brataas}}]{HuertasGB2006}%
  \BibitemOpen
  \bibfield  {author} {\bibinfo {author} {\bibfnamefont {D.}~\bibnamefont
  {Huertas-Hernando}}, \bibinfo {author} {\bibfnamefont {F.}~\bibnamefont
  {Guinea}}, \ and\ \bibinfo {author} {\bibfnamefont {A.}~\bibnamefont
  {Brataas}},\ }\href {\doibase 10.1103/PhysRevB.74.155426} {\bibfield
  {journal} {\bibinfo  {journal} {Phys. Rev. B}\ }\textbf {\bibinfo {volume}
  {74}},\ \bibinfo {pages} {155426} (\bibinfo {year} {2006})}\BibitemShut
  {NoStop}%
\bibitem [{\citenamefont {Jeong}\ and\ \citenamefont
  {Lee}(2009)}]{JeongJL2009}%
  \BibitemOpen
  \bibfield  {author} {\bibinfo {author} {\bibfnamefont {J.-S.}\ \bibnamefont
  {Jeong}}\ and\ \bibinfo {author} {\bibfnamefont {H.-W.}\ \bibnamefont
  {Lee}},\ }\href {\doibase 10.1103/PhysRevB.80.075409} {\bibfield  {journal}
  {\bibinfo  {journal} {Phys. Rev. B}\ }\textbf {\bibinfo {volume} {80}},\
  \bibinfo {pages} {075409} (\bibinfo {year} {2009})}\BibitemShut {NoStop}%
\bibitem [{\citenamefont {Izumida}\ \emph {et~al.}(2009)\citenamefont
  {Izumida}, \citenamefont {Sato},\ and\ \citenamefont {Saito}}]{Izumida2009}%
  \BibitemOpen
  \bibfield  {author} {\bibinfo {author} {\bibfnamefont {W.}~\bibnamefont
  {Izumida}}, \bibinfo {author} {\bibfnamefont {K.}~\bibnamefont {Sato}}, \
  and\ \bibinfo {author} {\bibfnamefont {R.}~\bibnamefont {Saito}},\ }\href
  {\doibase 10.1143/JPSJ.78.074707} {\bibfield  {journal} {\bibinfo  {journal}
  {J. Phys. Soc. Jpn.}\ }\textbf {\bibinfo {volume} {78}},\ \bibinfo {pages}
  {074707} (\bibinfo {year} {2009})}\BibitemShut {NoStop}%
\bibitem [{\citenamefont {Kuemmeth}\ \emph {et~al.}(2008)\citenamefont
  {Kuemmeth}, \citenamefont {Ilani}, \citenamefont {Ralph},\ and\ \citenamefont
  {McEuen}}]{Kuemmeth2008}%
  \BibitemOpen
  \bibfield  {author} {\bibinfo {author} {\bibfnamefont {F.}~\bibnamefont
  {Kuemmeth}}, \bibinfo {author} {\bibfnamefont {S.}~\bibnamefont {Ilani}},
  \bibinfo {author} {\bibfnamefont {D.}~\bibnamefont {Ralph}}, \ and\ \bibinfo
  {author} {\bibfnamefont {P.}~\bibnamefont {McEuen}},\ }\href@noop {}
  {\bibfield  {journal} {\bibinfo  {journal} {Nature}\ }\textbf {\bibinfo
  {volume} {452}},\ \bibinfo {pages} {448} (\bibinfo {year}
  {2008})}\BibitemShut {NoStop}%
\bibitem [{\citenamefont {Churchill}\ \emph
  {et~al.}(2009{\natexlab{a}})\citenamefont {Churchill}, \citenamefont
  {Kuemmeth}, \citenamefont {Harlow}, \citenamefont {Bestwick}, \citenamefont
  {Rashba}, \citenamefont {Flensberg}, \citenamefont {Stwertka}, \citenamefont
  {Taychatanapat}, \citenamefont {Watson},\ and\ \citenamefont
  {Marcus}}]{ChurchillFlensberg2009}%
  \BibitemOpen
  \bibfield  {author} {\bibinfo {author} {\bibfnamefont {H.~O.~H.}\
  \bibnamefont {Churchill}}, \bibinfo {author} {\bibfnamefont {F.}~\bibnamefont
  {Kuemmeth}}, \bibinfo {author} {\bibfnamefont {J.~W.}\ \bibnamefont
  {Harlow}}, \bibinfo {author} {\bibfnamefont {A.~J.}\ \bibnamefont
  {Bestwick}}, \bibinfo {author} {\bibfnamefont {E.~I.}\ \bibnamefont
  {Rashba}}, \bibinfo {author} {\bibfnamefont {K.}~\bibnamefont {Flensberg}},
  \bibinfo {author} {\bibfnamefont {C.~H.}\ \bibnamefont {Stwertka}}, \bibinfo
  {author} {\bibfnamefont {T.}~\bibnamefont {Taychatanapat}}, \bibinfo {author}
  {\bibfnamefont {S.~K.}\ \bibnamefont {Watson}}, \ and\ \bibinfo {author}
  {\bibfnamefont {C.~M.}\ \bibnamefont {Marcus}},\ }\href {\doibase
  10.1103/PhysRevLett.102.166802} {\bibfield  {journal} {\bibinfo  {journal}
  {Phys. Rev. Lett.}\ }\textbf {\bibinfo {volume} {102}},\ \bibinfo {pages}
  {166802} (\bibinfo {year} {2009}{\natexlab{a}})}\BibitemShut {NoStop}%
\bibitem [{\citenamefont {Jespersen}\ \emph {et~al.}(2011)\citenamefont
  {Jespersen}, \citenamefont {Grove-Rasmussen}, \citenamefont {Paaske},
  \citenamefont {Muraki}, \citenamefont {Fujisawa}, \citenamefont {Nygard},\
  and\ \citenamefont {Flensberg}}]{Jespersen2010}%
  \BibitemOpen
  \bibfield  {author} {\bibinfo {author} {\bibfnamefont {T.~S.}\ \bibnamefont
  {Jespersen}}, \bibinfo {author} {\bibfnamefont {K.}~\bibnamefont
  {Grove-Rasmussen}}, \bibinfo {author} {\bibfnamefont {J.}~\bibnamefont
  {Paaske}}, \bibinfo {author} {\bibfnamefont {K.}~\bibnamefont {Muraki}},
  \bibinfo {author} {\bibfnamefont {T.}~\bibnamefont {Fujisawa}}, \bibinfo
  {author} {\bibfnamefont {J.}~\bibnamefont {Nygard}}, \ and\ \bibinfo {author}
  {\bibfnamefont {K.}~\bibnamefont {Flensberg}},\ }\href@noop {} {\bibfield
  {journal} {\bibinfo  {journal} {Nat Phys}\ }\textbf {\bibinfo {volume} {7}},\
  \bibinfo {pages} {348} (\bibinfo {year} {2011})}\BibitemShut {NoStop}%
\bibitem [{\citenamefont {Bulaev}\ \emph {et~al.}(2008)\citenamefont {Bulaev},
  \citenamefont {Trauzettel},\ and\ \citenamefont {Loss}}]{Bulaev2008}%
  \BibitemOpen
  \bibfield  {author} {\bibinfo {author} {\bibfnamefont {D.~V.}\ \bibnamefont
  {Bulaev}}, \bibinfo {author} {\bibfnamefont {B.}~\bibnamefont {Trauzettel}},
  \ and\ \bibinfo {author} {\bibfnamefont {D.}~\bibnamefont {Loss}},\ }\href
  {\doibase 10.1103/PhysRevB.77.235301} {\bibfield  {journal} {\bibinfo
  {journal} {Phys. Rev. B}\ }\textbf {\bibinfo {volume} {77}},\ \bibinfo
  {pages} {1} (\bibinfo {year} {2008})}\BibitemShut {NoStop}%
\bibitem [{\citenamefont {Flensberg}\ and\ \citenamefont
  {Marcus}(2010)}]{FlensbergMarcus2010}%
  \BibitemOpen
  \bibfield  {author} {\bibinfo {author} {\bibfnamefont {K.}~\bibnamefont
  {Flensberg}}\ and\ \bibinfo {author} {\bibfnamefont {C.~M.}\ \bibnamefont
  {Marcus}},\ }\href {\doibase 10.1103/PhysRevB.81.195418} {\bibfield
  {journal} {\bibinfo  {journal} {Phys. Rev. B}\ }\textbf {\bibinfo {volume}
  {81}},\ \bibinfo {pages} {195418} (\bibinfo {year} {2010})}\BibitemShut
  {NoStop}%
\bibitem [{\citenamefont {Loss}\ and\ \citenamefont
  {DiVincenzo}(1998)}]{LossD98}%
  \BibitemOpen
  \bibfield  {author} {\bibinfo {author} {\bibfnamefont {D.}~\bibnamefont
  {Loss}}\ and\ \bibinfo {author} {\bibfnamefont {D.~P.}\ \bibnamefont
  {DiVincenzo}},\ }\href {\doibase 10.1103/PhysRevA.57.120} {\bibfield
  {journal} {\bibinfo  {journal} {Phys. Rev. A}\ }\textbf {\bibinfo {volume}
  {57}},\ \bibinfo {pages} {120} (\bibinfo {year} {1998})}\BibitemShut
  {NoStop}%
\bibitem [{\citenamefont {Petta}\ \emph {et~al.}(2005)\citenamefont {Petta},
  \citenamefont {Johnson}, \citenamefont {Taylor}, \citenamefont {Laird},
  \citenamefont {Yacoby}, \citenamefont {Lukin}, \citenamefont {Marcus},
  \citenamefont {Hanson},\ and\ \citenamefont {Gossard}}]{Petta2005}%
  \BibitemOpen
  \bibfield  {author} {\bibinfo {author} {\bibfnamefont {J.}~\bibnamefont
  {Petta}}, \bibinfo {author} {\bibfnamefont {A.}~\bibnamefont {Johnson}},
  \bibinfo {author} {\bibfnamefont {J.}~\bibnamefont {Taylor}}, \bibinfo
  {author} {\bibfnamefont {E.}~\bibnamefont {Laird}}, \bibinfo {author}
  {\bibfnamefont {A.}~\bibnamefont {Yacoby}}, \bibinfo {author} {\bibfnamefont
  {M.}~\bibnamefont {Lukin}}, \bibinfo {author} {\bibfnamefont
  {C.}~\bibnamefont {Marcus}}, \bibinfo {author} {\bibfnamefont
  {M.}~\bibnamefont {Hanson}}, \ and\ \bibinfo {author} {\bibfnamefont
  {A.}~\bibnamefont {Gossard}},\ }\href@noop {} {\bibfield  {journal} {\bibinfo
   {journal} {Science}\ }\textbf {\bibinfo {volume} {309}},\ \bibinfo {pages}
  {2180–2184} (\bibinfo {year} {2005})}\BibitemShut {NoStop}%
\bibitem [{\citenamefont {Hanson}\ \emph {et~al.}(2007)\citenamefont {Hanson},
  \citenamefont {Petta}, \citenamefont {Tarucha},\ and\ \citenamefont
  {Vandersypen}}]{HansonReview}%
  \BibitemOpen
  \bibfield  {author} {\bibinfo {author} {\bibfnamefont {R.}~\bibnamefont
  {Hanson}}, \bibinfo {author} {\bibfnamefont {J.~R.}\ \bibnamefont {Petta}},
  \bibinfo {author} {\bibfnamefont {S.}~\bibnamefont {Tarucha}}, \ and\
  \bibinfo {author} {\bibfnamefont {L.~M.~K.}\ \bibnamefont {Vandersypen}},\
  }\href {\doibase 10.1103/RevModPhys.79.1217} {\bibfield  {journal} {\bibinfo
  {journal} {Rev. Mod. Phys.}\ }\textbf {\bibinfo {volume} {79}},\ \bibinfo
  {pages} {1217} (\bibinfo {year} {2007})}\BibitemShut {NoStop}%
\bibitem [{\citenamefont {Churchill}\ \emph
  {et~al.}(2009{\natexlab{b}})\citenamefont {Churchill}, \citenamefont
  {Bestwick}, \citenamefont {Harlow}, \citenamefont {Marcos}, \citenamefont
  {Stwertka}, \citenamefont {Watson},\ and\ \citenamefont
  {Marcus}}]{Watson2009}%
  \BibitemOpen
  \bibfield  {author} {\bibinfo {author} {\bibfnamefont {H.}~\bibnamefont
  {Churchill}}, \bibinfo {author} {\bibfnamefont {A.}~\bibnamefont {Bestwick}},
  \bibinfo {author} {\bibfnamefont {J.}~\bibnamefont {Harlow}}, \bibinfo
  {author} {\bibfnamefont {D.}~\bibnamefont {Marcos}}, \bibinfo {author}
  {\bibfnamefont {C.}~\bibnamefont {Stwertka}}, \bibinfo {author}
  {\bibfnamefont {S.}~\bibnamefont {Watson}}, \ and\ \bibinfo {author}
  {\bibfnamefont {C.}~\bibnamefont {Marcus}},\ }\href {\doibase
  10.1038/NPHYS1247} {\bibfield  {journal} {\bibinfo  {journal} {Nat. Phys.}\
  }\textbf {\bibinfo {volume} {5}},\ \bibinfo {pages} {321} (\bibinfo {year}
  {2009}{\natexlab{b}})}\BibitemShut {NoStop}%
\bibitem [{\citenamefont {Wunsch}(2009)}]{WunschDQDCNT2009}%
  \BibitemOpen
  \bibfield  {author} {\bibinfo {author} {\bibfnamefont {B.}~\bibnamefont
  {Wunsch}},\ }\href {\doibase 10.1103/PhysRevB.79.235408} {\bibfield
  {journal} {\bibinfo  {journal} {Phys. Rev. B}\ }\textbf {\bibinfo {volume}
  {79}},\ \bibinfo {pages} {235408} (\bibinfo {year} {2009})}\BibitemShut
  {NoStop}%
\bibitem [{\citenamefont {von Stecher}\ \emph {et~al.}(2010)\citenamefont {von
  Stecher}, \citenamefont {Wunsch}, \citenamefont {Lukin}, \citenamefont
  {Demler},\ and\ \citenamefont {Rey}}]{WunschDQDCNT2010}%
  \BibitemOpen
  \bibfield  {author} {\bibinfo {author} {\bibfnamefont {J.}~\bibnamefont {von
  Stecher}}, \bibinfo {author} {\bibfnamefont {B.}~\bibnamefont {Wunsch}},
  \bibinfo {author} {\bibfnamefont {M.}~\bibnamefont {Lukin}}, \bibinfo
  {author} {\bibfnamefont {E.}~\bibnamefont {Demler}}, \ and\ \bibinfo {author}
  {\bibfnamefont {A.~M.}\ \bibnamefont {Rey}},\ }\href {\doibase
  10.1103/PhysRevB.82.125437} {\bibfield  {journal} {\bibinfo  {journal} {Phys.
  Rev. B}\ }\textbf {\bibinfo {volume} {82}},\ \bibinfo {pages} {125437}
  (\bibinfo {year} {2010})}\BibitemShut {NoStop}%
\bibitem [{\citenamefont {Weiss}\ \emph {et~al.}(2010)\citenamefont {Weiss},
  \citenamefont {Rashba}, \citenamefont {Kuemmeth}, \citenamefont {Churchill},\
  and\ \citenamefont {Flensberg}}]{WeissFlensberg2010}%
  \BibitemOpen
  \bibfield  {author} {\bibinfo {author} {\bibfnamefont {S.}~\bibnamefont
  {Weiss}}, \bibinfo {author} {\bibfnamefont {E.~I.}\ \bibnamefont {Rashba}},
  \bibinfo {author} {\bibfnamefont {F.}~\bibnamefont {Kuemmeth}}, \bibinfo
  {author} {\bibfnamefont {H.~O.~H.}\ \bibnamefont {Churchill}}, \ and\
  \bibinfo {author} {\bibfnamefont {K.}~\bibnamefont {Flensberg}},\ }\href
  {\doibase 10.1103/PhysRevB.82.165427} {\bibfield  {journal} {\bibinfo
  {journal} {Phys. Rev. B}\ }\textbf {\bibinfo {volume} {82}},\ \bibinfo
  {pages} {165427} (\bibinfo {year} {2010})}\BibitemShut {NoStop}%
\bibitem [{\citenamefont {P\'alyi}\ and\ \citenamefont
  {Burkard}(2010)}]{PalyiB10}%
  \BibitemOpen
  \bibfield  {author} {\bibinfo {author} {\bibfnamefont {A.}~\bibnamefont
  {P\'alyi}}\ and\ \bibinfo {author} {\bibfnamefont {G.}~\bibnamefont
  {Burkard}},\ }\href {\doibase 10.1103/PhysRevB.82.155424} {\bibfield
  {journal} {\bibinfo  {journal} {Phys. Rev. B}\ }\textbf {\bibinfo {volume}
  {82}},\ \bibinfo {pages} {155424} (\bibinfo {year} {2010})}\BibitemShut
  {NoStop}%
\bibitem [{\citenamefont {Schulten}\ and\ \citenamefont
  {Wolynes}(1978)}]{SchultenWolynes1978}%
  \BibitemOpen
  \bibfield  {author} {\bibinfo {author} {\bibfnamefont {K.}~\bibnamefont
  {Schulten}}\ and\ \bibinfo {author} {\bibfnamefont {P.~G.}\ \bibnamefont
  {Wolynes}},\ }\href {\doibase 10.1063/1.436135} {\bibfield  {journal}
  {\bibinfo  {journal} {J. Chem. Phys.}\ }\textbf {\bibinfo {volume} {68}},\
  \bibinfo {pages} {3292} (\bibinfo {year} {1978})}\BibitemShut {NoStop}%
\bibitem [{\citenamefont {Merkulov}\ \emph {et~al.}(2002)\citenamefont
  {Merkulov}, \citenamefont {Efros},\ and\ \citenamefont
  {Rosen}}]{MerkulovHyperfine}%
  \BibitemOpen
  \bibfield  {author} {\bibinfo {author} {\bibfnamefont {I.~A.}\ \bibnamefont
  {Merkulov}}, \bibinfo {author} {\bibfnamefont {A.~L.}\ \bibnamefont {Efros}},
  \ and\ \bibinfo {author} {\bibfnamefont {M.}~\bibnamefont {Rosen}},\ }\href
  {\doibase 10.1103/PhysRevB.65.205309} {\bibfield  {journal} {\bibinfo
  {journal} {Phys. Rev. B}\ }\textbf {\bibinfo {volume} {65}},\ \bibinfo
  {pages} {205309} (\bibinfo {year} {2002})}\BibitemShut {NoStop}%
\bibitem [{\citenamefont {Coish}\ and\ \citenamefont {Loss}(2005)}]{CoishDQD}%
  \BibitemOpen
  \bibfield  {author} {\bibinfo {author} {\bibfnamefont {W.~A.}\ \bibnamefont
  {Coish}}\ and\ \bibinfo {author} {\bibfnamefont {D.}~\bibnamefont {Loss}},\
  }\href {\doibase 10.1103/PhysRevB.72.125337} {\bibfield  {journal} {\bibinfo
  {journal} {Phys. Rev. B}\ }\textbf {\bibinfo {volume} {72}},\ \bibinfo
  {pages} {125337} (\bibinfo {year} {2005})}\BibitemShut {NoStop}%
\bibitem [{\citenamefont {Taylor}\ \emph {et~al.}(2007)\citenamefont {Taylor},
  \citenamefont {Petta}, \citenamefont {Johnson}, \citenamefont {Yacoby},
  \citenamefont {Marcus},\ and\ \citenamefont
  {Lukin}}]{TaylorDephasingTheory2007}%
  \BibitemOpen
  \bibfield  {author} {\bibinfo {author} {\bibfnamefont {J.~M.}\ \bibnamefont
  {Taylor}}, \bibinfo {author} {\bibfnamefont {J.~R.}\ \bibnamefont {Petta}},
  \bibinfo {author} {\bibfnamefont {A.~C.}\ \bibnamefont {Johnson}}, \bibinfo
  {author} {\bibfnamefont {A.}~\bibnamefont {Yacoby}}, \bibinfo {author}
  {\bibfnamefont {C.~M.}\ \bibnamefont {Marcus}}, \ and\ \bibinfo {author}
  {\bibfnamefont {M.~D.}\ \bibnamefont {Lukin}},\ }\href {\doibase
  10.1103/PhysRevB.76.035315} {\bibfield  {journal} {\bibinfo  {journal} {Phys.
  Rev. B}\ }\textbf {\bibinfo {volume} {76}},\ \bibinfo {pages} {035315}
  (\bibinfo {year} {2007})}\BibitemShut {NoStop}%
\bibitem [{\citenamefont {\ifmmode~\mbox{\c{C}}\else
  \c{C}\fi{}ak\ifmmode\imath\else\i\fi{}r}\ and\ \citenamefont
  {Takagahara}(2008)}]{Cakr2008}%
  \BibitemOpen
  \bibfield  {author} {\bibinfo {author} {\bibfnamefont {O.}~\bibnamefont
  {\ifmmode~\mbox{\c{C}}\else \c{C}\fi{}ak\ifmmode\imath\else\i\fi{}r}}\ and\
  \bibinfo {author} {\bibfnamefont {T.}~\bibnamefont {Takagahara}},\ }\href
  {\doibase 10.1103/PhysRevB.77.115304} {\bibfield  {journal} {\bibinfo
  {journal} {Phys. Rev. B}\ }\textbf {\bibinfo {volume} {77}},\ \bibinfo
  {pages} {115304} (\bibinfo {year} {2008})}\BibitemShut {NoStop}%
\bibitem [{\citenamefont {S\"arkk\"a}\ and\ \citenamefont
  {Harju}(2009)}]{Harju2009}%
  \BibitemOpen
  \bibfield  {author} {\bibinfo {author} {\bibfnamefont {J.}~\bibnamefont
  {S\"arkk\"a}}\ and\ \bibinfo {author} {\bibfnamefont {A.}~\bibnamefont
  {Harju}},\ }\href {\doibase 10.1103/PhysRevB.80.045323} {\bibfield  {journal}
  {\bibinfo  {journal} {Phys. Rev. B}\ }\textbf {\bibinfo {volume} {80}},\
  \bibinfo {pages} {045323} (\bibinfo {year} {2009})}\BibitemShut {NoStop}%
\bibitem [{\citenamefont {Culcer}\ \emph {et~al.}(2010)\citenamefont {Culcer},
  \citenamefont {Cywi\ifmmode~\acute{n}\else \'{n}\fi{}ski}, \citenamefont
  {Li}, \citenamefont {Hu},\ and\ \citenamefont
  {Das~Sarma}}]{CulcerSarmaQD2010}%
  \BibitemOpen
  \bibfield  {author} {\bibinfo {author} {\bibfnamefont {D.}~\bibnamefont
  {Culcer}}, \bibinfo {author} {\bibfnamefont {L.}~\bibnamefont
  {Cywi\ifmmode~\acute{n}\else \'{n}\fi{}ski}}, \bibinfo {author}
  {\bibfnamefont {Q.}~\bibnamefont {Li}}, \bibinfo {author} {\bibfnamefont
  {X.}~\bibnamefont {Hu}}, \ and\ \bibinfo {author} {\bibfnamefont
  {S.}~\bibnamefont {Das~Sarma}},\ }\href {\doibase 10.1103/PhysRevB.82.155312}
  {\bibfield  {journal} {\bibinfo  {journal} {Phys. Rev. B}\ }\textbf {\bibinfo
  {volume} {82}},\ \bibinfo {pages} {155312} (\bibinfo {year}
  {2010})}\BibitemShut {NoStop}%
\bibitem [{\citenamefont {Fischer}\ \emph {et~al.}(2009)\citenamefont
  {Fischer}, \citenamefont {Trauzettel},\ and\ \citenamefont
  {Loss}}]{Fischer2009}%
  \BibitemOpen
  \bibfield  {author} {\bibinfo {author} {\bibfnamefont {J.}~\bibnamefont
  {Fischer}}, \bibinfo {author} {\bibfnamefont {B.}~\bibnamefont {Trauzettel}},
  \ and\ \bibinfo {author} {\bibfnamefont {D.}~\bibnamefont {Loss}},\ }\href
  {\doibase 10.1103/PhysRevB.80.155401} {\bibfield  {journal} {\bibinfo
  {journal} {Phys. Rev. B}\ }\textbf {\bibinfo {volume} {80}},\ \bibinfo
  {pages} {155401} (\bibinfo {year} {2009})}\BibitemShut {NoStop}%
\bibitem [{\citenamefont {P\'alyi}\ and\ \citenamefont
  {Burkard}(2009)}]{PalyiB09}%
  \BibitemOpen
  \bibfield  {author} {\bibinfo {author} {\bibfnamefont {A.}~\bibnamefont
  {P\'alyi}}\ and\ \bibinfo {author} {\bibfnamefont {G.}~\bibnamefont
  {Burkard}},\ }\href {\doibase 10.1103/PhysRevB.80.201404} {\bibfield
  {journal} {\bibinfo  {journal} {Phys. Rev. B}\ }\textbf {\bibinfo {volume}
  {80}},\ \bibinfo {pages} {201404} (\bibinfo {year} {2009})}\BibitemShut
  {NoStop}%
\bibitem [{\citenamefont {Reynoso}\ and\ \citenamefont
  {Flensberg}(2010)}]{PosterKonstanz2010}%
  \BibitemOpen
  \bibfield  {author} {\bibinfo {author} {\bibfnamefont {A.~A.}\ \bibnamefont
  {Reynoso}}\ and\ \bibinfo {author} {\bibfnamefont {K.}~\bibnamefont
  {Flensberg}},\ }\href@noop {} {} (\bibinfo {year} {2010}),\ \bibinfo {note}
  {{\rm Poster at} {\it Spin-based quantum information processing}, {\rm
  Konstanz, Germany.}}\BibitemShut {Stop}%
\bibitem [{\citenamefont {Saito}\ \emph {et~al.}(1998)\citenamefont {Saito},
  \citenamefont {Dresselhaus},\ and\ \citenamefont
  {Dresselhaus}}]{DresselhausBook}%
  \BibitemOpen
  \bibfield  {author} {\bibinfo {author} {\bibfnamefont {R.}~\bibnamefont
  {Saito}}, \bibinfo {author} {\bibfnamefont {G.}~\bibnamefont {Dresselhaus}},
  \ and\ \bibinfo {author} {\bibfnamefont {M.~S.}\ \bibnamefont
  {Dresselhaus}},\ }\href@noop {} {\emph {\bibinfo {title} {{Physical
  Properties of Carbon Nanotubes}}}}\ (\bibinfo  {publisher} {{World Scientific
  Publishing Company}},\ \bibinfo {year} {1998})\BibitemShut {NoStop}%
\bibitem [{\citenamefont {Ando}(2005)}]{AndoReviewCNT}%
  \BibitemOpen
  \bibfield  {author} {\bibinfo {author} {\bibfnamefont {T.}~\bibnamefont
  {Ando}},\ }\href {\doibase 10.1143/JPSJ.74.777} {\bibfield  {journal}
  {\bibinfo  {journal} {J. Phys. Soc. Jpn.}\ }\textbf {\bibinfo {volume}
  {74}},\ \bibinfo {pages} {777} (\bibinfo {year} {2005})}\BibitemShut
  {NoStop}%
\bibitem [{\citenamefont {Minot}\ \emph {et~al.}(2004)\citenamefont {Minot},
  \citenamefont {Yaish}, \citenamefont {Sazonova},\ and\ \citenamefont
  {McEuen}}]{Minot2004}%
  \BibitemOpen
  \bibfield  {author} {\bibinfo {author} {\bibfnamefont {E.~D.}\ \bibnamefont
  {Minot}}, \bibinfo {author} {\bibfnamefont {Y.}~\bibnamefont {Yaish}},
  \bibinfo {author} {\bibfnamefont {V.}~\bibnamefont {Sazonova}}, \ and\
  \bibinfo {author} {\bibfnamefont {P.~L.}\ \bibnamefont {McEuen}},\
  }\href@noop {} {\bibfield  {journal} {\bibinfo  {journal} {Nature}\ }\textbf
  {\bibinfo {volume} {428}},\ \bibinfo {pages} {536} (\bibinfo {year}
  {2004})}\BibitemShut {NoStop}%
\bibitem [{\citenamefont {Recher}\ \emph {et~al.}(2009)\citenamefont {Recher},
  \citenamefont {Nilsson}, \citenamefont {Burkard},\ and\ \citenamefont
  {Trauzettel}}]{RecherNBT09}%
  \BibitemOpen
  \bibfield  {author} {\bibinfo {author} {\bibfnamefont {P.}~\bibnamefont
  {Recher}}, \bibinfo {author} {\bibfnamefont {J.}~\bibnamefont {Nilsson}},
  \bibinfo {author} {\bibfnamefont {G.}~\bibnamefont {Burkard}}, \ and\
  \bibinfo {author} {\bibfnamefont {B.}~\bibnamefont {Trauzettel}},\ }\href
  {\doibase 10.1103/PhysRevB.79.085407} {\bibfield  {journal} {\bibinfo
  {journal} {Phys. Rev. B}\ }\textbf {\bibinfo {volume} {79}},\ \bibinfo
  {pages} {085407} (\bibinfo {year} {2009})}\BibitemShut {NoStop}%
\end{thebibliography}%
\end{document}